\documentclass[a4paper,11pt]{article}
\usepackage{jheppub}
\usepackage[utf8]{inputenc}
\usepackage{graphicx,cancel,float}
\usepackage{amssymb}
\usepackage{textcomp}
\usepackage{amsmath}
\usepackage{tabularx}
\usepackage{bm}
\usepackage{times}
\usepackage{color}
\usepackage{slashed}
\usepackage{multirow}
\usepackage{verbatim}
\usepackage{epsfig,epstopdf}
\usepackage{subfigure}
\usepackage{youngtab}
\allowdisplaybreaks[4]

\def\lsim{\mathrel{\raise.3ex\hbox{$<$\kern-.75em\lower1ex\hbox{$\sim$}}}}
\def\gsim{\mathrel{\raise.3ex\hbox{$>$\kern-.75em\lower1ex\hbox{$\sim$}}}}

\definecolor{orange}{rgb}{1,0.5,0}

\preprint{CPPC-2025-02}
\subheader{\it \today}

\title{\Large{\bf Baryon-number-violating nucleon decays in sterile neutrino effective field theories}}

\author[a]{Tong Li}
\emailAdd{litong@nankai.edu.cn}
\author[b]{Michael A. Schmidt}
\emailAdd{m.schmidt@unsw.edu.au}
\author[c]{Chang-Yuan Yao}
\emailAdd{c.yao@soton.ac.uk}

\affiliation[a]{School of Physics, Nankai University, 94 Weijin Road, Tianjin 300071, China}
\affiliation[b]{
Sydney Consortium for Particle Physics and Cosmology,\\
School of Physics, The University of New South Wales, Sydney, New South Wales 2052, Australia
}
\affiliation[c]{
School of Physics and Astronomy, University of Southampton, Southampton SO17 1BJ, United Kingdom}

\abstract{
The search for baryon-number-violating (BNV) nucleon decay provides an intriguing probe of new physics beyond the Standard Model (SM). Future neutrino experiments will improve the sensitivity to BNV nucleon decays and can serve to search for dark particles.
In this work, we study the sterile neutrino effective field theories (EFTs) with baryon number violation and the impact of light sterile neutrino on BNV nucleon decays. We revisit the dimension-6 and dimension-7 EFT operator bases with $|\Delta (B-L)|=2$ or $|\Delta (B-L)|=0$. They are then matched to the baryon chiral perturbation theory. We obtain the effective chiral Lagrangian at low energies and the BNV interactions between the sterile neutrino and baryons and mesons. The rates of nucleon decay to SM neutrinos or a sterile neutrino are calculated. We then show the constraints on the ultraviolet scale from nucleon decay search at Super-K. The correlation of two EFT operators and the dependence on the sterile neutrino mass are also investigated.
}

\begin{document}

\maketitle
\setcounter{page}{2}

\newpage

\section{Introduction}
\label{sec:Intro}

The baryon number ($B$) is an accidental global symmetry in the Standard Model (SM) and its conservation guarantees the stability of the proton.
The observation of baryon number violation in proton decay would imply the existence of new physics (NP) beyond the SM. 
The neutrino experiments with a huge number of nucleons in their detector provide an advantageous environment to probe baryon number violation in nucleon decays. Super-K~\cite{Takhistov:2016eqm} has performed many searches for nucleon decay in an SM lepton plus a meson, e.g.~proton decay $p\to e^+ (\mu^+) \pi^0$, but no evidence has been found to date~\cite{Super-Kamiokande:2020wjk,Super-Kamiokande:2014otb,Super-Kamiokande:2017gev,Super-Kamiokande:2013rwg,Super-Kamiokande:2005lev,Super-Kamiokande:2012zik}. The future neutrino experiments, such as Hyper-K~\cite{Hyper-Kamiokande:2018ofw}, DUNE~\cite{DUNE:2016evb,DUNE:2020ypp} and JUNO~\cite{JUNO:2015zny}, will improve the search sensitivity based on larger exposure and different detection technologies.

Many theoretical frameworks predict baryon-number-violating (BNV) nucleon decays, including Grand Unified Theories (GUTs)~\cite{Georgi:1974sy,Fritzsch:1974nn,Langacker:1980js,Nath:2006ut} and supersymmetry (SUSY)~\cite{Farrar:1978xj,Dimopoulos:1981zb,Sakai:1981gr}. BNV nucleon decays are also motivated from the viewpoint of effective field theories (EFTs)~\cite{deGouvea:2014lva,Heeck:2019kgr,Girmohanta:2019xya,Antusch:2020ztu,He:2021sbl,Fajfer:2023gfi,Beneito:2023xbk,Gargalionis:2024nij,Beneke:2024hox,Gisbert:2024sjw} (see Refs.~\cite{FileviezPerez:2022ypk,Dev:2022jbf,Ohlsson:2023ddi} for recent reviews), since the baryon number is an accidental symmetry of the SM. The large lower bounds on the lifetime of the proton place stringent limits on the very massive mediators and the NP energy scale of the BNV EFT operators.
The lower limits on the NP energy scale are found to be $\sim 10^{15}$ GeV and $10^{11}$ GeV for dimension-6 or dimension-7 BNV operators in the SM effective field theory (SMEFT)~\cite{Beneito:2023xbk}, respectively.

In addition to baryon number violation through the exchange of heavy particles in GUT or SUSY frameworks, NP frameworks with light states are also well motivated in light of the search for noncanonical nucleon decays~\cite{Davoudiasl:2014gfa,Helo:2018bgb,Barducci:2018rlx,Heeck:2020nbq,Davoudiasl:2023peu,Liang:2023yta,Fridell:2023tpb,Li:2024liy}.
A new state with feeble interaction and negligible mass may escape the detector and mimic the standard BNV decay modes with SM neutrinos in the final states. The decay to a new massive state has distinct kinematics, which makes it distinguishable from the canonical BNV nucleon decay~\cite{Davoudiasl:2014gfa,Fridell:2023tpb,Li:2024liy}. These decays are best described within SMEFT extended with a new light state.
In recent years, the EFT operator bases with a weakly interacting light particle have been widely studied~\cite{Song:2023lxf,Song:2023jqm,Grojean:2023tsd,delAguila:2008ir,Aparici:2009fh,Bhattacharya:2015vja,Liao:2016qyd,Chala:2020vqp,Li:2020lba,Li:2021tsq}. The sterile neutrino is one of the well-studied light states beyond the SM. One (or more than one) light SM singlet fermion may be introduced to explain non-zero neutrino masses. The small mixing with the SM neutrinos leads to its feeble interaction with ordinary matter in the SM. SMEFT extended with sterile neutrinos (named $\nu$SMEFT) and the corresponding low-energy effective field theory (LEFT) were constructed in a series of papers much earlier~\cite{delAguila:2008ir,Aparici:2009fh,Bhattacharya:2015vja,Liao:2016qyd,Chala:2020vqp,Li:2020lba,Li:2021tsq}. The BNV operators in $\nu$SMEFT induce nucleon decays to a meson and a massive sterile neutrino beyond the modes with SM neutrinos~\cite{Helo:2018bgb,Fridell:2023tpb}~\footnote{In R-parity violating SUSY theories similar decay modes have been discussed with a light neutralino~\cite{Hall:1983id,Chang:1996sw,Domingo:2024qoj}, light axino or gravitino~\cite{Choi:1996nk,Choi:1998ak} instead of a massive sterile neutrino.}.

In this work, we performed a comprehensive study of the impact of light (sterile) neutrinos on BNV nucleon decays within SMEFT and $\nu$SMEFT. We revisit their BNV operator bases and focus on the dimension-6 and dimension-7 SMEFT/$\nu$SMEFT operators with $|\Delta (B-L)|=2$ or $|\Delta (B-L)|=0$.
In particular, we corrected some long-standing mistakes about their symmetry properties in the literature and in a public code. The calculation of BNV nucleon decays is based on a tower of EFTs. The BNV SMEFT ($\nu$SMEFT) operators are defined at the ultraviolet (UV) scale $\Lambda$, matched to LEFT at the EW scale and to baryon chiral perturbation theory (BChPT)~\cite{Claudson:1981gh,Jenkins:1991ne} at the nuclear scale based on their representations under the quark flavor group $\rm SU(3)_L\times SU(3)_R$. The leading quantum corrections from gauge loops and the top Yukawa coupling are included through renormalization group (RG) evolution of the BNV operators between the UV scale $\Lambda$ and the nuclear scale $\sim 2$ GeV.
The BChPT interactions induce two-body nucleon decays with either SM neutrinos or the sterile neutrino in the final states and thus the experimental signatures are a meson and missing energy. Table~\ref{tab:process} lists the relevant decay modes. The nucleon lifetime bounds place upper (lower) limits on the corresponding WCs (UV scale) of SMEFT and $\nu$SMEFT for certain two-body decays.

\begin{table}[tb!]
\centering\def\arraystretch{1.2}
\begin{tabular}{c|c||c|c}
\hline
$N\to P\nu$  &
lifetime ($10^{33}$ years)
&
$N\to P\nu$ &
lifetime ($10^{33}$ years)
\\
\hline
$p\to \pi^+ \nu$ &  $0.39$~\cite{Super-Kamiokande:2013rwg}
&
$p\to K^+ \nu$  & $6.61$\cite{Super-Kamiokande:2014otb}
\\
$n\to \pi^0 \nu$  & $1.1$~\cite{Super-Kamiokande:2013rwg}
&
$n\to K_S^0 \nu$  & $0.26$~\cite{Super-Kamiokande:2005lev}
\\
$n\to \eta^0 \nu$ & $0.158$~\cite{McGrew:1999nd}
&
&
\\
\hline
\end{tabular}
\caption{
Lower limits on the lifetimes of baryon-number-violating two-body nucleon decays with a neutrino $\nu$ or antineutrino $\bar\nu$ and a pseudoscalar meson $P$ in the final states. As the nature of the final state neutrino is not experimentally determined, we collectively denote them by $\nu$. All limits are at 90\% C.L. .
}
\label{tab:process}
\end{table}

This paper is organized as follows. In Sec.~\ref{sec:nuEFT}, we describe the effective field theory frameworks for baryon number violation at the relevant energy scales.
The formulas of BNV nucleon decay rates are presented in Sec.~\ref{sec:BNV}. In Sec.~\ref{sec:Results} we discuss the nucleon decay searches with missing energy and show the numerical results for the two-body BNV nucleon decays to neutrino. Our main conclusions are summarized in Sec.~\ref{sec:Con}.

\section{Sterile neutrinos and BNV operators in effective field theories}
\label{sec:nuEFT}

The neutrino Yukawa interaction generally induces mixing between the active left-handed neutrinos $\nu_L$ and the sterile right-handed neutrinos $N$. The mixing is described by a unitary $(3+n)\times (3+n)$ matrix $\mathbb{N}$. The gauge interaction eigenstates and the mass eigenstates of neutrinos are related through the matrix $\mathbb{N}$ as follows~\cite{Atre:2009rg}
\begin{eqnarray}
\left(
  \begin{array}{c}
    \nu_L \\
    (N^c)_L \\
  \end{array}
\right) = \mathbb{N}\left(
  \begin{array}{c}
    \nu_L \\
    (N^c)_L \\
  \end{array}
\right)_{mass}, \qquad\qquad \mathbb{N}_{(3+n)\times (3+n)}= \left(
  \begin{array}{cc}
    U_{3\times 3} & K_{3\times n} \\
    X_{n\times 3} & Y_{n\times n} \\
  \end{array}
\right)\;,
\label{eq:nuMixDefs}
\end{eqnarray}
where $n$ denotes the degree of freedom of the sterile neutrino. We consider the case in which only one massive sterile neutrino is kinematically accessible. The unitarity condition for $\mathbb{N}$ results in the following relations
\begin{eqnarray}
&&UU^\dagger + KK^\dagger = U^\dagger U + X^\dagger X = I_{3\times 3}\;,\\
&&XX^\dagger + YY^\dagger = K^\dagger K + Y^\dagger Y = I_{n\times n}\;,
\end{eqnarray}
where $U$ and $Y$ parametrically satisfy $UU^\dagger, YY^\dagger\sim I$ with $I$ being the identity matrix, whereas the off-diagonal entries are suppressed $KK^\dagger, X^\dagger X\sim m_\nu/m_N$. In the following analysis, we approximately take $U$ and $Y$ as unitary matrices and neglect $K$ and $X$ matrices.

There is an active experimental program dedicated to the search for sterile neutrinos or heavy neutral leptons, as recently summarized in Ref.~\cite{Abdullahi:2022jlv}. Demanding the sterile neutrinos to contribute to active neutrino masses via the seesaw mechanism~\cite{Minkowski:1977sc} translates into a lower bound on the mass of sterile neutrinos. Big bang nucleosynthesis (BBN) imposes an upper bound on the lifetime of sterile neutrinos of $\tau_N<0.02s$~\cite{Boyarsky:2020dzc} which translates into a lower bound on the active-sterile mixing angle. Beam dump and collider experiments provide an upper limit on the active-sterile mixing angle. The BBN constraint together with the kinematic threshold of $K\to \pi \nu N$ excludes sterile neutrinos lighter than 330--360 MeV except for a small mass window between 120--140 MeV.~\cite{Bondarenko:2021cpc}

\subsection{BNV operators in SMEFT, $\nu$SMEFT and LEFT}
\label{sec:SMEFT}

We consider the BNV operators in the SMEFT and the $\nu$SMEFT extended with sterile neutrinos, as shown in Tab.~\ref{tab:nuSMEFT}. The effective Lagrangian reads as
\begin{eqnarray}
\mathcal{L}_{\rm SMEFT/\nu SMEFT}=\sum_i C_{i,prst} \mathcal{O}_{i,prst}+h.c.\;,
\end{eqnarray}
where $C_i$ denotes the WCs for the BNV operators and $p, r, s, t$ refer to the generation indices of the fermions to the order in the operators of Tab.~\ref{tab:nuSMEFT}. We work in the up-type quark basis~\footnote{In the up-type quark basis, the up-type
quark mass matrix is assumed to be diagonal and the CKM matrix diagonalizes the down-type quark mass matrix.} and include the dimension-6 SMEFT operators with $|\Delta (B-L)|=0$ (top-left)~\cite{Jenkins:2017jig}, dimension-7 operators with $|\Delta (B-L)|=2$ (bottom-left)~\cite{Liao:2020zyx}, dimension-6
$\nu$SMEFT operators with $|\Delta (B-L)|=0$ (top-right)~\cite{Li:2020lba,Li:2021tsq}
and dimension-7 $\nu$SMEFT operators with $|\Delta (B-L)|=2$ (bottom-right)~\cite{Liao:2016qyd,Li:2020lba,Li:2021tsq}. The operators with checkmark ``$\checkmark$'' are relevant for the nucleon decay to neutrino and we will focus on them below.

\begin{table}[htb!]
\centering
\renewcommand{\arraystretch}{1.2}
\begin{tabular}{c|c|c|c}
\hline
Name & SMEFT Operator
&
Name & $\nu$SMEFT Operator\\
\hline
$\mathcal{O}_{duLQ}$ &
$\epsilon^{\alpha\beta\gamma}\epsilon^{ij}(\overline{d^{c}_{\alpha}} u_{\beta})(\overline{L^c_j}Q_{\gamma i})$ $\checkmark$
&
$\mathcal{O}_{QQNd}$ \Yboxdim{6pt}\yng(2)&
$\epsilon_{ij}\epsilon^{\alpha\beta\gamma}(\overline{Q^{ic}_{\alpha}} Q^j_{\beta})(\overline{N^c} d_\gamma)$ $\checkmark$
\\
$\mathcal{O}_{QQLQ}$ \Yboxdim{6pt}\yng(1,1,1)+\yng(2,1)+\yng(3)  &
$\epsilon^{\alpha\beta\gamma}\epsilon^{il}\epsilon^{jk}(\overline{Q^{c}_{\alpha i}} Q_{\beta j})(\overline{L^c_l} Q_{\gamma k})$ $\checkmark$
&
$\mathcal{O}_{udNd}$ \Yboxdim{6pt}\yng(2)+\yng(1,1)&
$\epsilon^{\alpha\beta\gamma}(\overline{u^{c}_{\alpha}} d_{\beta})(\overline{N^c} d_\gamma)$
$\checkmark$
\\
\hline
\hline
$\mathcal{O}_{dDdLQ}$ \Yboxdim{6pt}\yng(2)&
$\epsilon^{\alpha\beta\gamma}(\overline{d^c_\alpha}i \overleftrightarrow{D}^\mu d_\beta)(\overline{L}\gamma_\mu Q_\gamma)$
&
$\mathcal{O}_{dDNu}$ \Yboxdim{6pt} \yng(2)&
$\epsilon^{\alpha\beta\gamma}(\overline{d^c_\alpha}i \overleftrightarrow{D}^\mu d_\beta)(\overline{N}\gamma_\mu u_\gamma)$
\\
$\mathcal{O}_{dDded}$ \Yboxdim{6pt} \yng(3)&
$\epsilon^{\alpha\beta\gamma}(\overline{d^c_\alpha}i \overleftrightarrow{D}^\mu d_\beta)(\overline{e}\gamma_\mu d_\gamma)$
&
$\mathcal{O}_{QDNd}$ \Yboxdim{6pt} \yng(1,1)&
$\epsilon_{ij}\epsilon^{\alpha\beta\gamma} (\overline{Q^{ic}_\alpha} i \overleftrightarrow{D}^\mu Q^j_\beta)(\overline{N}\gamma_\mu d_\gamma)$
\\
$\mathcal{O}_{udLdH}$ \Yboxdim{6pt} \yng(2)+\yng(1,1) &
$\epsilon^{\alpha\beta\gamma}(\overline{u^c_\alpha}d_\beta)(\overline{L}d_\gamma)\tilde{H}$ $\checkmark$
&
$\mathcal{O}_{HdNQ}$ \Yboxdim{6pt} \yng(1,1)&
$\epsilon^{\alpha\beta\gamma}\tilde{H}^\dagger(\overline{d^c_\alpha}d_\beta)(\overline{N}Q_\gamma)$ $\checkmark$
\\
$\mathcal{O}_{ddLdH}$ \Yboxdim{6pt} \yng(2,1) &
$\epsilon^{\alpha\beta\gamma}(\overline{d^c_\alpha}d_\beta)(\overline{L}d_\gamma)H$
&
$\mathcal{O}_{HQNQ}$ \Yboxdim{6pt} \yng(1,1,1)+\yng(2,1)+\yng(3)&
$\epsilon_{ij}\epsilon^{\alpha\beta\gamma}H^\dagger(\overline{Q^c_\alpha}Q^i_\beta)(\overline{N}Q^j_\gamma)$ $\checkmark$
\\
$\mathcal{O}_{ddeQH}$ \Yboxdim{6pt} \yng(1,1) &
$\epsilon^{\alpha\beta\gamma}\epsilon_{ij}(\overline{d^c_\alpha}d_\beta)(\overline{e}Q^i_\gamma)\tilde{H}^j$
&
$\mathcal{O}_{HduNQ}$ &
$\epsilon^{\alpha\beta\gamma}H^\dagger(\overline{d^c_\alpha}u_\beta)(\overline{N}Q_\gamma)$ $\checkmark$
\\
$\mathcal{O}_{QQLdH}$ \Yboxdim{6pt} \yng(1,1) + \yng(2)  &
$\epsilon^{\alpha\beta\gamma}\epsilon_{ij}(\overline{Q^c_\alpha}Q^i_\beta)(\overline{L}d_\gamma)\tilde{H}^j$ $\checkmark$
&&
\\
\hline
\end{tabular}
\caption{BNV operators with (sterile) neutrinos in the SMEFT (left) and the $\nu$SMEFT (right).
The dimension-6 (7) operators with $|\Delta (B-L)|=0~(2)$ are above (below) the horizontal double-line.
We define $\tilde{H}^i=\epsilon^{ij}H^{\ast j}$.
Operators marked with a checkmark are relevant for the discussion below.
}
\label{tab:nuSMEFT}
\end{table}

\begin{table}[htb!]
\centering
\renewcommand{\arraystretch}{1.2}
\begin{tabular}{c|c|c|c}
\hline
Name & Operator 
& Matching & $SU(3)_L\times SU(3)_R$\\
\hline
$\mathcal{O}^{S,LL}_{udd}$ \Yboxdim{6pt}\yng(2)+\yng(1,1) & 
$\epsilon^{\alpha\beta\gamma}(\overline{u^{c}_{L\alpha}} d_{L\beta})(\overline{\nu^c_{L}} d_{L\gamma})$ & $\mathcal{O}_{QQLQ}$, $\mathcal{O}_{HQNQ}$ & $(8,1)$\\
$\mathcal{O}^{S,RL}_{dud}$ & 
$\epsilon^{\alpha\beta\gamma}(\overline{d^{c}_{R\alpha}} u_{R\beta})(\overline{\nu^c_{L}} d_{L\gamma})$ & $\mathcal{O}_{duLQ}$, $\mathcal{O}_{HduNQ}$ & $(3,\bar{3})$ \\
$\mathcal{O}^{S,RL}_{ddu}$ \Yboxdim{6pt}\yng(1,1) & 
$\epsilon^{\alpha\beta\gamma}(\overline{d^{c}_{R\alpha}} d_{R\beta})(\overline{\nu^c_{L}} u_{L\gamma})$ & $\mathcal{O}_{HdNQ}$ & $(3,\bar{3})$\\
$\mathcal{O}^{S,LR}_{udd}$ & $\epsilon^{\alpha\beta\gamma}(\overline{u^{c}_{L\alpha}} d_{L\beta})(\overline{\nu_L} d_{R\gamma})$ & $\mathcal{O}_{QQLdH}$, $\mathcal{O}_{QQNd}$ & $(\bar{3},3)$\\
$\mathcal{O}^{S,RR}_{udd}$ \Yboxdim{6pt} \yng(2)+\yng(1,1) & $\epsilon^{\alpha\beta\gamma}(\overline{u^{c}_{R\alpha}} d_{R\beta})(\overline{\nu_L} d_{R\gamma})$ & $\mathcal{O}_{udLdH}$, $\mathcal{O}_{udNd}$ & $(1,8)$\\
$\mathcal{O}^{S,LR}_{ddu}$ \Yboxdim{6pt} \yng(1,1) & $\epsilon^{\alpha\beta\gamma}(\overline{d^{c}_{L\alpha}} d_{L\beta})(\overline{\nu_L} u_{R\gamma})$ & $-$ & $(\bar{3},3)$\\
\hline
\end{tabular}
\caption{Dimension-6
BNV LEFT operators~\cite{Jenkins:2017jig}.
}
\label{tab:nuLEFTBL}
\end{table}

Tab.~\ref{tab:nuLEFTBL} displays the dimension-6 BNV LEFT operators below the EW scale~\cite{Jenkins:2017jig}.
The low-energy effective Lagrangian reads as
\begin{eqnarray}
\mathcal{L}_{\rm LEFT}=\sum_i L_{i,prst} \mathcal{O}_{i,prst}+h.c.\;,
\end{eqnarray}
where $L_i$ denotes the WCs in LEFT.
The indices of neutrinos in LEFT operators are denoted by $\nu_{s=1\sim 4}=(\nu_{s=1\sim 3}, N^c)$. The active neutrinos as a part of the lepton doublet $L$ are labeled with $s=1\sim 3$ and the only one sterile neutrino $N$ is labeled with $s=4$. Here we also show the transformation properties of the LEFT operators under the quark flavor group $\rm SU(3)_L \times SU(3)_R$. They can be matched to BChPT at leading order of chiral power counting based on these transformation properties. The third column of Tab.~\ref{tab:nuLEFTBL} indicates which SMEFT/$\nu$SMEFT operators can be matched to the corresponding LEFT operator, which will be discussed in next subsection.

In Tabs.~\ref{tab:nuSMEFT} and \ref{tab:nuLEFTBL},
we present the flavor symmetries of the repeated quark fields indicated by Young tableaux using Sym2Int~\cite{Fonseca:2017lem,Fonseca:2019yya}.
The operators with two repeated fermion fields have in principle $n_f^2$ different combinations. Their Young tableaux include
\begin{itemize}
\item
\Yboxdim{6pt} \yng(2) implies that there are only $n_f (n_f+1)/2$ combinations (e.g. 6 for $n_f=3$) and thus $n_f^2-n_f(n_f+1)/2$ are zero (3 for $n_f=3$).
\item
\Yboxdim{6pt} \yng(1,1) implies that there are only $n_f (n_f-1)/2$ combinations (e.g. 3 for $n_f=3$) and thus $n_f^2-n_f(n_f-1)/2$ are zero (6 for $n_f=3$).
\end{itemize}
The operators with 3 repeated fermion fields have in principle $n_f^3$ different combinations. Their Young tableaux include
\begin{itemize}
\item
\Yboxdim{6pt} \yng(3) implies that there are only $n_f (n_f+1)(n_f+2)/2/3$ combinations (e.g. 10 for $n_f=3$) and thus $n_f^3-n_f(n_f+1)(n_f+2)/2/3$ are zero (17 for $n_f=3$).
\item
\Yboxdim{6pt} \yng(1,1,1) implies that there are only $n_f (n_f-1)(n_f-2)/2/3$ combinations (e.g. 1 for $n_f=3$) and thus $n_f^3-n_f(n_f-1)(n_f-2)/2/3$ are zero (26 for $n_f=3$).
\item
\Yboxdim{6pt} \yng(2,1) implies that there are only $8$ combinations for $n_f=3$ .
\end{itemize}
Concretely, we find the following results for the redundant operators (assuming $n_f=3$)
\begin{itemize}
\item \Yboxdim{6pt} \yng(2) 3 vanishing operators\;,
\item \Yboxdim{6pt} \yng(1,1) 6 vanishing operators\;,
\item \Yboxdim{6pt} \yng(3) 17 vanishing operators\;,   \item \Yboxdim{6pt} \yng(1,1,1) + \yng(2,1) + \yng(3) 8 vanishing operators\;.
\end{itemize}
For illustration, the operator $QQLQ$ has an index symmetry described by the sum of the 3 Young tableaux. It allows $1+8+10=19$ possible index combinations, where we used the same order as the Young tableaux. Hence, there have to be $3^3-19=8$ vanishing operators.

\subsection{Matching of SMEFT and $\nu$SMEFT to LEFT}

After the EW symmetry breaking, the operators of SMEFT and $\nu$SMEFT are matched to the LEFT operators. Next we give the matching results for the operators of interest.

\subsubsection{SMEFT operators}

\begin{align}
[L^{S,RL}_{dud}]_{prst}&=-V_{t't}[C_{duLQ}]_{prst'}\;,\\
[L^{S,LL}_{udd}]_{prst}&=V_{r'r} V_{t't}([C_{QQLQ}]_{r'pst'}+[C_{QQLQ}]_{t'r'sp}-[C_{QQLQ}]_{r't'sp})\;,
\\
[L^{S,RR}_{udd}]_{prst}&=[C_{udLdH}]_{prst}{v\over \sqrt{2}}\;,\\
[L^{S,LR}_{udd}]_{prst}&=-V_{r'r}[C_{QQLdH}]_{pr'st}{v\over \sqrt{2}}\;,
\end{align}
where $V$ denotes the CKM matrix, $v$ is the EW vacuum expectation value, and $s=1\sim 3$ is the active neutrino index. The two operators with covariant derivative do not contribute to dimension-6 LEFT operators, but only dimension-7 LEFT operators. They are further suppressed by $m_p/v$ with $m_p$ being the proton mass. We thus focus on the SMEFT/$\nu$SMEFT operators without covariant derivative.

Note that the above matching results are under the assumption of unitary matrix $U$ and vanishing matrix $K$. More precisely, taking the matching between $C_{duLQ}$ and $L_{dud}^{S,RL}$ as an example, the matching result becomes
\begin{align}
[L^{S,RL}_{dud}]_{prst}&=-V_{t't}U_{s's}[C_{duLQ}]_{prs't'}
\;,~s=1\sim 3\;,\\
[L^{S,RL}_{dud}]_{prst}&=-V_{t't}K_{s's}[C_{duLQ}]_{prs't'}\sim 0\;,~s=4\;.
\end{align}
Suppose the $K$ matrix is negligible, the SMEFT operators can only be matched to the LEFT operators with active neutrinos. When we sum over the active neutrinos in the final states of nucleon decay, the dependence of mixing matrix $U$ vanishes
\begin{align}
\label{eq:sum}
\sum_s\nu_s\propto \sum_s L_{prst} L_{prst}^\ast\propto \sum_{s,s^\prime,s^{\prime\prime}} U_{s's} C_{prs't} U^\ast_{s^{\prime\prime}s}C^\ast_{prs^{\prime\prime}t}=\sum_{s',s^{\prime\prime}}\delta_{s's^{\prime\prime}}C_{prs't} C^\ast_{prs^{\prime\prime}t}=\sum_{s'} C_{prs't} C^\ast_{prs't}\;.
\end{align}

\subsubsection{$\nu$SMEFT operators}

\begin{align}
    [L^{S,LR}_{udd}]_{prst}&=V_{r'r}([C_{QQNd}]_{pr'st}+[C_{QQNd}]_{r'pst})\;,\\
     [L^{S,RR}_{udd}]_{prst}&=[C_{udNd}]_{prst}\;,
\\
[L^{S,RL}_{ddu}]_{prst}&=[C_{HdNQ}]_{prst}{v\over \sqrt{2}}\;,
\\
[L^{S,LL}_{udd}]_{prst}&=V_{r'r} V_{t't}(-[C_{HQNQ}]_{r'pst'}-[C_{HQNQ}]_{t'r'sp}+[C_{HQNQ}]_{r't'sp}){v\over \sqrt{2}}\;,
\\
[L^{S,RL}_{dud}]_{prst}&=V_{t't}[C_{HduNQ}]_{prst'}{v\over \sqrt{2}}\;,
\end{align}
where $s=4$. The $\nu$SMEFT operators can only be matched to the LEFT operators with the sterile neutrino.
For the matching, we used the Fierz identity
\begin{equation}
(\overline{\psi_{1L}^c}\,  \psi_{2L})(\overline{\psi_{3L}^c}\, \psi_{4L})
=-(\overline{\psi_{1L}^c}\, \psi_{3L})(\overline{\psi_{4L}^c}\,  \psi_{2L})-(\overline{\psi_{1L}^c}\, \psi_{4L})(\overline{\psi_{3L}^c}\,  \psi_{2L})\,,
\end{equation}
and $\overline{\psi^c_i}\Gamma\psi^{cj} = -(-1)^A \eta_\Gamma^c \overline{\psi^j}\Gamma\psi_i$, where~\cite{Dreiner:2008tw}
\begin{align}
C^{-1}\Gamma C &= \eta^c_\Gamma \Gamma^T\;, &
\eta^c_\Gamma = \begin{cases} +1 & \text{for}\;\Gamma=1,\gamma_5,\gamma^\mu\gamma_5\\
-1 & \text{for}\; \Gamma=\gamma^\mu, \sigma^{\mu\nu}, \sigma^{\mu\nu}\gamma_5
\end{cases}
\end{align}
and $(-1)^A$ is $-1$ ($+1$) for anticommuting (commuting) spinors.

In summary, the four BNV SMEFT operators of interest can be matched to LEFT operators $\mathcal{O}_{udd}^{S,LL}$, $\mathcal{O}_{dud}^{S,RL}$, $\mathcal{O}_{udd}^{S,LR}$ and $\mathcal{O}_{udd}^{S,RR}$ with neutrino index $s=1\sim 3$. The five BNV $\nu$SMEFT operators can be matched to LEFT operators  $\mathcal{O}_{udd}^{S,LL}$, $\mathcal{O}_{dud}^{S,RL}$, $\mathcal{O}_{ddu}^{S,RL}$, $\mathcal{O}_{udd}^{S,LR}$ and $\mathcal{O}_{udd}^{S,RR}$ with neutrino index $s=4$. The last LEFT operator $\mathcal{O}_{ddu}^{S,LR}$ in Tab.~\ref{tab:nuLEFTBL} cannot be generated by the considered operators in Tab.~\ref{tab:nuSMEFT}. We show the dependence of this operator in our analytical formula, but do not include it in the numerical results below.

\subsection{Renormalization group evolution}

The RG evolution is essential between two different scales. Here we collect the RG equations (RGEs) for the WCs. The general RGE follows
\begin{eqnarray}
\dot{C}_i = 16\pi^2 \mu {dC_i\over d\mu}=\gamma_{ij} C_j\;,
\end{eqnarray}
where $\gamma$ denotes the anomalous dimension function.

We start from the RGEs of the SMEFT/$\nu$SMEFT WCs which are used to run the operators from UV scale $\Lambda$ down to the EW scale $m_Z$.
The approximate RGEs of the dimension-6 SMEFT WCs are~\footnote{We ignore Yukawa terms of lepton, light quarks and sterile neutrino, and only keep gauge contributions and top Yukawa terms in RGEs.}~\cite{Alonso:2014zka}
\begin{eqnarray}
\dot{C}_{duLQ,prst}&=&-(4g_3^2+{9\over 2}g_2^2+{11\over 6}g_1^2)C_{duLQ,prst}\nonumber\\
&&-C_{duLQ,pvsw}(y_u)_{vt}(y_u^\dagger)_{wr}+C_{duLQ,pvst}(y_u y_u^\dagger)_{vr}+{1\over 2}C_{duLQ,prsv}(y_u^\dagger y_u)_{vt}\;,\\
\dot{C}_{QQLQ,prst}&=&-(4g_3^2+3g_2^2+{1\over 3}g_1^2)C_{QQLQ,prst}\nonumber\\
&&-4(C_{QQLQ,rpst}+C_{QQLQ,trsp}+C_{QQLQ,ptsr})g_2^2\nonumber\\
&&+{1\over 2}C_{QQLQ,vrst}(y_u^\dagger y_u)_{vp}+{1\over 2}C_{QQLQ,pvst}(y_u^\dagger y_u)_{vr}+{1\over 2}C_{QQLQ,prsv}(y_u^\dagger y_u)_{vt}\;.
\label{eq:RGE_QQLQ}
\end{eqnarray}
We neglect light quark in the RGEs and choose $y_u={\rm diag}(0,0,y_t)$ for the up-type quark basis with $y_t$ being the top quark Yukawa coupling.
As discussed above, some of the WCs respect a flavor index symmetry. This has to be taken into account when defining the initial conditions.
In particular, care has to be taken with the operator $\mathcal{O}_{QQLQ}$. The electroweak symmetry structure exhibits two antisymmetric $\epsilon$ tensors in the $\mathcal{O}_{QQLQ}$ operator which can be rewritten using their Schouten identity
\begin{eqnarray}
\epsilon^{ij}\epsilon^{kl}+\epsilon^{jk}\epsilon^{il}+\epsilon^{ki}\epsilon^{jl}=0
\end{eqnarray}
to derive the operator relation~\cite{Abbott:1980zj}
\begin{eqnarray}
\mathcal{O}_{QQLQ,prst}+\mathcal{O}_{QQLQ,rpst}=\mathcal{O}_{QQLQ,trsp}+\mathcal{O}_{QQLQ,tpsr}\;.
\label{eq:Osym}
\end{eqnarray}
After inserting this relation to the product $C_{QQLQ,prst} \mathcal{O}_{QQLQ,prst}$, it is straightforward to obtain
\begin{eqnarray}
C_{QQLQ,prst} \mathcal{O}_{QQLQ,prst}
&=&C_{QQLQ,prst} (\mathcal{O}_{QQLQ,trsp}+\mathcal{O}_{QQLQ,tpsr}-\mathcal{O}_{QQLQ,rpst})\nonumber\\
&=&(C_{QQLQ,trsp}+C_{QQLQ,rtsp}-C_{QQLQ,rpst})\mathcal{O}_{QQLQ,prst}\;.
\end{eqnarray}
We find the WC satisfies the following relation
\begin{eqnarray}
C_{QQLQ,prst}+C_{QQLQ,rpst}=C_{QQLQ,rtsp}+C_{QQLQ,trsp}\;.
\label{eq:Csym}
\end{eqnarray}
It is worth mentioning that the ordering of quark flavor indices in the WC relations differs from that in the operator relation Eq.~\eqref{eq:Osym}. At first glance, one might conclude that the symmetry of WCs matches that of their corresponding operators, as seen in Ref.~\cite{Alonso:2014zka} and implemented in tools like the Mathematica package DSixTools~\cite{Celis:2017hod,Fuentes-Martin:2020zaz}, the Python package Wilson~\cite{Aebischer:2018bkb}, and other literature. However, as demonstrated in our derivation, this intuition is incorrect. This is also highlighted in Ref.~\cite{Li:2021tsq}, which explicitly demonstrates how the symmetry relations of WCs can be systematically derived from those of the operators.
The RGEs of the dimension-6 $\nu$SMEFT WCs are~\cite{Alonso:2014zka}~\footnote{The neutrino Yukawa couplings are small for sub-GeV sterile neutrinos with small active-sterile mixing and thus can be neglected in the RGEs.}
\begin{eqnarray}
\dot{C}_{QQNd,prst}&=&-(4g_3^2+{9\over 2}g_2^2-{1\over 6} g_1^2)C_{QQNd,prst} \nonumber\\
&&+{1\over 2}C_{QQNd,vrst}(y_u^\dagger y_u)_{vp}+{1\over 2}C_{QQNd,pvst}(y_u^\dagger y_u)_{vr}\;,\\
\dot{C}_{udNd,prst}&=&-(4g_3^2+{2\over 3}g_1^2)C_{udNd,prst}-{4\over 3}g_1^2 C_{udNd,ptsr}\nonumber\\
&&+C_{udNd,vrst}(y_u y_u^\dagger)_{vp}\;,
\end{eqnarray}
where $C_{QQNd,prst}=C_{QQNd,rpst}$ and the coefficient $C_{udNd,prst}$ can be decomposed
into the symmetric and antisymmetric combinations ${1\over 2}(C_{udNd,prst}\pm C_{udNd,ptsr})$.

The RGEs of the dimension-7 SMEFT WCs are~\cite{Liao:2016hru,Zhang:2023ndw}
\begin{eqnarray}
\dot{C}_{udLdH,prst}&=&-[4g_3^2+{9\over 4}g_2^2+{17\over 12}g_1^2-3{\rm tr}(y_u^\dagger y_u)]C_{udLdH,prst}-{10\over 3}g_1^2C_{udLdH,ptsr}\nonumber\\
&&+2(y_u^\dagger y_u)_{vp}C_{udLdH,vrst}-\frac{1}{6}(11 g_1^2+24 g_3^2)(y_u)_{vp}C_{dDdLQ,trsv}\nonumber\\
&&+\frac{1}{6}(13 g_1^2+48 g_3^2)(y_u)_{vp}C_{dDdLQ,rtsv}\;,\\
\dot{C}_{QQLdH,prst}&=&-[4g_3^2+{15\over 4}g_2^2+{19\over 12}g_1^2-3{\rm tr}(y_u^\dagger y_u)]C_{QQLdH,prst}-3g_2^2C_{QQLdH,rpst}\nonumber\\
&&+2(y_u y_u^\dagger)_{vr}C_{QQLdH,vpst}+{5\over 2}(y_u y_u^\dagger)_{vp}C_{QQLdH,vrst}-{3\over 2}(y_u y_u^\dagger)_{vr}C_{QQLdH,pvst}\;.
\end{eqnarray}
Although there is a linear dependence of Yukawa $y_u$ associated with coefficient $C_{dDdLQ}$ in the RGE of $C_{udLdH}$, we will not consider it in our calculation as it can only be generated at loop level.
Both of them can be decomposed
into the symmetric and antisymmetric combinations:
${1\over 2}(C_{udLdH,prst}\pm C_{udLdH,ptsr})$ and ${1\over 2}(C_{QQLdH,prst}\pm C_{QQLdH,rpst})$.
The RGEs for the dimension-7 $\nu$SMEFT WCs are not available in the literature. We include the leading contributions from gauge interactions and top quark loop corrections to the Higgs self-energy, but refrain from obtaining the full set of RGEs, since the other contributions are not numerically significant. It is straightforward to infer QCD corrections and top quark corrections to the Higgs field renormalization from the existing dimension-7 SMEFT RGEs. Details for EW corrections are summarised in App.~\ref{app:EWdim7nuSMEFT}.
We obtain the following RGEs for the dimension-7 $\nu$SMEFT WCs
\begin{eqnarray}
\dot{C}_{HdNQ,prst}&=&-[4g_3^2+\frac94 g_2^2+\frac{1}{12} g_1^2-3{\rm tr}(y_u^\dagger y_u)]C_{HdNQ,prst}\;,\\
\dot{C}_{HQNQ,prst}
&=&-[4g_3^2 + \frac{15}{4} g_2^2 + \frac{7}{12} g_1^2 -3{\rm tr}(y_u^\dagger y_u)]C_{HQNQ,prst}
\\\nonumber &&
    - 2g_2^2 \left(C_{HQNQ,rpst} + C_{HQNQ,ptsr} + C_{HQNQ,rtsp} \right)
\;,\\
\dot{C}_{HduNQ,prst}&=&-[4g_3^2 + \frac94 g_2^2 +\frac{25}{12} g_1^2 -3{\rm tr}(y_u^\dagger y_u)]C_{HduNQ,prst}\;,
\end{eqnarray}
where $C_{HdNQ,prst}=-C_{HdNQ,rpst}$.
The WC of the $C_{HQNQ,prst}$ operator follows a similar relation to that of $C_{QQLQ,prst}$
\begin{eqnarray}
C_{HQNQ,prst}+C_{HQNQ,rpst}=C_{HQNQ,trsp}+C_{HQNQ,rtsp}\;,
\end{eqnarray}
which has been used to simplify the RG equation.

In LEFT the dominant contributions to the RGEs are from gauge interactions and thus parity symmetric. The RGEs of the dimension-6 LEFT WCs are as follows~\cite{Jenkins:2017dyc}
\begin{eqnarray}
\dot{L}^{S,LL}_{udd,prst}&=&(-4g_3^2+e^2[6q_uq_d-2(q_u-q_d)q_d])L^{S,LL}_{udd,prst}+4e^2(q_u-q_d)q_d L^{S,LL}_{udd,ptsr}\;,\\
\dot{L}^{S,RL}_{dud,prst}&=&(-4g_3^2+6e^2q_dq_u)L^{S,RL}_{dud,prst}\;,\\
\dot{L}^{S,RL}_{ddu,prst}&=&(-4g_3^2+6e^2q_d^2)L^{S,RL}_{ddu,prst}\;,
\end{eqnarray}
and
\begin{eqnarray}
\dot{L}^{S,LR}_{udd,prst}&=&(-4g_3^2+6e^2q_dq_u)L^{S,LR}_{udd,prst}\;,\\
\dot{L}^{S,RR}_{udd,prst}&=&(-4g_3^2+e^2[6q_dq_u-2(q_u-q_d)q_d])L^{S,RR}_{udd,prst}+4e^2(q_u-q_d)q_dL^{S,RR}_{udd,ptsr}\;,\\
\dot{L}^{S,LR}_{ddu,prst}&=&(-4g_3^2+6e^2q_d^2)L^{S,LR}_{ddu,prst}\;,
\end{eqnarray}
where $q_u$ ($q_d$) denotes the electric charge of up-type (down-type) quark.
We use them to evolve the WCs down to the nucleon scale $\mu=2$ GeV which is used in lattice calculation for hadronic parameters~\cite{Yoo:2021gql}.

\subsection{Matching to the BChPT}

In this section, we match the relevant LEFT operators to BChPT and derive the neutrino interactions with baryons and mesons.

The baryon $B$, $\overline{B}$ and meson $M$ fields in BChPT are given by
\begin{eqnarray}
&&B=\left(
  \begin{array}{ccc}
    {\Sigma^0\over \sqrt{2}}+{\Lambda^0\over \sqrt{6}} & \Sigma^+ & p \\
    \Sigma^- & -{\Sigma^0\over \sqrt{2}}+{\Lambda^0\over \sqrt{6}} & n \\
    \Xi^- & \Xi^0 & -\sqrt{2\over 3}\Lambda^0 \\
  \end{array}
\right)\;,~~\overline{B}=\left(
  \begin{array}{ccc}
    {\overline{\Sigma^0}\over \sqrt{2}}+{\overline{\Lambda^0}\over \sqrt{6}} & \overline{\Sigma^-} & \overline{\Xi^-} \\
    \overline{\Sigma^+} & -{\overline{\Sigma^0}\over \sqrt{2}}+{\overline{\Lambda^0}\over \sqrt{6}} & \overline{\Xi^0} \\
    \overline{p} & \overline{n} & -\sqrt{2\over 3}\overline{\Lambda^0} \\
  \end{array}
\right)\;,\\
&&M=\left(
  \begin{array}{ccc}
    {\pi^0\over \sqrt{2}}+{\eta^0\over \sqrt{6}} & \pi^+ & K^+ \\
    \pi^- & -{\pi^0\over \sqrt{2}}+{\eta^0\over \sqrt{6}} & K^0 \\
    K^- & \bar{K}^0 & -\sqrt{2\over 3}\eta^0 \\
  \end{array}
\right)\;,~~M^\dagger=M\;,~~\Sigma\equiv e^{2iM/f_\pi}\;,~~\xi\equiv e^{iM/f_\pi}\;.
\end{eqnarray}
In principle, the neutral components of the baryon or meson octet mix to form the physical states after isospin breaking. Here we ignore the quark mass splitting effects for both baryons and mesons because they are generally small. The leading-order chiral Lagrangian for baryons and mesons then becomes~\cite{Claudson:1981gh}
\begin{eqnarray}
\mathcal{L}={f_\pi^2\over 8} {\rm tr}(\partial_\mu \Sigma \partial^\mu \Sigma^\dagger) + {\rm tr}(\overline{B}(i\cancel{D}-M_B)B)-{D\over 2}{\rm tr}(\overline{B}\gamma^\mu \gamma_5 \{\xi_\mu,B\})-{F\over 2}{\rm tr}(\overline{B}\gamma^\mu \gamma_5 [\xi_\mu,B])\;,
\label{eq:ChPT}
\end{eqnarray}
where $D_\mu B=\partial_\mu B + [\Gamma_\mu,B]$ with
\begin{eqnarray}
\Gamma_\mu={1\over 2}(\xi^\dagger \partial_\mu \xi+\xi \partial_\mu \xi^\dagger)\;,~~~\xi_\mu=i(\xi^\dagger \partial_\mu \xi-\xi \partial_\mu \xi^\dagger)\;.
\end{eqnarray}
This Lagrangian results in the couplings of two baryons to a meson as shown in Refs.~\cite{Claudson:1981gh,Nath:2006ut,Beneito:2023xbk}~\footnote{Some typos in the past literature are corrected in Ref.~\cite{Beneito:2023xbk}.}.

To match the LEFT operators to BChPT, we make use of the flavor symmetry and match the operators which have the same quark flavor symmetry properties~\cite{Claudson:1981gh}. The flavor symmetry transformations of the relevant operators in BChPT are
\begin{align}
    \xi B \xi \sim (3,\bar 3)& \to L \xi B \xi R^\dagger \;, &
    \xi^\dagger B \xi^\dagger  \sim (\bar 3,3) & \to R \xi^\dagger B \xi^\dagger L^\dagger \;,
    \\\nonumber
    \xi B \xi^\dagger \sim(8,1) & \to L \xi B \xi^\dagger L^\dagger \;, &
    \xi^\dagger B \xi \sim (1,8)& \to R \xi^\dagger B \xi R^\dagger \;,
\end{align}
where $L$ ($R$) is an element of $\rm SU(3)_L$ ($\rm SU(3)_R$).
In addition, we need the leading-order nuclear matrix elements which can be related to the two independent low-energy constants $\alpha$ and $\beta$ defined by
\begin{eqnarray}
&&\langle 0|\epsilon^{\alpha\beta\gamma} (\bar{u}^c_{R,\alpha} d_{R,\beta}) u_{L,\gamma}|p^{(s)}\rangle = \alpha u^{(s)}_{pL}\;,~~\langle 0|\epsilon^{\alpha\beta\gamma} (\bar{u}^c_{L,\alpha} d_{L,\beta}) u_{L,\gamma}|p^{(s)}\rangle = \beta u^{(s)}_{pL}\;.
\end{eqnarray}
The matrix elements of other baryons in the baryon octet can be related to these two by imposing $\rm SU(3)$ flavor symmetry and requiring parity conservation
\begin{eqnarray}
&&\langle 0|\epsilon^{\alpha\beta\gamma} (\bar{q}^c_{R,i\alpha} q_{R,j\beta}) q_{L,k\gamma}|B^{(s)}\rangle = \alpha u^{(s)}_{BL}\;,
~~\langle 0|\epsilon^{\alpha\beta\gamma} (\bar{q}^c_{L,i\alpha} q_{L,j\beta}) q_{L,k\gamma}|B^{(s)}\rangle = \beta u^{(s)}_{BL}\;,
\\
&&\langle 0|\epsilon^{\alpha\beta\gamma} (\bar{q}^c_{L,i\alpha} q_{L,j\beta}) q_{R,k\gamma}|B^{(s)}\rangle = -\alpha u^{(s)}_{BR}\;,
~~\langle 0|\epsilon^{\alpha\beta\gamma} (\bar{q}^c_{R,i\alpha} q_{R,j\beta}) q_{R,k\gamma}|B^{(s)}\rangle = -\beta u^{(s)}_{BR}\;.
\end{eqnarray}
Here and above, $u_{pL}$ and $u_{BL}$ ($u_{BR}$) stand for the baryon states projected by chirality projector $P_L$ ($P_R$).

In the following results, we only keep relevant terms for the calculation of BNV nucleon decay at leading order, \emph{i.e.}, we only keep terms at order $f_\pi^{-1}$ with protons and neutrons and drop terms with other baryons.
The projection operator $P_{ij}$ ($\tilde{P}_{ij}$) is given by setting the $(i,j)$ entry to $1 \; (-1)$ and all other entries to zero.
The matching of $|\Delta(B-L)|=0$ LEFT operators to BChPT becomes
\begin{align}
[\mathcal{O}^{S,LL}_{udd}]_{11s1}&\to -\beta\overline{\nu^c_{Ls}}{\rm tr}(\xi B\xi^\dagger P_{32})\supset \overline{\nu^c_{Ls}} [-\beta n - {i\beta\over f_\pi}(\sqrt{3\over 2}n\eta^0-{1\over \sqrt{2}}n\pi^0+p\pi^-) ]\;,\\
[\mathcal{O}^{S,LL}_{udd}]_{12s1}&\to -\beta\overline{\nu^c_{Ls}}{\rm tr}(\xi B\xi^\dagger \tilde{P}_{22})\supset \overline{\nu^c_{Ls}}[-\beta(-{\Lambda^0\over \sqrt{6}}+{\Sigma^0\over \sqrt{2}})-{i\beta\over f_\pi}n\bar{K}^0]\;,\\
[\mathcal{O}^{S,LL}_{udd}]_{11s2}&\to -\beta\overline{\nu^c_{Ls}}{\rm tr}(\xi B\xi^\dagger P_{33})\supset \overline{\nu^c_{Ls}} [\beta\sqrt{2\over 3}\Lambda^0-{i\beta\over f_\pi}(n\bar{K}^0+pK^-)]\;,\\
[\mathcal{O}^{S,LL}_{udd}]_{12s2}&\to -\beta\overline{\nu^c_{Ls}}{\rm tr}(\xi B\xi^\dagger \tilde{P}_{23})\supset \overline{\nu^c_{Ls}} [\beta\Xi^0]\;,
\end{align}
\begin{align}
[\mathcal{O}^{S,RL}_{dud}]_{11s1}&\to \alpha\overline{\nu^c_{Ls}}{\rm tr}(\xi B\xi \tilde{P}_{32})\supset \overline{\nu^c_{Ls}} [-\alpha n+{i\alpha\over f_\pi}({1\over\sqrt{6}}n\eta^0+{1\over \sqrt{2}}n\pi^0-p\pi^-)]\;,\\
[\mathcal{O}^{S,RL}_{dud}]_{21s1}&\to \alpha\overline{\nu^c_{Ls}}{\rm tr}(\xi B\xi P_{22})\supset \overline{\nu^c_{Ls}} [\alpha ({1\over\sqrt{6}}\Lambda^0-{1\over\sqrt{2}}\Sigma^0)+{i\alpha\over f_\pi}n\bar{K}^0]\;,\\
[\mathcal{O}^{S,RL}_{dud}]_{11s2}&\to \alpha\overline{\nu^c_{Ls}}{\rm tr}(\xi B\xi \tilde{P}_{33})\supset \overline{\nu^c_{Ls}} [\alpha\sqrt{2\over 3}\Lambda^0-{i\alpha\over f_\pi}(n\bar{K}^0+pK^-)]\;,\\
[\mathcal{O}^{S,RL}_{dud}]_{21s2}&\to \alpha\overline{\nu^c_{Ls}}{\rm tr}(\xi B\xi P_{23})\supset \overline{\nu^c_{Ls}} [\alpha \Xi^0]\;,
\end{align}
\begin{align}
[\mathcal{O}^{S,RL}_{ddu}]_{12s1}&\to -\alpha\overline{\nu^c_{Ls}}{\rm tr}(\xi B\xi P_{11})\supset \overline{\nu^c_{Ls}} [-\alpha ({1\over\sqrt{6}}\Lambda^0+{1\over\sqrt{2}}\Sigma^0)-{i\alpha\over f_\pi} pK^-]\;.
\end{align}
The matching of $|\Delta(B-L)|=2$ LEFT operators to BChPT becomes
\begin{align}
[\mathcal{O}^{S,LR}_{udd}]_{11s1}&\to \alpha\overline{\nu_{Ls}}{\rm tr}(\xi^\dagger B\xi^\dagger P_{32})\supset \overline{\nu_{Ls}} [\alpha n + {i\alpha\over f_\pi}({1\over \sqrt{6}}n\eta^0+{1\over \sqrt{2}}n\pi^0-p\pi^-)]\;,\\
[\mathcal{O}^{S,LR}_{udd}]_{12s1}&\to \alpha\overline{\nu_{Ls}}{\rm tr}(\xi^\dagger B\xi^\dagger \tilde{P}_{22})\supset \overline{\nu_{Ls}} [\alpha({1\over \sqrt{2}}\Sigma^0-{1\over \sqrt{6}}\Lambda^0)+{i\alpha\over f_\pi}n\bar{K}^0]\;,\\
[\mathcal{O}^{S,LR}_{udd}]_{11s2}&\to \alpha\overline{\nu_{Ls}}{\rm tr}(\xi^\dagger B\xi^\dagger P_{33})\supset \overline{\nu_{Ls}} [-\alpha \sqrt{2\over 3}\Lambda^0-{i\alpha\over f_\pi}(n\bar{K}^0+pK^-)]\;,\\
[\mathcal{O}^{S,LR}_{udd}]_{12s2}&\to \alpha\overline{\nu_{Ls}}{\rm tr}(\xi^\dagger B\xi^\dagger \tilde{P}_{23})\supset \overline{\nu_{Ls}} [-\alpha \Xi^0]\;,
\end{align}
\begin{align}
[\mathcal{O}^{S,RR}_{udd}]_{11s1}&\to \beta\overline{\nu_{Ls}}{\rm tr}(\xi^\dagger B\xi P_{32})\supset \overline{\nu_{Ls}} [\beta n-{i\beta\over f_\pi}(\sqrt{3\over 2}n\eta^0-{1\over \sqrt{2}}n\pi^0+p\pi^-)]\;,\\
[\mathcal{O}^{S,RR}_{udd}]_{12s1}&\to \beta\overline{\nu_{Ls}}{\rm tr}(\xi^\dagger B\xi \tilde{P}_{22})\supset \overline{\nu_{Ls}} [\beta({1\over \sqrt{2}}\Sigma^0-{1\over \sqrt{6}}\Lambda^0)-{i\beta\over f_\pi}n\bar{K}^0]\;,\\
[\mathcal{O}^{S,RR}_{udd}]_{11s2}&\to \beta\overline{\nu_{Ls}}{\rm tr}(\xi^\dagger B\xi P_{33})\supset \overline{\nu_{Ls}} [-\beta \sqrt{2\over 3}\Lambda^0-{i\beta\over f_\pi}(n\bar{K}^0+pK^-)]\;,\\
[\mathcal{O}^{S,RR}_{udd}]_{12s2}&\to \beta\overline{\nu_{Ls}}{\rm tr}(\xi^\dagger B\xi \tilde{P}_{23})\supset \overline{\nu_{Ls}} [-\beta \Xi^0]\;,
\end{align}
\begin{align}
[\mathcal{O}^{S,LR}_{ddu}]_{12s1}&\to \alpha\overline{\nu_{Ls}}{\rm tr}(\xi^\dagger B\xi^\dagger P_{11})\supset \overline{\nu_{Ls}} [\alpha ({1\over \sqrt{2}}\Sigma^0+{1\over \sqrt{6}}\Lambda^0)-{i\alpha\over f_\pi}pK^-]\;.
\end{align}

\subsection{Discussion on UV models}

As a phenomenological study, we remain agnostic about the origin of the effective BNV operators. The WCs of BNV operators were simply taken as independent parameters in an effective framework. Nevertheless, the complete tree-level UV dictionary for up to
dimension-7 operators in SMEFT was studied in Refs.~\cite{deBlas:2017xtg,Li:2022abx,Li:2023cwy}.
We refer the UV completions for dimension-6 and dimension-7 BNV operators in $\nu$SMEFT to their tree-level decompositions in the Tab.~10 of Ref.~\cite{Beltran:2023ymm}. The skeleton diagrams were shown in their Figs.~1 and 2. For illustration, the two dimension-6 $\nu$SMEFT operators of interest can only be induced by the exchange of a scalar $\rm SU(2)$ singlet $\omega_1\sim (3,1,-{1\over 3})$ (or $\omega_2\sim (3,1,{2\over 3})$) or a vector doublet $\mathcal{Q}_1\sim (3,2,{1\over 6})$. The three dimension-7 operators can be opened through
three different categories of diagram.

\section{Baryon-number-violating Lagrangian and nucleon decay rates}
\label{sec:BNV}

\subsection{Effective Lagrangian and nucleon decay rate}

From the baryon-meson chiral Lagrangian and the above matching results from LEFT to BChPT, one can read out the necessary couplings for BNV nucleon decay calculation.
In Tabs.~\ref{tab:2bodyBL0} and \ref{tab:2bodyBL2}, we collect the couplings for BNV decays $N\to M\bar{\nu}$ with $|\Delta (B-L)|=0$ and $N\to M \nu$ with $|\Delta (B-L)|=2$, respectively. They are given by the LEFT WCs and the hadronic parameters. We find that all decays can be generated by either SMEFT or $\nu$SMEFT operators.
In particular, $p\to K^+\bar{\nu}$ and $n\to K^0\bar{\nu}$ can be generated by only $\nu$SMEFT operator $\mathcal{O}_{HdNQ}$ through LEFT operator $\mathcal{O}_{ddu}^{S,RL}$.

\begin{table}[htbp!]
\renewcommand{\arraystretch}{1.2}
\hspace*{-0.05\columnwidth}
\resizebox{1.1\columnwidth}{!}{
\begin{tabular}{c|c|c|c}
\hline
Process & $g_{MB}^N$ & $m_{Bs,X}$ & $y_{Ms,X}^N$\\
\hline
$p\to \pi^+ \bar{\nu}$ & $g_{\pi n}^p={D+F\over f_\pi}$ & $m_{ns,R}=[L^{S,LL}_{ udd}]_{11s1} (-\beta)+[L^{S,RL}_{ dud}]_{11s1}(-\alpha)$ & $y_{\pi s,R}^p=[L^{S,LL}_{ udd}]_{11s1}(-{\beta\over f_\pi})+[L^{S,RL}_{ dud}]_{11s1}(-{\alpha\over f_\pi})$\\
$p\to K^+ \bar{\nu}$ & $g_{K\Sigma}^p={D-F\over \sqrt{2}f_\pi}$ & $m_{\Sigma s,R}=[L^{S,LL}_{ udd}]_{12s1} (-{\beta\over \sqrt{2}})+[L^{S,RL}_{ dud}]_{21s1} (-{\alpha\over \sqrt{2}})$ &  $y_{K s,R}^p=[L^{S,LL}_{ udd}]_{11s2}(-{\beta\over f_\pi})+[L^{S,RL}_{ dud}]_{11s2}(-{\alpha\over f_\pi})$ \\
&& $+[L^{S,RL}_{ ddu}]_{12s1} {-\alpha\over \sqrt{2}}$ & $+[L^{S,RL}_{ ddu}]_{12s1}(-{\alpha\over f_\pi})$\\
& $g_{K\Lambda}^p=-{D+3F\over \sqrt{6}f_\pi}$ & $m_{\Lambda s,R}=[L^{S,LL}_{ udd}]_{12s1} {\beta\over \sqrt{6}}+[L^{S,LL}_{ udd}]_{11s2}\beta\sqrt{2\over 3}$ & \\
& & $+[L^{S,RL}_{dud}]_{21s1}{\alpha\over \sqrt{6}}+[L^{S,RL}_{dud}]_{11s2}\alpha\sqrt{2\over 3}+[L^{S,RL}_{ddu}]_{12s1} {-\alpha\over \sqrt{6}}$ &\\
\hline
$n\to \pi^0\bar{\nu}$ & $g_{\pi n}^n=-{D+F\over \sqrt{2}f_\pi}$ & $m_{ns,R}=[L^{S,LL}_{ udd}]_{11s1} (-\beta)+[L^{S,RL}_{ dud}]_{11s1}(-\alpha)$ & $y_{\pi s,R}^n=[L^{S,LL}_{ udd}]_{11s1}({\beta\over \sqrt{2}f_\pi})+[L^{S,RL}_{dud}]_{11s1}({\alpha\over \sqrt{2}f_\pi})$\\
$n\to \eta^0\bar{\nu}$ & $g_{\eta n}^n={3F-D\over \sqrt{6}f_\pi}$ & $m_{ns,R}=[L^{S,LL}_{ udd}]_{11s1} (-\beta)+[L^{S,RL}_{ dud}]_{11s1}(-\alpha)$ & $y_{\eta s,R}^n=[L^{S,LL}_{ udd}]_{11s1}(-{\beta\over f_\pi}\sqrt{3\over 2})+[L^{S,RL}_{dud}]_{11s1}({\alpha\over \sqrt{6}f_\pi})$\\
$n\to K^0\bar{\nu}$ & $g_{K\Sigma}^n=-{D-F\over \sqrt{2}f_\pi}$ &  $m_{\Sigma s,R}=[L^{S,LL}_{ udd}]_{12s1} (-{\beta\over \sqrt{2}})+[L^{S,RL}_{ dud}]_{21s1} (-{\alpha\over \sqrt{2}})$  & $y_{Ks,R}^n=[L^{S,LL}_{udd}]_{12s1} (-{\beta\over f_\pi})+[L^{S,LL}_{udd}]_{11s2} (-{\beta\over f_\pi})$\\
&& $+[L^{S,RL}_{ ddu}]_{12s1} {-\alpha\over \sqrt{2}}$ & $+[L^{S,RL}_{dud}]_{21s1} ({\alpha\over f_\pi})+[L^{S,RL}_{dud}]_{11s2} (-{\alpha\over f_\pi})$\\
& $g_{K\Lambda}^n=-{D+3F\over \sqrt{6}f_\pi}$ & $m_{\Lambda s,R}=[L^{S,LL}_{ udd}]_{12s1} {\beta\over \sqrt{6}}+[L^{S,LL}_{ udd}]_{11s2}\beta\sqrt{2\over 3}$ & \\
& & $+[L^{S,RL}_{dud}]_{21s1}{\alpha\over \sqrt{6}}+[L^{S,RL}_{dud}]_{11s2}\alpha\sqrt{2\over 3}+[L^{S,RL}_{ddu}]_{12s1} {-\alpha\over \sqrt{6}}$ &\\
\hline
\end{tabular}
}
\caption{
The couplings for BNV decays $N\to M\bar{\nu}$ with $|\Delta (B-L)|=0$. One should additionally multiply the $g$ and $y$ couplings by $-{1\over \sqrt{2}}$ for $n\to K^0_S \bar{\nu}$ decay.
}
\label{tab:2bodyBL0}
\end{table}

\begin{table}[htbp!]
\renewcommand{\arraystretch}{1.2}
\hspace*{-0.05\columnwidth}
\resizebox{1.1\columnwidth}{!}{
\begin{tabular}{c|c|c|c}
\hline
Process & $g_{MB}^N$ & $m_{Bs,X}$ & $y_{Ms,X}^N$\\
\hline
$p\to \pi^+ \nu$ & $g_{\pi n}^p={D+F\over f_\pi}$ & $m_{ns,L}=[L^{S,LR}_{ udd}]_{11s1} (\alpha)+[L^{S,RR}_{ udd}]_{11s1}(\beta)$ & $y_{\pi s,L}^p=[L^{S,LR}_{ udd}]_{11s1}(-{\alpha\over f_\pi})+[L^{S,RR}_{ udd}]_{11s1}(-{\beta\over f_\pi})$\\
$p\to K^+ \nu$ & $g_{K\Sigma}^p={D-F\over \sqrt{2}f_\pi}$ & $m_{\Sigma s,L}=[L^{S,LR}_{ udd}]_{12s1} ({\alpha\over \sqrt{2}})+[L^{S,RR}_{ udd}]_{12s1} ({\beta\over \sqrt{2}})$ &  $y_{K s,L}^p=[L^{S,LR}_{ udd}]_{11s2}(-{\alpha\over f_\pi})+[L^{S,RR}_{ udd}]_{11s2}(-{\beta\over f_\pi})$ \\
&& $+[L^{S,LR}_{ ddu}]_{12s1} {\alpha\over \sqrt{2}}$ & $+[L^{S,LR}_{ ddu}]_{12s1}(-{\alpha\over f_\pi})$\\
& $g_{K\Lambda}^p=-{D+3F\over \sqrt{6}f_\pi}$ & $m_{\Lambda s,L}=[L^{S,LR}_{ udd}]_{12s1} {-\alpha\over \sqrt{6}}+[L^{S,LR}_{ udd}]_{11s2}(-\alpha\sqrt{2\over 3})$ & \\
& & $+[L^{S,RR}_{udd}]_{12s1}{-\beta\over \sqrt{6}}+[L^{S,RR}_{udd}]_{11s2}(-\beta\sqrt{2\over 3})+[L^{S,LR}_{ddu}]_{12s1} {\alpha\over \sqrt{6}}$ &\\
\hline
$n\to \pi^0\nu$ & $g_{\pi n}^n=-{D+F\over \sqrt{2}f_\pi}$ & $m_{ns,L}=[L^{S,LR}_{ udd}]_{11s1} (\alpha)+[L^{S,RR}_{ udd}]_{11s1}(\beta)$ & $y_{\pi s,L}^n=[L^{S,LR}_{ udd}]_{11s1}({\alpha\over \sqrt{2}f_\pi})+[L^{S,RR}_{udd}]_{11s1}({\beta\over \sqrt{2}f_\pi})$\\
$n\to \eta^0\nu$ & $g_{\eta n}^n={3F-D\over \sqrt{6}f_\pi}$ & $m_{ns,L}=[L^{S,LR}_{ udd}]_{11s1} (\alpha)+[L^{S,RR}_{ udd}]_{11s1}(\beta)$ & $y_{\eta s,L}^n=[L^{S,LR}_{ udd}]_{11s1}({\alpha\over \sqrt{6}f_\pi})+[L^{S,RR}_{udd}]_{11s1}(-{\beta\over f_\pi}\sqrt{3\over 2})$\\
$n\to K^0\nu$ & $g_{K\Sigma}^n=-{D-F\over \sqrt{2}f_\pi}$ &  $m_{\Sigma s,L}=[L^{S,LR}_{ udd}]_{12s1} ({\alpha\over \sqrt{2}})+[L^{S,RR}_{ udd}]_{12s1} ({\beta\over \sqrt{2}})$  & $y_{Ks,L}^n=[L^{S,LR}_{udd}]_{12s1} ({\alpha\over f_\pi})+[L^{S,LR}_{udd}]_{11s2} (-{\alpha\over f_\pi})$\\
&& $+[L^{S,LR}_{ ddu}]_{12s1} {\alpha\over \sqrt{2}}$ & $+[L^{S,RR}_{udd}]_{12s1} (-{\beta\over f_\pi})+[L^{S,RR}_{udd}]_{11s2} (-{\beta\over f_\pi})$\\
& $g_{K\Lambda}^n=-{D+3F\over \sqrt{6}f_\pi}$ & $m_{\Lambda s,L}=[L^{S,LR}_{ udd}]_{12s1} {-\alpha\over \sqrt{6}}+[L^{S,LR}_{ udd}]_{11s2}(-\alpha\sqrt{2\over 3})$ & \\
& & $+[L^{S,RR}_{udd}]_{12s1}{-\beta\over \sqrt{6}}+[L^{S,RR}_{udd}]_{11s2}(-\beta\sqrt{2\over 3})+[L^{S,LR}_{ddu}]_{12s1} {\alpha\over \sqrt{6}}$ &\\
\hline
\end{tabular}
}
\caption{
The couplings for BNV decays $N\to M\nu$ with $|\Delta (B-L)|=2$. One should additionally multiply the $g$ and $y$ couplings by $-{1\over \sqrt{2}}$ for $n\to K^0_S \nu$ decay.
}
\label{tab:2bodyBL2}
\end{table}

\begin{figure}[tb!]
\centering
\includegraphics[width=0.7\textwidth]{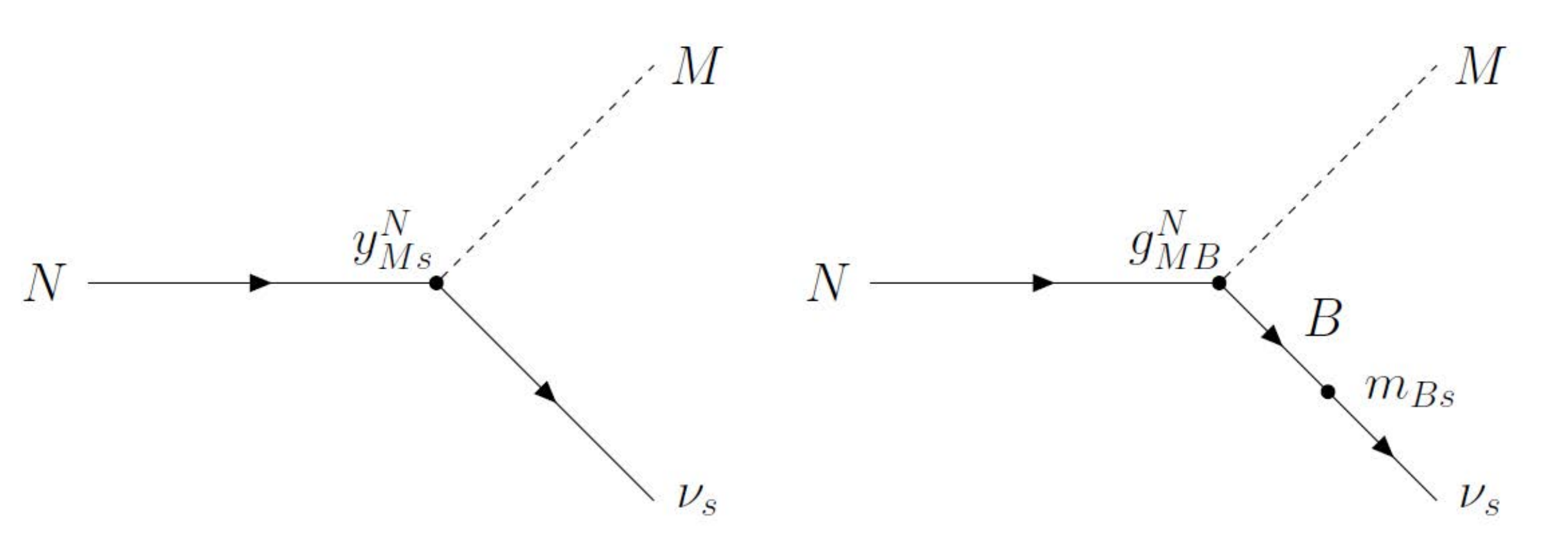}
\caption{The Feynman diagrams for the two-body nucleon decay $N\to M \nu_s$.
}
\label{fig:Feynman}
\end{figure}

The $|\Delta(B-L)|=2$ effective Lagrangian of relevant interactions is
\begin{eqnarray}
\mathcal{L}=g_{MB}^N \overline{B} \gamma^\mu \gamma_5 N \partial_\mu M + m_{Bs,L} \overline{\nu_s} P_R  B + iy_{Ms,L}^N \overline{\nu_s} P_R N M \;.
\end{eqnarray}
The matrix element of two-body decay $N\to M\nu$ then becomes
\begin{eqnarray}
i\mathcal{M}=\overline{u_{\nu_s}} P_R (-y_{Ms,L}^N+\sum_B m_{Bs,L}{\cancel{k}+m_B\over k^2-m_B^2} g_{MB}^N \cancel{p}_M \gamma_5) u_N\;,
\end{eqnarray}
where $p_M$ denotes the outgoing meson momentum in the final state. The spin-averaged squared matrix element thus becomes
\begin{eqnarray}
\overline{|\mathcal{M}|^2}/m_N^2&=&{1\over 2}|y_{Ms,L}^N|^2 (1+x_s^2-x_M^2)+{1\over 2}\sum_{B,B'}g_{MB}^N g_{MB'}^{N*} m_{Bs,L} m_{B's,L}^\ast g(x_B,x_{B'})\nonumber\\
&&-\sum_B {\rm Re}(y_{Ms,L}m_{Bs,L}^\ast g_{MB}^{N\ast})h(x_B)\;,
\label{eq:matrixelement}
\end{eqnarray}
where $x_s=m_{\nu_s}/m_N$, $x_M=m_M/m_N$, $x_{B^{(\prime)}}=m_{B^{(\prime)}}/m_N$, and
\begin{eqnarray}
g(x_1,x_2)&=&{(x_s^2+x_1x_2)(1-x_M^2-x_s^2(2+x_M^2)+x_s^4)-2x_s^2x_M^2(x_1+x_2)\over (x_1^2-x_s^2)(x_2^2-x_s^2)}\;,\\
h(x_B)&=&{x_B(1-x_M^2-x_s^2)-x_s^2(1+x_M^2-x_s^2)\over x_B^2-x_s^2}\;.
\end{eqnarray}
The decay width is given by
\begin{eqnarray}
\Gamma (N\to M \nu_s)=m_N{\lambda^{1/2}(1,x_M^2,x_s^2)\over 16\pi}{\overline{|\mathcal{M}|^2}\over m_N^2}\;,
\end{eqnarray}
where $\lambda(x,y,z)=x^2+y^2+z^2-2xy-2xz-2yz$. We show the illustrative Feynman diagrams for the two-body nucleon decay $N\to M \nu_s$ in Fig.~\ref{fig:Feynman}.

For decay $N\to M\bar{\nu}$ with $|\Delta(B-L)|=0$, we have the substitution of $\nu\to \nu^c$ in the Lagrangian, $R\leftrightarrow L$ in both the Lagrangian and matrix element, and $L\to R$ in the spin-averaged squared matrix element as well as $-\to +$ for the sign in the last term of Eq.~\eqref{eq:matrixelement}.

\subsection{Description of our calculation method}

Next we describe our numerical calculation method of BNV decay rate in the EFT framework.

For EFT calculation, there exist several packages (e.g., Wilson~\cite{Aebischer:2018bkb}, DSixTools~\cite{Celis:2017hod,Fuentes-Martin:2020zaz}) but restricted to SMEFT operators up to dimension 6. Although a recently developed version of the Wilson package~\cite{Aebischer:2024csk} supports $\nu$SMEFT operators, it remains limited to dimension~6 and does not accommodate dimension-7 operators. We have developed our own Mathematica code to perform the numerical calculations of all operators discussed in this paper.

For the running of the gauge couplings, we use the one-loop results from Ref.~\cite{Mihaila:2012pz}.
The RGE can involve WCs with different flavor indices, forming a coupled system of differential equations. Some of these coupled equations can be solved using diagonalization methods. For example, we find that the RGEs for dimension-6 LEFT WCs are amenable to this approach. However, for certain SMEFT RGEs, the equations cannot be diagonalized using a constant matrix, rendering the diagonalization method inapplicable. Consequently, we solve the RGEs for the SMEFT/$\nu$SMEFT WCs directly using the Mathematica function \texttt{NDSolve}.

Our constraints are presented in two forms. The first scenario fixes the dimensionless WC at unity and explores the corresponding energy scale. The second one establishes the energy scale at a fixed point, specifically $\Lambda = 10^{15}\,\text{GeV}$ for dimension-6 operators and $\Lambda = 10^{11}\,\text{GeV}$ for dimension-7 operators, and investigates the permissible range of the dimensionless WCs. The initial conditions for RGEs are set accordingly in these two scenarios.
Special care must be taken when operators and WCs possess flavor index symmetries, as this leads to redundancies in both the operators and the WCs. To address this, symmetry relations are applied to eliminate redundant degrees of freedom. The resulting independent WCs are expressed as linear combinations of the original WCs, and the RGE initial conditions are formulated in terms of these independent combinations. By solving the corresponding linear relationships, these can then be converted to initial conditions for the original WCs.
For example, we consider the symmetry of the WC $C_{QQLQ,prst}$ as an illustration. By setting $p = r = 1$ and $t = 2$, the three WCs $C_{QQLQ,11s2}$, $C_{QQLQ,12s1}$, and $C_{QQLQ,21s1}$ couple in the RGE of Eq.~\eqref{eq:RGE_QQLQ}. From the operator and WC symmetry relations in Eq.~\eqref{eq:Osym} and Eq.~\eqref{eq:Csym}, we have
\begin{align}
  \mathcal{O}_{QQLQ,11s2} &= \mathcal{O}_{QQLQ,12s1}, \\
  2C_{QQLQ,11s2} &= C_{QQLQ,12s1} + C_{QQLQ,21s1}.
\end{align}
The non-redundant basis is expressed as
\begin{equation}
  \begin{split}
    &C_{QQLQ,11s2}\mathcal{O}_{QQLQ,11s2} + C_{QQLQ,12s1}\mathcal{O}_{QQLQ,12s1} + C_{QQLQ,21s1}\mathcal{O}_{QQLQ,21s1} \\
    =& (C_{QQLQ,11s2} + C_{QQLQ,21s1})\mathcal{O}_{QQLQ,21s1} + (2C_{QQLQ,11s2} - C_{QQLQ,21s1})\mathcal{O}_{QQLQ,12s1}.
  \end{split}
\end{equation}
Now we consider a scenario in which only $\mathcal{O}_{QQLQ,21s1}$ is turned on in the non-redundant basis, while all other operators are turned off. We set the WCs in the new basis as $C'_{QQLQ,21s1} = C_{QQLQ,11s2} + C_{QQLQ,21s1} = c$ and $C'_{QQLQ,12s1} = 2C_{QQLQ,11s2} - C_{QQLQ,21s1} = 0$. Solving these relations yields the initial conditions for the original WCs
\begin{equation}
  C_{QQLQ,11s2} = \frac{c}{3}, \quad C_{QQLQ,21s1} = \frac{2c}{3}, \quad C_{QQLQ,12s1} = 0.
\end{equation}

With these initial conditions defined at high-energy scale $\Lambda$, we evolve the WCs through the RGEs to systematically connect UV physics to low-energy observables. This enables us to investigate BNV nucleon decay within a model-independent framework using EFTs. Assuming NP occurs at a high scale $\Lambda \gg m_Z$, we employ the SMEFT with dimension-6 and dimension-7 BNV operators to describe physics between $\Lambda$ and the electroweak scale $m_Z$. Below $m_Z$, we transit to the LEFT governed by $\mathrm{SU}(3)_C \times \mathrm{U}(1)_{\text{EM}}$ symmetry, and finally match to BChPT below 2~GeV to compute nucleon decay amplitudes.
Our analysis proceeds in three stages:
\begin{itemize}
\item RG evolution of SMEFT WCs from $\Lambda$ to $m_Z$ at one-loop accuracy
\item Tree-level matching to LEFT at $m_Z$, followed by RG running to hadronic scales with sequential decoupling of heavy quarks and lepton ($t$ at $m_Z$, $b$ at 5~GeV, $c$ and $\tau$ at 2~GeV)
\item Matching to BChPT operators to calculate two-body nucleon decay rates, supplemented by lattice QCD results where necessary
\end{itemize}
Consequently, by employing this framework, we are able to connect low-energy constraints on BNV nucleon decay to derive bounds on the WCs governing NP at high-energy scales.

\section{Numerical results}
\label{sec:Results}

\subsection{Overview of nucleon decay searches with missing energy}
\label{sec:exp}

Before performing a numerical analysis, we review the relevant experimental searches in light of the different decay kinematics for massive sterile neutrinos. The most sensitive searches for BNV two-body nucleon decays into a pseudoscalar meson and a neutrino $\nu$ or antineutrino $\bar\nu$ ($N\to P + \nu$) are listed in Tab.~\ref{tab:process}. As neither the flavor nor the nature of the final state neutrino is determined experimentally, we collectively denote them $\nu$. Decay rates have to be summed over all possible neutrino and antineutrino flavors. Neglecting active-sterile neutrino mixing, this is equivalent to considering neutrino flavor eigenstates for active neutrinos.
The operators with first generation quarks will be mostly constrained by $N\to \pi\nu$ and the operators with strange quarks by $N\to K\nu$. Decays to heavier mesons like $\eta^0$ are less sensitive. We do not discuss decays to vector mesons, e.g. $K^*$, $\rho$ and $\omega$ which are heavier and thus the decays are limited by the available phase space.

Super-K searched for BNV nucleon decay $N\to \pi \nu$ using data from a combined exposure of 172.8 $\rm kton\cdot years$ from the Super-Kamiokande-I, -II and -III~\cite{Super-Kamiokande:2013rwg}. The analyses for both decays are searching for a single pion signal above the pion background from atmospheric neutrinos. Super-K demands two showering (electron-like) Cherenkov rings from neutral pion decay without electrons from muon decay for $n\to\pi\nu$, and one non-showering (muon-like) Cherenkov ring and at most one electron from muon decay for $p\to\pi^+\nu$. The reconstructed momentum is required to be less than 1 GeV and the reconstructed invariant $\pi^0$ mass lies in $[85,185]$ MeV. All of the selection criteria are also satisfied for decays to massive sterile neutrinos, but the reconstruction efficiency depends on the $\pi^0$ and $\mu^+$ momenta as illustrated in figure 1 in Ref.~\cite{Super-Kamiokande:2013rwg}.

Accounting for this momentum dependence requires to carry out a detector simulation including the effects of bound state nucleons which is beyond the scope of this paper. In this paper we take a simplified approach. A massive sterile neutrino generally results in a smaller meson momentum. For $n\to \pi^0 \nu$ this results in a higher event selection efficiency due to the smaller $\pi^0$ momentum and thus we expect the constraint with a sterile neutrino to become slightly more stringent. For non-showering events in $p\to \pi^+ \nu$, the event selection efficiency is suppressed for momenta $\lesssim 200$ MeV, while it can be treated as approximately constant for larger momenta. We thus apply the constraint for $p\to \pi^+\nu$ for pion momenta larger than 200 MeV or sterile neutrinos lighter than $665$ MeV.

In Ref.~\cite{Super-Kamiokande:2014otb}, Super-K reported the results for the search of the proton decay mode $p\to K^+\nu$. The analysis uses three different methods. The first two search for $K^+\to \mu^+\nu_\mu$, where the kaon decays at rest but with different tagging strategies. The first method requires a prompt photon~\footnote{The energy of the photon depends on the excited nuclear state. The most probably state is $p_{3/2}$ with a photon energy of $6.3$ MeV.} from the nuclear de-excitation after the decay $p\to K^+\nu$ with a time difference which is consistent with the kaon lifetime and demands the muon momentum to lie in the range $[215,260]$ MeV. The second method does not require those two constraints and performs a spectral fit to the muon momentum instead. The third method considers the decay $K^+\to \pi^+ \pi^0$. All three methods search for kaon decay at rest and thus are largely independent of the kaon momentum. As long as the decay is kinematically allowed, the limit can be directly translated to decays with a sterile neutrino.

The most stringent constraint on the related neutron decay mode $n\to K^0_S \nu$ has been placed by Super-K in Ref.~\cite{Super-Kamiokande:2005lev}~\footnote{Decays to $K_L^0$ are not considered since their lifetime is long and many will scatter before decaying.}. Similar to $p\to K^+\nu$, the analysis searches for kaon decay at rest to two pions. Thus the limits also apply to massive sterile neutrinos, as long as the decays are kinematically allowed.

The Irvine-Michigan-Brookhaven-3 (IMB-3) experiment placed the most stringent limit on $n\to \eta^0 \nu$~\cite{McGrew:1999nd}.
It is not straightforward to recast the limit to sterile neutrinos, because it requires a detector simulation and Ref.~\cite{McGrew:1999nd} does not provide enough details about the applied cuts. We naively apply the limits for light kinematically accessible sterile neutrinos, but caution the reader that the obtained limits may only be trusted for negligible sterile neutrino masses.

\subsection{Numerical results for the two-body BNV nucleon decays}

In this section, we show the numerical results for the two-body BNV decay rates $N\to M \nu (\bar \nu)$.
We adopt the following hadronic parameters for numerical calculations
\begin{eqnarray}
&&\alpha = -0.01257(111)~{\rm GeV}^3~\text{\cite{Yoo:2021gql}}\;,~~~\beta= 0.01269(107)~{\rm GeV}^3~\text{\cite{Yoo:2021gql}}\;, \\
&&D=0.730(11)~\text{\cite{Bali:2022qja}}\;,~~~F=0.447^{(6)}_{(7)}~\text{\cite{Bali:2022qja}}\;,~~~f_\pi= 130.41(20)~{\rm MeV}~\text{\cite{Workman:2022ynf}}\;.
\end{eqnarray}
For convenience, in SMEFT and $\nu$SMEFT, we redefine the dimensionful WC for effective operator $\mathcal{O}_i$ with dimension $d_i$ as
\begin{eqnarray}
C_i = {c_i\over \Lambda^{d_i-4}}\;,
\end{eqnarray}
where $c_i$ denotes a corresponding dimensionless WC and $\Lambda$ is the UV scale. We assume that the dimensionless WCs are real unless otherwise stated.

\subsubsection{SMEFT only}

We first show the lower bounds on the UV energy scale for one dimension-6 (left) or dimension-7 (right) SMEFT operator at a time in Tab.~\ref{tab:SMEFTdim67}.
The second column represents the decay mode giving the most stringent limit shown in the third column. For dimension-6 operator $\mathcal{O}_{duLQ}$, the two most stringent bounds are $2.98\times 10^{15}~{\rm GeV}$ for $\mathcal{O}_{duLQ,11s1}$ from $n\to \pi^0\bar{\nu}$ and $4.07\times 10^{15}~{\rm GeV}$ for $\mathcal{O}_{duLQ,11s2}$ from $p\to K^+\bar{\nu}$. For operator $\mathcal{O}_{QQLQ}$, they are $4.15\times 10^{15}~{\rm GeV}$ for $\mathcal{O}_{QQLQ,11s1}$ from $n\to \pi^0\bar{\nu}$ and $4.58\times 10^{15}~{\rm GeV}$ for $\mathcal{O}_{QQLQ,11s2}$ and $\mathcal{O}_{QQLQ,21s1}$ from $p\to K^+\bar{\nu}$. For dimension-7 operator $\mathcal{O}_{udLdH}$, the two most stringent bounds are $1.01\times 10^{11}~{\rm GeV}$ for $\mathcal{O}_{udLdH,11s1}$ from $n\to \pi^0 \nu$ and $1.22\times 10^{11}~{\rm GeV}$ for $\mathcal{O}_{udLdH,11s2}$ from $p\to K^+ \nu$. For operator $\mathcal{O}_{QQLdH}$, they are $1.05\times 10^{11}~{\rm GeV}$ for $\mathcal{O}_{QQLdH,11s1}$ from $n\to \pi^0 \nu$ and $1.29\times 10^{11}~{\rm GeV}$ for $\mathcal{O}_{QQLdH,11s2}$ from $p\to K^+ \nu$.

\begin{table}[htb!]
\centering
\renewcommand{\arraystretch}{1.2}
\begin{minipage}{0.5\textwidth}
    \centering
\begin{tabular}{c|c|c}
\hline
Wilson coefficients & Processes & Bounds [GeV] \\
\hline
$\Lambda/\sqrt[4]{\sum_s c^2_{duLQ,11s1}}$ & $n \to \pi^0 \bar{\nu}$ & $2.98\times 10^{15}$\\ \hline
$\Lambda/\sqrt[4]{\sum_s c^2_{duLQ,11s2}}$ & $p \to K^+ \bar{\nu}$ & $4.07\times 10^{15}$\\ \hline
$\Lambda/\sqrt[4]{\sum_s c^2_{duLQ,11s3}}$ & $p \to K^+ \bar{\nu}$ & $8.11\times 10^{14}$\\ \hline
$\Lambda/\sqrt[4]{\sum_s c^2_{duLQ,21s1}}$ & $p \to K^+ \bar{\nu}$ & $2.04\times 10^{15}$\\ \hline
$\Lambda/\sqrt[4]{\sum_s c^2_{duLQ,21s2}}$ & $p \to K^+ \bar{\nu}$ & $9.73\times 10^{14}$\\ \hline
$\Lambda/\sqrt[4]{\sum_s c^2_{duLQ,21s3}}$ & $p \to K^+ \bar{\nu}$ & $1.88\times 10^{14}$\\ \hline
\hline
$\Lambda/\sqrt[4]{\sum_s c^2_{QQLQ,11s1}}$ & $n \to \pi^0 \bar{\nu}$ & $4.15\times 10^{15}$\\ \hline
$\Lambda/\sqrt[4]{\sum_s c^2_{QQLQ,11s2}}$ & $p \to K^+ \bar{\nu}$ & $4.58\times 10^{15}$\\ \hline
$\Lambda/\sqrt[4]{\sum_s c^2_{QQLQ,21s1}}$ & $p \to K^+ \bar{\nu}$ & $4.58\times 10^{15}$\\ \hline
$\Lambda/\sqrt[4]{\sum_s c^2_{QQLQ,12s1}}$ & $p \to K^+ \bar{\nu}$ & $2.02\times 10^{15}$\\ \hline
$\Lambda/\sqrt[4]{\sum_s c^2_{QQLQ,11s3}}$ & $p \to K^+ \bar{\nu}$ & $9.07\times 10^{14}$\\ \hline
$\Lambda/\sqrt[4]{\sum_s c^2_{QQLQ,31s1}}$ & $p \to K^+ \bar{\nu}$ & $9.07\times 10^{14}$\\ \hline
$\Lambda/\sqrt[4]{\sum_s c^2_{QQLQ,13s1}}$ & $p \to K^+ \bar{\nu}$ & $4.19\times 10^{14}$\\ \hline
$\Lambda/\sqrt[4]{\sum_s c^2_{QQLQ,21s2}}$ & $p \to K^+ \bar{\nu}$ & $2.39\times 10^{15}$\\ \hline
$\Lambda/\sqrt[4]{\sum_s c^2_{QQLQ,22s1}}$ & $p \to K^+ \bar{\nu}$ & $1.27\times 10^{15}$\\ \hline
$\Lambda/\sqrt[4]{\sum_s c^2_{QQLQ,12s2}}$ & $p \to K^+ \bar{\nu}$ & $1.27\times 10^{15}$\\ \hline
$\Lambda/\sqrt[4]{\sum_s c^2_{QQLQ,21s3}}$ & $p \to K^+ \bar{\nu}$ & $4.62\times 10^{14}$\\ \hline
$\Lambda/\sqrt[4]{\sum_s c^2_{QQLQ,32s1}}$ & $p \to K^+ \bar{\nu}$ & $2.44\times 10^{14}$\\ \hline
$\Lambda/\sqrt[4]{\sum_s c^2_{QQLQ,31s2}}$ & $p \to K^+ \bar{\nu}$ & $4.59\times 10^{14}$\\ \hline
$\Lambda/\sqrt[4]{\sum_s c^2_{QQLQ,23s1}}$ & $p \to K^+ \bar{\nu}$ & $2.44\times 10^{14}$\\ \hline
$\Lambda/\sqrt[4]{\sum_s c^2_{QQLQ,13s2}}$ & $p \to K^+ \bar{\nu}$ & $2.38\times 10^{14}$\\ \hline
$\Lambda/\sqrt[4]{\sum_s c^2_{QQLQ,12s3}}$ & $p \to K^+ \bar{\nu}$ & $2.44\times 10^{14}$\\ \hline
$\Lambda/\sqrt[4]{\sum_s c^2_{QQLQ,31s3}}$ & $p \to K^+ \bar{\nu}$ & $9.08\times 10^{13}$\\ \hline
$\Lambda/\sqrt[4]{\sum_s c^2_{QQLQ,33s1}}$ & $p \to K^+ \bar{\nu}$ & $4.59\times 10^{13}$\\ \hline
$\Lambda/\sqrt[4]{\sum_s c^2_{QQLQ,13s3}}$ & $p \to K^+ \bar{\nu}$ & $4.59\times 10^{13}$\\ \hline
\end{tabular}
\end{minipage}%
  \hspace{0.05\textwidth}
\begin{minipage}{0.35\textwidth}
    \centering
\begin{tabular}{c|c|c}
\hline
Wilson coefficients & Processes & Bounds [GeV] \\
\hline
$\Lambda/\sqrt[6]{\sum_s c^2_{udLdH,11s1}}$ & $n \to \pi^0 \nu$ & $1.01\times 10^{11}$\\ \hline
$\Lambda/\sqrt[6]{\sum_s c^2_{udLdH,11s2}}$ & $p \to K^+ \nu$ & $1.22\times 10^{11}$\\ \hline
$\Lambda/\sqrt[6]{\sum_s c^2_{udLdH,12s1}}$ & $p \to K^+ \nu$ & $8.28\times 10^{10}$\\ \hline
\hline
$\Lambda/\sqrt[6]{\sum_s c^2_{QQLdH,11s1}}$ & $n \to \pi^0 \nu$ & $1.05\times 10^{11}$\\ \hline
$\Lambda/\sqrt[6]{\sum_s c^2_{QQLdH,11s2}}$ & $p \to K^+ \nu$ & $1.29\times 10^{11}$\\ \hline
$\Lambda/\sqrt[6]{\sum_s c^2_{QQLdH,12s1}}$ & $p \to K^+ \nu$ & $7.79\times 10^{10}$\\ \hline
$\Lambda/\sqrt[6]{\sum_s c^2_{QQLdH,21s1}}$ & $p \to K^+ \nu$ & $4.01\times 10^{10}$\\ \hline
$\Lambda/\sqrt[6]{\sum_s c^2_{QQLdH,12s2}}$ & $p \to K^+ \nu$ & $7.53\times 10^{10}$\\ \hline
$\Lambda/\sqrt[6]{\sum_s c^2_{QQLdH,21s2}}$ & $p \to K^+ \nu$ & $3.88\times 10^{10}$\\ \hline
$\Lambda/\sqrt[6]{\sum_s c^2_{QQLdH,13s1}}$ & $p \to K^+ \nu$ & $2.78\times 10^{10}$\\ \hline
$\Lambda/\sqrt[6]{\sum_s c^2_{QQLdH,31s1}}$ & $p \to K^+ \nu$ & $5.24\times 10^9$\\ \hline
$\Lambda/\sqrt[6]{\sum_s c^2_{QQLdH,13s2}}$ & $p \to K^+ \nu$ & $2.64\times 10^{10}$\\ \hline
$\Lambda/\sqrt[6]{\sum_s c^2_{QQLdH,31s2}}$ & $p \to K^+ \nu$ & $4.94\times 10^9$\\ \hline
\end{tabular}
\end{minipage}
\caption{The bounds on the UV energy scale for each relevant dimension-6 (left) or dimension-7 (right) SMEFT operator.
}
\label{tab:SMEFTdim67}
\end{table}

\begin{figure}[htb!]
    \centering
    \includegraphics[width=\linewidth]{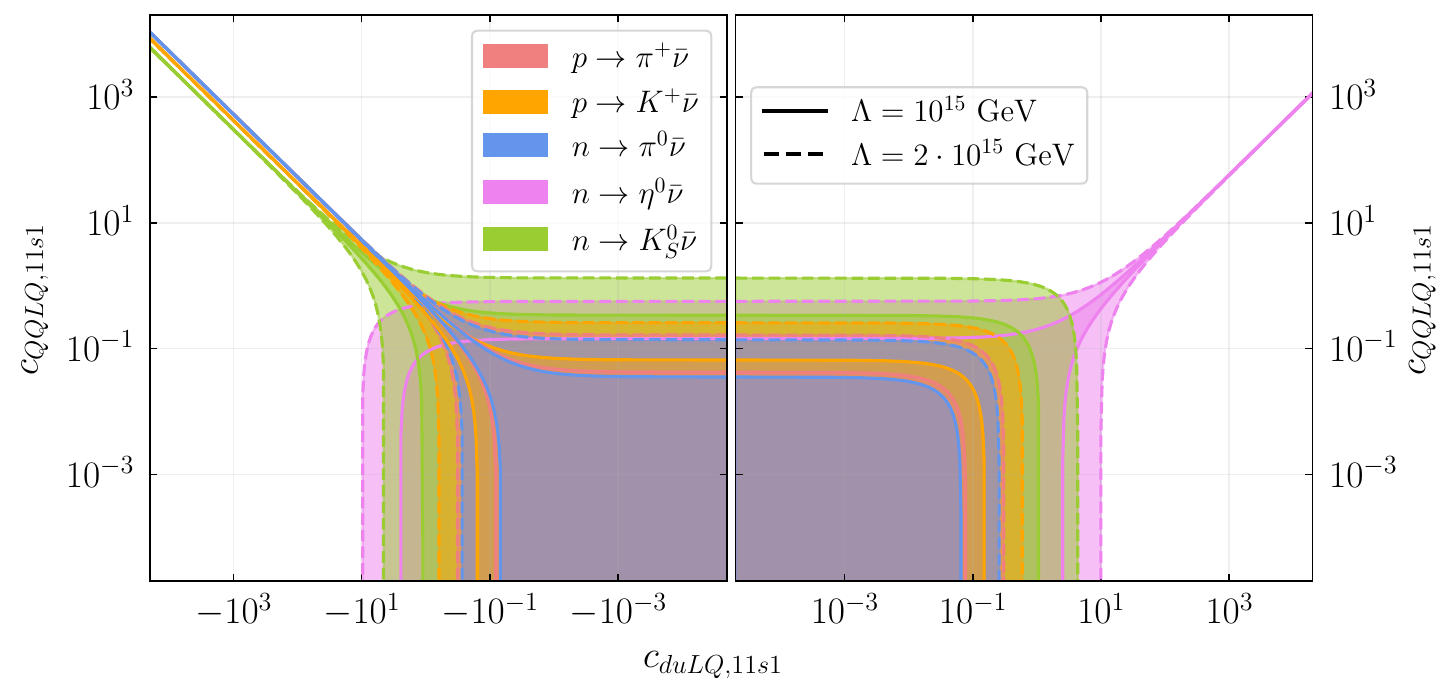}
    \caption{The allowed parameter spaces for the dimensionless WCs $c_{duLQ,11s1}$ and $c_{QQLQ,11s1}$ at the UV scale, illustrated for two cutoff scales: $\Lambda=10^{15}$~GeV (solid line) and $\Lambda=2\times10^{15}$~GeV (dashed line). In this analysis, we assume that $c_{duLQ,11s1}$ or $c_{QQLQ,11s1}$ is universal for $s=1,2,3$. The constraints on these coefficients are determined by applying the current experimental limits, as provided in Tab.~\ref{tab:process}, to a range of relevant processes.}
    \label{fig:dim6_two_wc}
\end{figure}

\begin{figure}[htb!]
    \centering
    \includegraphics[width=\linewidth]{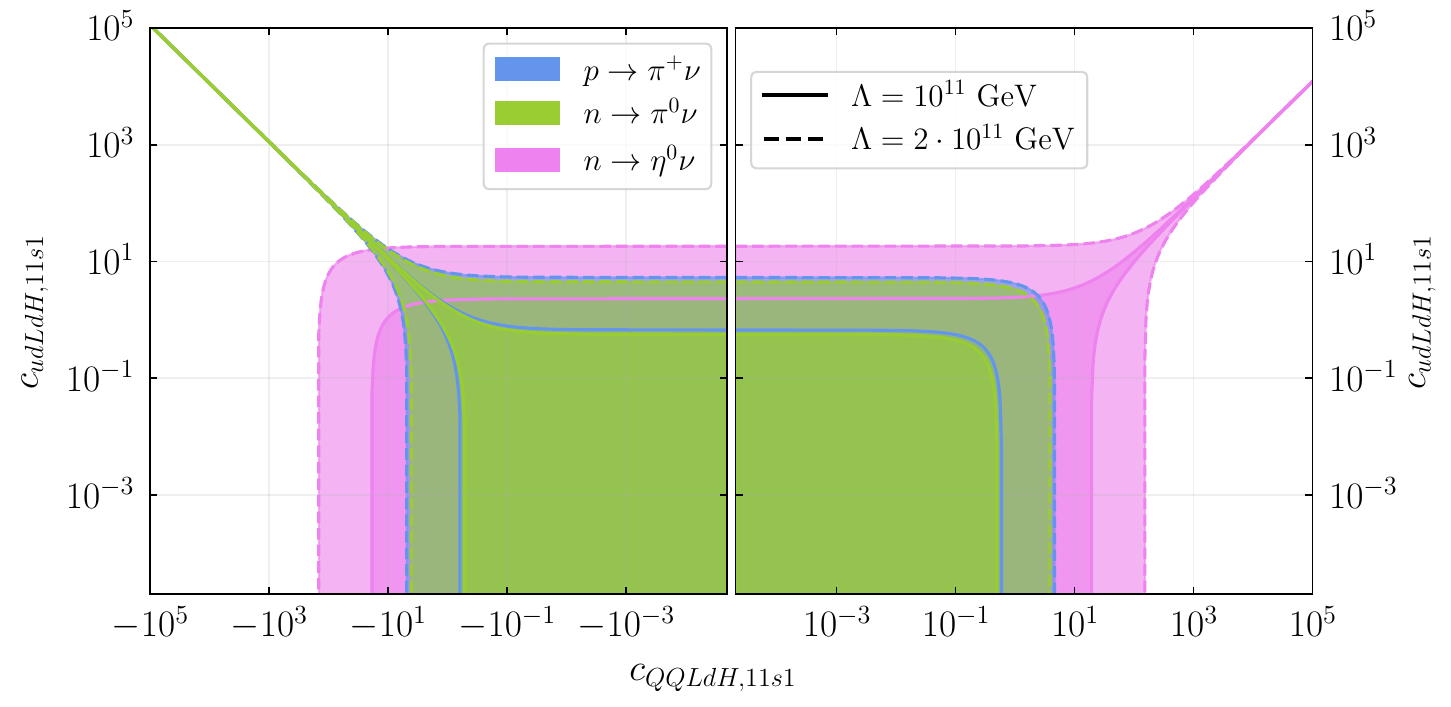}
    \caption{The allowed parameter spaces for the dimensionless WCs $c_{QQLdH,11s1}$ and $c_{udLdH,11s1}$ at the UV scale, illustrated for two cutoff scales: $\Lambda=10^{11}$~GeV (solid line) and $\Lambda=2\times10^{11}$~GeV (dashed line). In this analysis, we assume that $c_{QQLdH,11s1}$ or $c_{udLdH,11s1}$ is universal for $s=1,2,3$. The constraints on these coefficients are determined by applying the current experimental limits, as provided in Tab.~\ref{tab:process}, to a range of relevant processes.}
    \label{fig:dim7_two_wc}
\end{figure}

Next we discuss the case in the presence of two SMEFT operators. For illustration, Fig.~\ref{fig:dim6_two_wc}~(Fig.~\ref{fig:dim7_two_wc}) shows the allowed region in the parameter space of both dimensionless WCs $c_{duLQ,11s1}$ and $c_{QQLQ,11s1}$ for dimension-6 operators ($c_{udLdH,11s1}$ and $c_{QQLdH,11s1}$ for dimension-7 operators). We choose two illustrative UV cutoff scales: $\Lambda=10^{15}$~GeV (solid line) and $\Lambda=2\times10^{15}$~GeV (dashed line).
In this analysis, we assume that each WC is universal for $s=1,2,3$.
The constraints on these coefficients are determined by applying the current experimental limits from the relevant nucleon decays in the legend. They exhibit flat directions in the parameter space of logarithmic coordinates. The decay $n\to \eta^0 \bar{\nu}~(\nu)$ loses sensitivity in the correlated regime of $c_{QQLQ,11s1}\approx 0.1 c_{duLQ,11s1}$ ($c_{udLdH,11s1}\approx 0.1 c_{QQLdH,11s1}$).
Other relevant decays are not sensitive to $c_{QQLQ,11s1}\approx -0.5 c_{duLQ,11s1}$ or $c_{udLdH,11s1}\approx -c_{QQLdH,11s1}$.

\begin{figure}[htb!]
    \centering
    \includegraphics[width=0.8\linewidth]{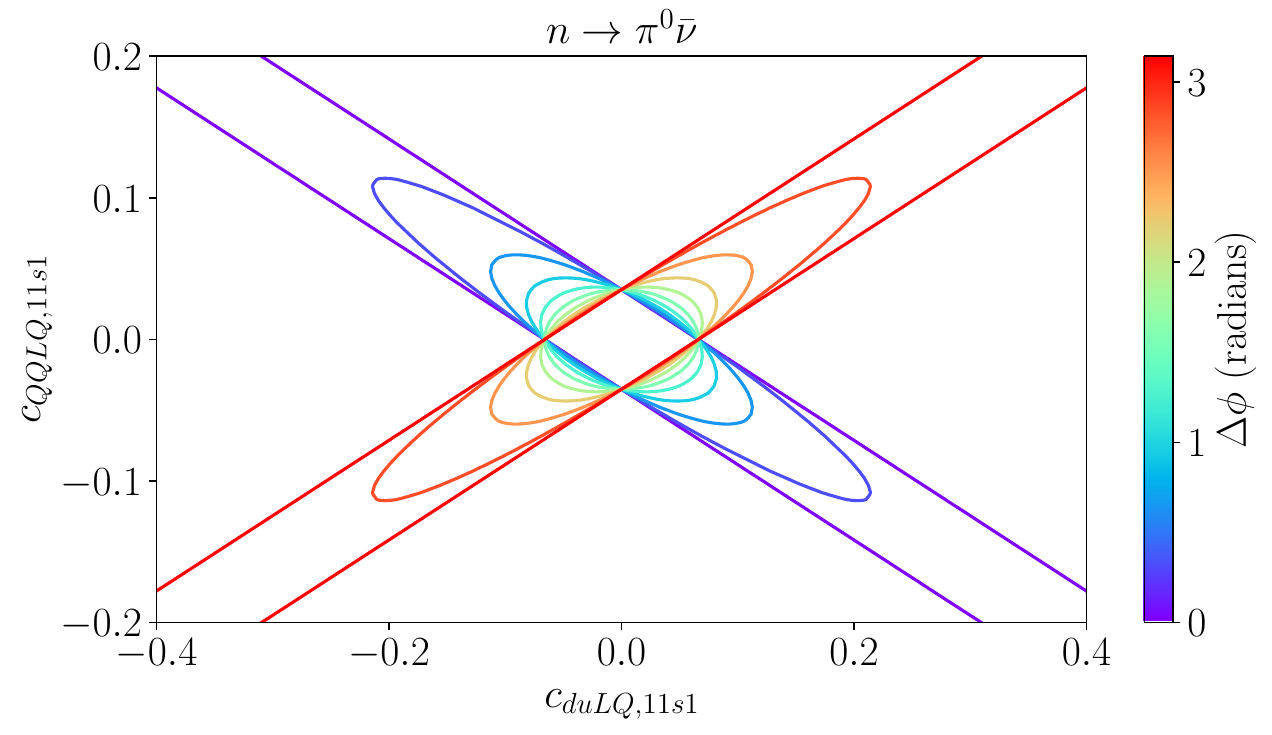}
    \caption{
    The allowed parameter spaces for the dimensionless WCs $c_{duLQ,11s1}$ and $c_{QQLQ,11s1}$ at the UV scale $\Lambda=10^{15}$~GeV, constrained by the process $n\to\pi^0\bar{\nu}$. The different colors represent varying phase difference between these two WCs, as indicated by the color bar legend. In this analysis, we assume that the parameter $c_{duLQ,11s1}$ or $c_{QQLQ,11s1}$ is universal for $s=1,2,3$.}
    \label{fig:dim6_two_wc_phase}
\end{figure}

In Fig.~\ref{fig:dim6_two_wc_phase}, we also display the impact of the phase difference between two WCs. The regions inside the contours denote the allowed parameter spaces for the dimensionless WCs $c_{duLQ,11s1}$ and $c_{QQLQ,11s1}$ at the UV scale $\Lambda=10^{15}$~GeV, constrained by the process $n\to\pi^0\bar{\nu}$.
The different colors represent varying phase difference between these two WCS defined by the substitution of real coefficients $c_{duLQ,11s1}\to c_{duLQ,11s1}$, $c_{QQLQ,11s1}\to c_{QQLQ,11s1}e^{i\Delta\phi}$ in our calculation, as indicated by the color bar legend. When the phase is zero, one can see that the allowed band is oriented along the northwest-southeast direction. As the phase difference increases, the allowed elliptical space narrows. At a phase difference of $\pi/2$, the preferred orientation shifts to the southwest-northeast direction, and the region continues to expand until it forms a parallel band when the phase reaches $\pi$.

\subsubsection{$\nu$SMEFT only}

Here we show the results for $\nu$SMEFT and the impact of sterile neutrino on nucleon decays. Tab.~\ref{tab:nuSMEFTdim67} summarizes the bounds on the UV energy scale for each dimension-6 (left) or dimension-7 (right) $\nu$SMEFT operator, assuming negligible sterile neutrino mass. The most stringent bound for dimension-6 (dimension-7) operator is $5.6\times 10^{15}~{\rm GeV}$ ($1.22\times 10^{11}~{\rm GeV}$) from $p\to K^+ \nu$ ($p\to K^+ \bar{\nu}$).

Note that the couplings and decay rates for BNV decays are both given with respect to the LEFT WCs, as shown in Tabs.~\ref{tab:2bodyBL0} and \ref{tab:2bodyBL2}. However, Tab.~\ref{tab:nuSMEFT} shows that SMEFT and $\nu$SMEFT operators with the same $B-L$ number have opposite neutrino chiralities. The LEFT operators are all expressed by $\nu_L$ and its charge conjugate state $\nu_L^c$. Thus, as shown in Tab.~\ref{tab:nuLEFTBL}, dimension-6 (7) SMEFT operators with $\Delta(B-L) = 0$ ($|\Delta(B-L)| = 2$) and dimension-7 (6) $\nu$SMEFT operators with $|\Delta(B-L)| = 2$ ($\Delta(B-L) = 0$) are matched to the same LEFT operators but with different neutrino flavor indices. Note that $\mathcal{O}_{ddu}^{S,RL}$ can only be induced by $\mathcal{O}_{HdNQ}$ operator. One can see that $\Delta(B-L) = 0$ decays $N \to M \overline{\nu}$ ($|\Delta(B-L)| = 2$
decays $N \to M\nu$) are indeed induced by LEFT operators with $\Delta(B-L) = 0$ ($|\Delta(B-L)| = 2$).

\begin{table}[htb!]
\centering
\renewcommand{\arraystretch}{1.2}
\begin{minipage}{0.5\textwidth}
    \centering
\begin{tabular}{c|c|c}
\hline
Wilson coefficients & Processes & Bounds [GeV] \\
\hline
$\Lambda/\sqrt{|c_{QQNd,11s1}|}$ & $n \to \pi^0 \nu$ & $4.1\times 10^{15}$\\ \hline
$\Lambda/\sqrt{|c_{QQNd,11s2}|}$ & $p \to K^+ \nu$ & $5.6\times 10^{15}$\\ \hline
$\Lambda/\sqrt{|c_{QQNd,12s1}|}$ & $p \to K^+ \nu$ & $1.98\times 10^{15}$\\ \hline
$\Lambda/\sqrt{|c_{QQNd,12s2}|}$ & $p \to K^+ \nu$ & $1.88\times 10^{15}$\\ \hline
$\Lambda/\sqrt{|c_{QQNd,13s1}|}$ & $p \to K^+ \nu$ & $3.95\times 10^{14}$\\ \hline
$\Lambda/\sqrt{|c_{QQNd,13s2}|}$ & $p \to K^+ \nu$ & $3.64\times 10^{14}$\\ \hline
$\Lambda/\sqrt{|c_{QQNd,21s1}|}$ & $p \to K^+ \nu$ & $1.98\times 10^{15}$\\ \hline
$\Lambda/\sqrt{|c_{QQNd,21s2}|}$ & $p \to K^+ \nu$ & $1.88\times 10^{15}$\\ \hline
$\Lambda/\sqrt{|c_{QQNd,31s1}|}$ & $p \to K^+ \nu$ & $3.95\times 10^{14}$\\ \hline
$\Lambda/\sqrt{|c_{QQNd,31s2}|}$ & $p \to K^+ \nu$ & $3.64\times 10^{14}$\\ \hline\hline
$\Lambda/\sqrt{|c_{udNd,11s1}|}$ & $n \to \pi^0 \nu$ & $2.61\times 10^{15}$\\ \hline
$\Lambda/\sqrt{|c_{udNd,11s2}|}$ & $p \to K^+ \nu$ & $3.51\times 10^{15}$\\ \hline
$\Lambda/\sqrt{|c_{udNd,12s1}|}$ & $p \to K^+ \nu$ & $1.9\times 10^{15}$\\ \hline
\end{tabular}
\end{minipage}%
  \hspace{0.05\textwidth}
\begin{minipage}{0.35\textwidth}
    \centering
\begin{tabular}{c|c|c}
\hline
Wilson coefficients & Processes & Bounds [GeV] \\
\hline
$\Lambda/\sqrt[3]{|c_{HdNQ,12s1}|}$ & $p \to K^+ \bar{\nu}$ & $9.6\times 10^{10}$\\ \hline\hline
$\Lambda/\sqrt[3]{|c_{HduNQ,11s1}|}$ & $n \to \pi^0 \bar{\nu}$ & $9.78\times 10^{10}$\\ \hline
$\Lambda/\sqrt[3]{|c_{HduNQ,11s2}|}$ & $p \to K^+ \bar{\nu}$ & $1.2\times 10^{11}$\\ \hline
$\Lambda/\sqrt[3]{|c_{HduNQ,11s3}|}$ & $p \to K^+ \bar{\nu}$ & $4.18\times 10^{10}$\\ \hline
$\Lambda/\sqrt[3]{|c_{HduNQ,21s1}|}$ & $p \to K^+ \bar{\nu}$ & $7.6\times 10^{10}$\\ \hline
$\Lambda/\sqrt[3]{|c_{HduNQ,21s2}|}$ & $p \to K^+ \bar{\nu}$ & $4.65\times 10^{10}$\\ \hline
$\Lambda/\sqrt[3]{|c_{HduNQ,21s3}|}$ & $p \to K^+ \bar{\nu}$ & $1.59\times 10^{10}$\\ \hline\hline
$\Lambda/\sqrt[3]{|c_{HQNQ,11s1}|}$ & $n \to \pi^0 \bar{\nu}$ & $1.09\times 10^{11}$\\ \hline
$\Lambda/\sqrt[3]{|c_{HQNQ,11s2}|}$ & $p \to K^+ \bar{\nu}$ & $1.22\times 10^{11}$\\ \hline
$\Lambda/\sqrt[3]{|c_{HQNQ,21s1}|}$ & $p \to K^+ \bar{\nu}$ & $1.22\times 10^{11}$\\ \hline
$\Lambda/\sqrt[3]{|c_{HQNQ,12s1}|}$ & $p \to K^+ \bar{\nu}$ & $9.18\times 10^{10}$\\ \hline
$\Lambda/\sqrt[3]{|c_{HQNQ,11s3}|}$ & $p \to K^+ \bar{\nu}$ & $4.23\times 10^{10}$\\ \hline
$\Lambda/\sqrt[3]{|c_{HQNQ,31s1}|}$ & $p \to K^+ \bar{\nu}$ & $4.23\times 10^{10}$\\ \hline
$\Lambda/\sqrt[3]{|c_{HQNQ,13s1}|}$ & $p \to K^+ \bar{\nu}$ & $3.22\times 10^{10}$\\ \hline
$\Lambda/\sqrt[3]{|c_{HQNQ,21s2}|}$ & $p \to K^+ \bar{\nu}$ & $7.99\times 10^{10}$\\ \hline
$\Lambda/\sqrt[3]{|c_{HQNQ,22s1}|}$ & $p \to K^+ \bar{\nu}$ & $4.39\times 10^{10}$\\ \hline
$\Lambda/\sqrt[3]{|c_{HQNQ,12s2}|}$ & $p \to K^+ \bar{\nu}$ & $4.39\times 10^{10}$\\ \hline
$\Lambda/\sqrt[3]{|c_{HQNQ,21s3}|}$ & $p \to K^+ \bar{\nu}$ & $2.74\times 10^{10}$\\ \hline
$\Lambda/\sqrt[3]{|c_{HQNQ,32s1}|}$ & $p \to K^+ \bar{\nu}$ & $1.47\times 10^{10}$\\ \hline
$\Lambda/\sqrt[3]{|c_{HQNQ,31s2}|}$ & $p \to K^+ \bar{\nu}$ & $2.71\times 10^{10}$\\ \hline
$\Lambda/\sqrt[3]{|c_{HQNQ,23s1}|}$ & $p \to K^+ \bar{\nu}$ & $1.71\times 10^{10}$\\ \hline
$\Lambda/\sqrt[3]{|c_{HQNQ,13s2}|}$ & $p \to K^+ \bar{\nu}$ & $1.48\times 10^{10}$\\ \hline
$\Lambda/\sqrt[3]{|c_{HQNQ,12s3}|}$ & $p \to K^+ \bar{\nu}$ & $1.47\times 10^{10}$\\ \hline
$\Lambda/\sqrt[3]{|c_{HQNQ,31s3}|}$ & $p \to K^+ \bar{\nu}$ & $9.42\times 10^9$\\ \hline
$\Lambda/\sqrt[3]{|c_{HQNQ,33s1}|}$ & $p \to K^+ \bar{\nu}$ & $5.01\times 10^9$\\ \hline
$\Lambda/\sqrt[3]{|c_{HQNQ,13s3}|}$ & $p \to K^+ \bar{\nu}$ & $5.01\times 10^9$\\ \hline
\end{tabular}
\end{minipage}
\caption{The bounds on the UV energy scale for each relevant dimension-6 (left) or dimension-7 (right) $\nu$SMEFT operator with $s=4$.
}
\label{tab:nuSMEFTdim67}
\end{table}

\begin{figure}[htb!]
    \centering
    \includegraphics[width=0.45\linewidth]{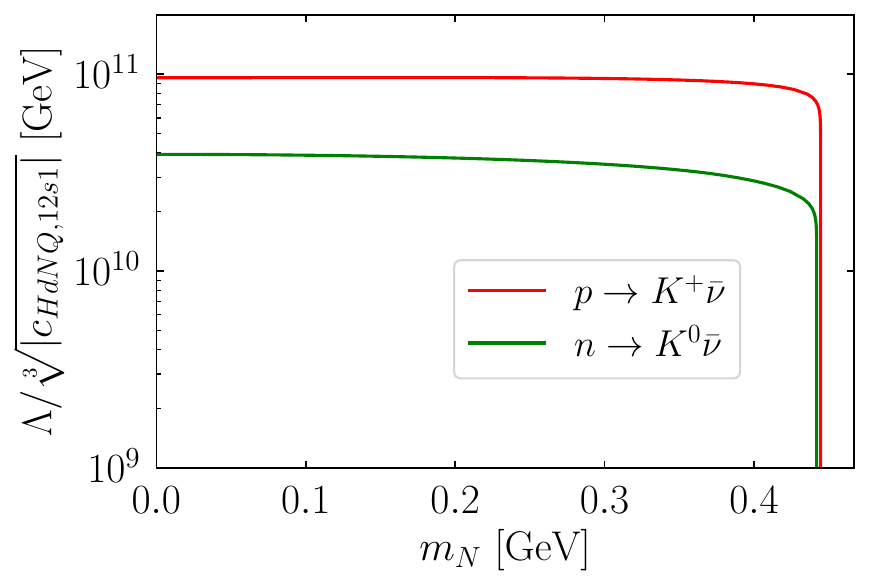}
    \includegraphics[width=0.45\linewidth]{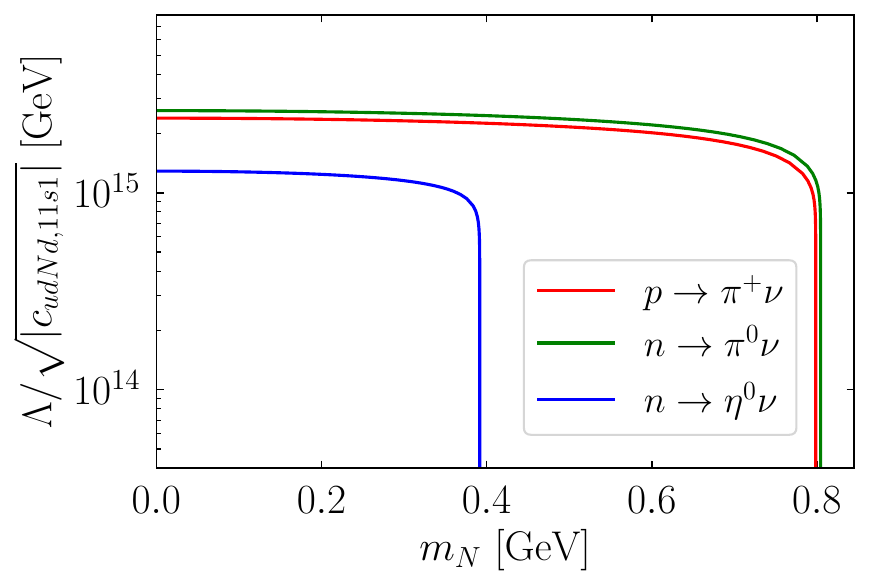}
    \includegraphics[width=0.45\linewidth]{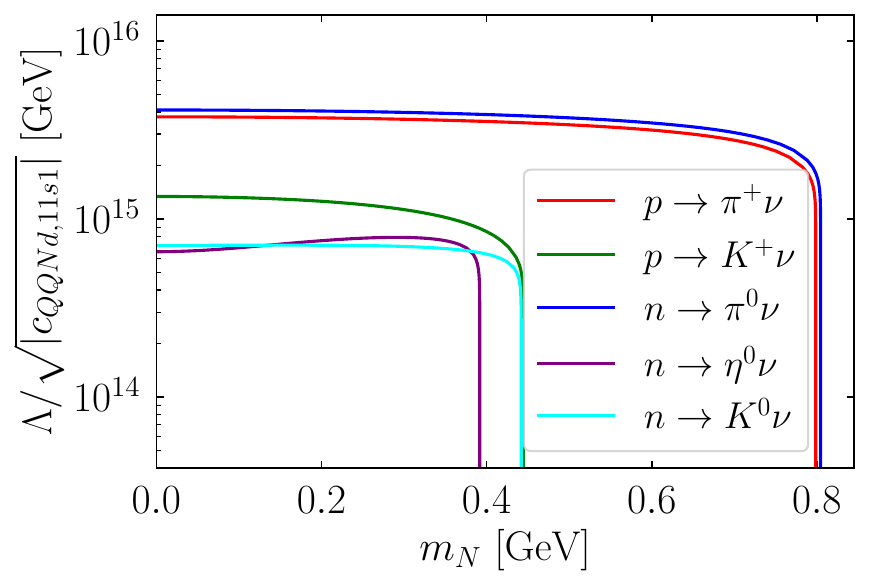}
    \caption{Bounds on three illustrative WCs and UV energy scales in $\nu$SMEFT with $s=4$ derived from different processes as functions of the sterile neutrino mass $m_N$.
    See the discussion in Sec.~\ref{sec:exp} for caveats in reinterpreting the experimental limits. In particular, the $p\to\pi^+\nu$ limit actually doesn't apply for sterile neutrino mass above 665 MeV.
    }
    \label{fig:bounds_numas}
\end{figure}

The illustrative dependence of sterile neutrino mass is shown in Fig.~\ref{fig:bounds_numas}.
As stated before, the dimension-7 operator $\mathcal{O}_{HdNQ}$ can only induce $p\to K^+\bar{\nu}$ and $n\to K^+\bar{\nu}$. It is the only dimension-6 operator $\mathcal{O}_{udNd,11s1}$ which can exclusively induce the three decays to pion/eta mesons: $p\to \pi^+\nu$, $n\to \pi^0\nu$ and $n\to \eta^0 \nu$. Here we show the mass dependence for these two particular operators together with $\mathcal{O}_{QQNd,11s1}$ which can induce all five decays. The results for other operators are collected in App.~\ref{app:mNdep}. One can see that the dependence of sterile neutrino mass relies on different $\nu$SMEFT operators and decay processes. Generally, the UV scale bound decreases as the sterile neutrino mass increases.

\subsubsection{Both SMEFT and $\nu$SMEFT}

\begin{figure}[htb!]
    \centering
    \includegraphics[width=0.446\linewidth]{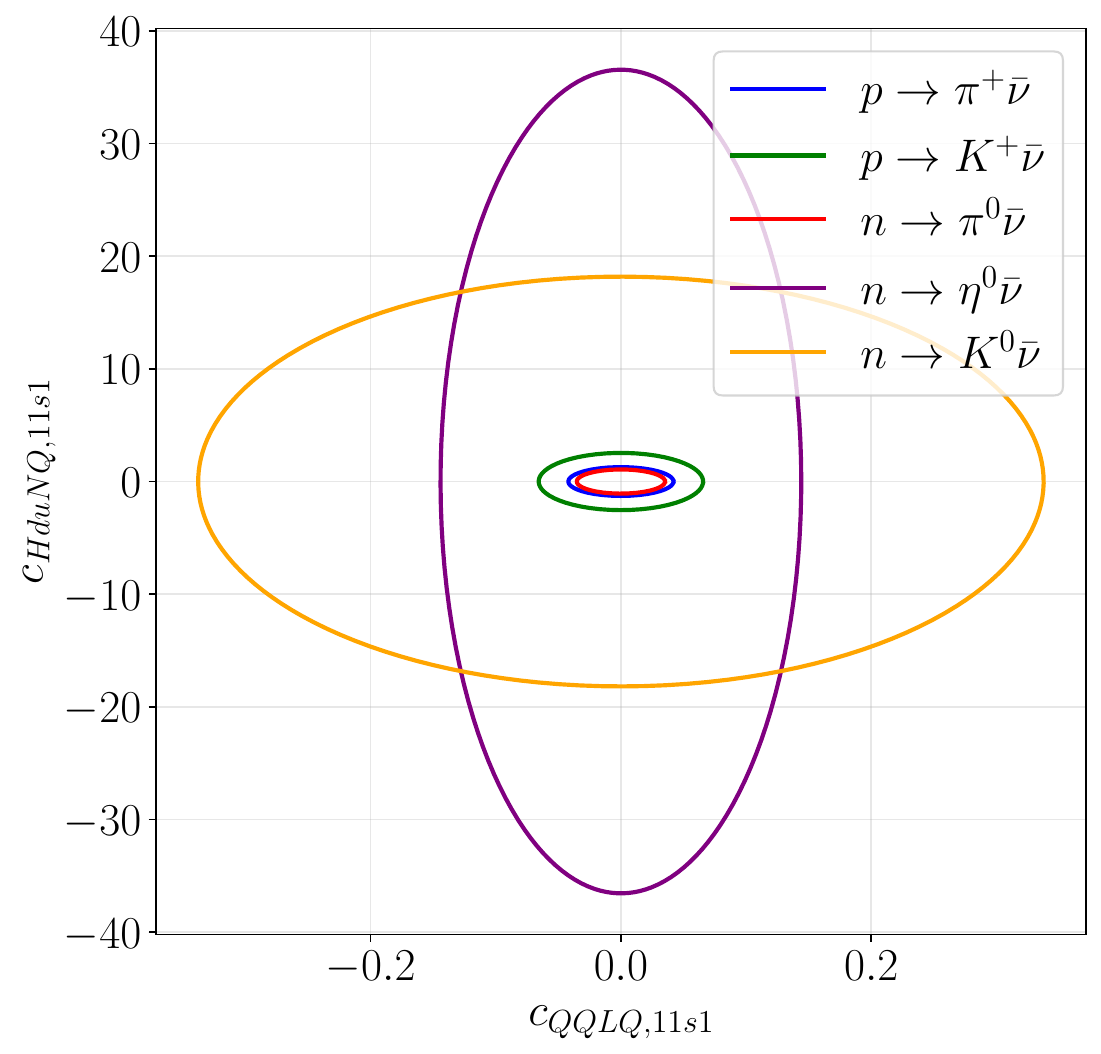}
    \includegraphics[width=0.45\linewidth]{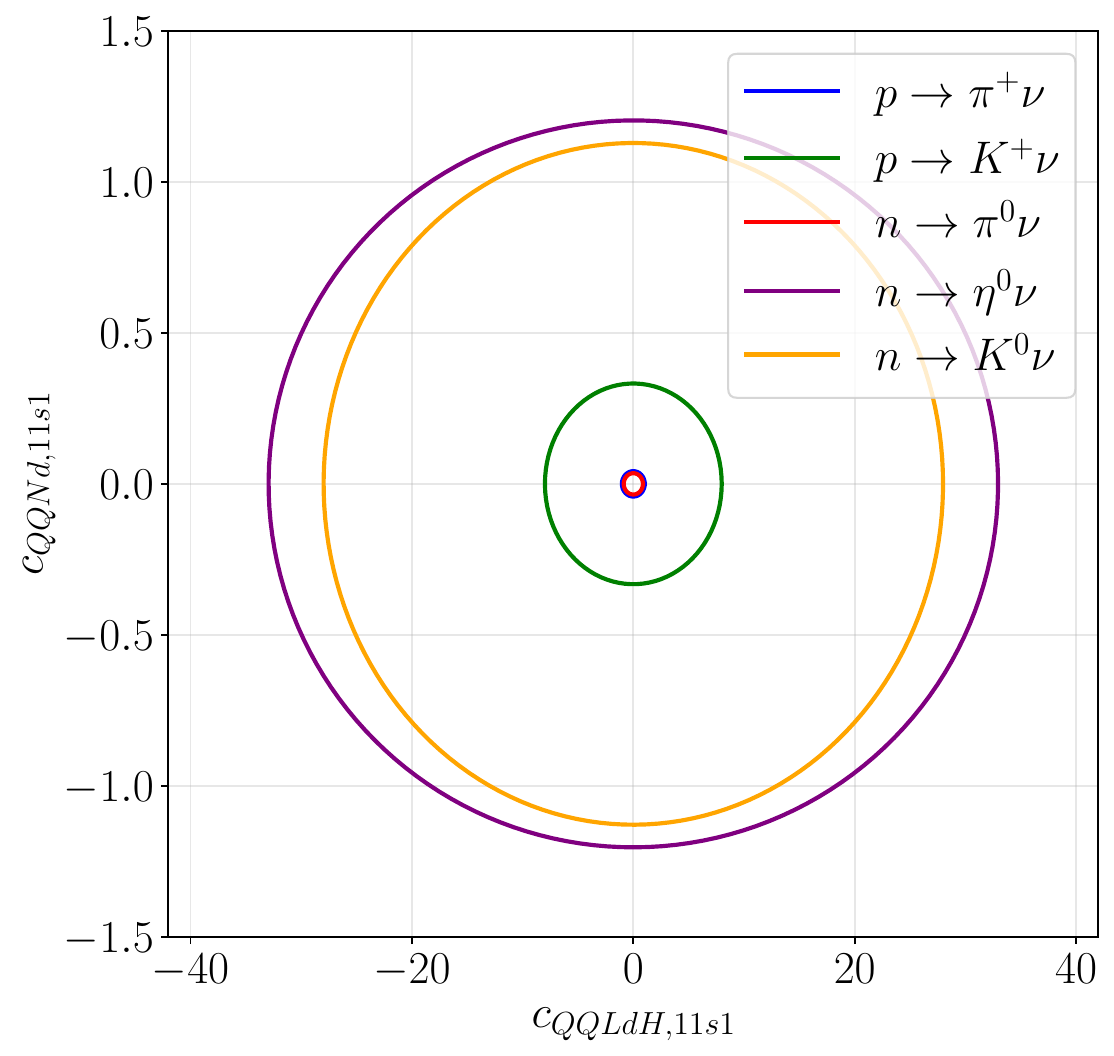}
    \caption{
    The allowed parameter space of both SMEFT and $\nu$SMEFT WCs for $c_{QQLQ,11s1}$ vs. $c_{HduNQ,11s1}$ (left) and $c_{QQLdH,11s1}$ vs. $c_{QQNd,11s1}$ (right), constrained by the current experimental limits in Tab.~\ref{tab:process}.
    For SMEFT, we assume the WC is universal for $s=1-3$.
    For $\nu$SMEFT, we take $m_N = 0.1$ GeV and $s=4$. The UV scales are set as $\Lambda = 10^{15}$ GeV for dimension-6 WCs and $\Lambda = 10^{11}$ GeV for dimension-7 WCs, respectively.
    }
    \label{fig:bounds_SMEFT_nuSMEFT}
\end{figure}

\begin{figure}[htb!]
    \centering
    \includegraphics[width=0.44\linewidth]{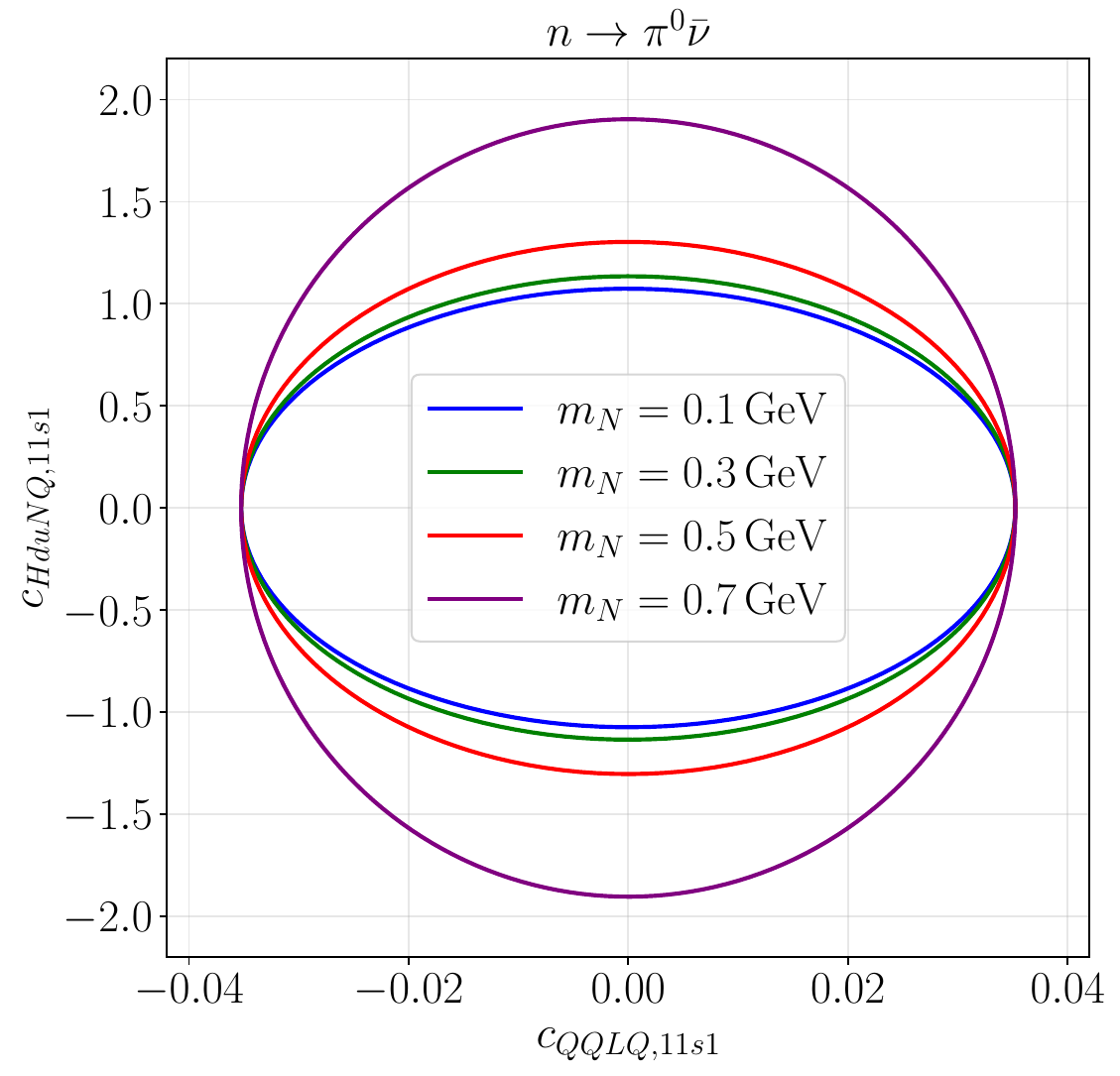}
    \includegraphics[width=0.45\linewidth]{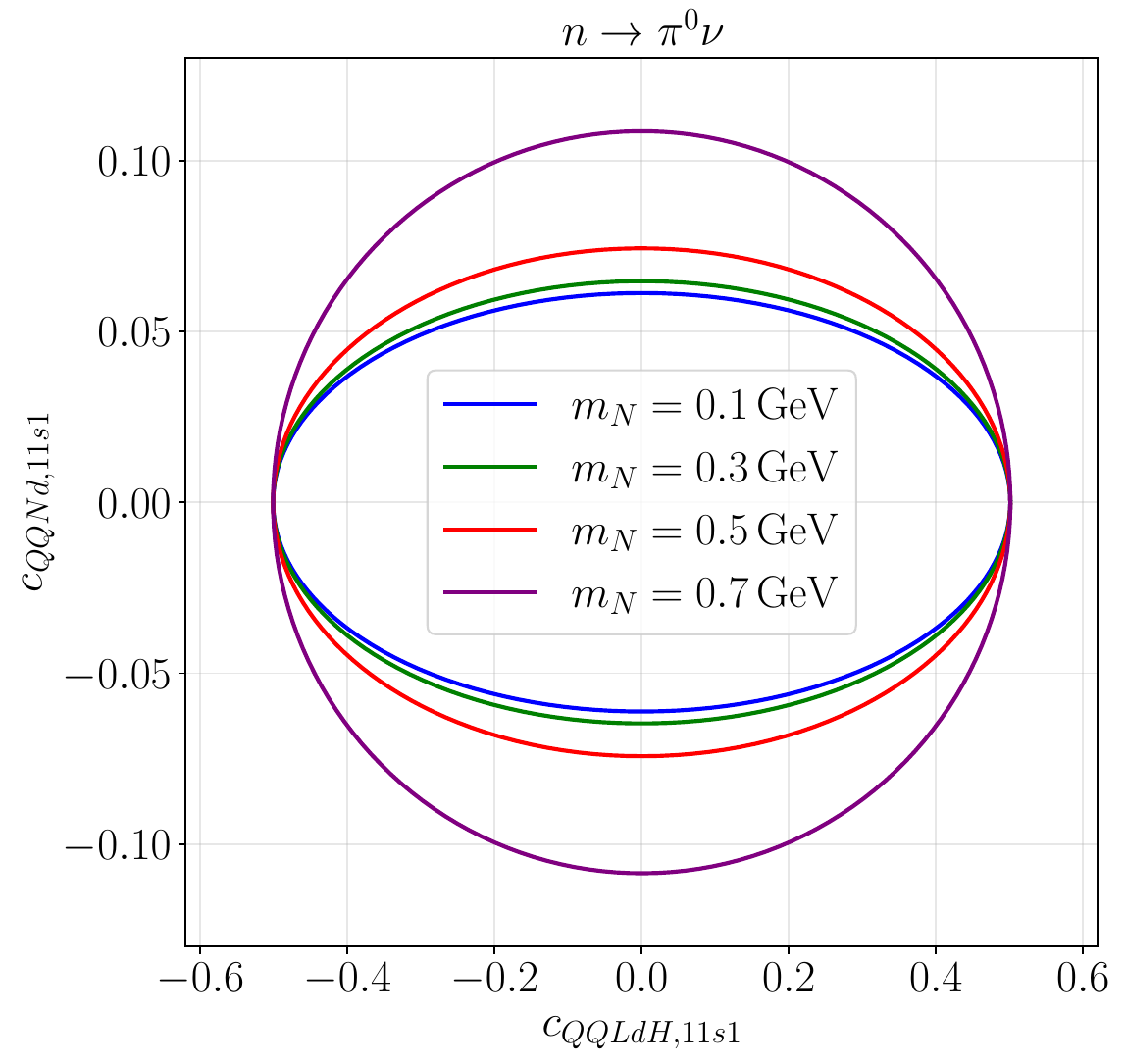}
    \caption{
    The allowed parameter space of both SMEFT and $\nu$SMEFT WCs for $c_{QQLQ,11s1}$ vs. $c_{HduNQ,11s1}$ constrained by $n\to \pi^0\bar{\nu}$ (left) and $c_{QQLdH,11s1}$ vs. $c_{QQNd,11s1}$ by $n\to \pi^0\nu$ (right), given different sterile neutrino masses.
    }
    \label{fig:bounds_SMEFT_nuSMEFT_mass}
\end{figure}

Finally, we discuss the correlation of both SMEFT and $\nu$SMEFT WCs. The left (right) panel of Fig.~\ref{fig:bounds_SMEFT_nuSMEFT} shows the allowed parameter space of dimensionless WCs $c_{QQLQ,11s1}$ and $c_{HduNQ,11s1}$ ($c_{QQLdH,11s1}$ and $c_{QQNd,11s1}$). The constraints on these WCs are derived by applying the current experimental limits, as listed in Tab.~\ref{tab:process}, to a range of relevant processes. For SMEFT, we assume the WC is universal for $s=1-3$. For $\nu$SMEFT, we take $m_N = 0.1$ GeV and $s=4$. The UV scales are set as $\Lambda = 10^{15}$ GeV for dimension-6 WCs and $\Lambda = 10^{11}$ GeV for dimension-7 WCs, respectively. As the two WCs do not have interference, the allowed parameter space exhibits a elliptical region. One can see that the involvement of non-zero $\nu$SMEFT (SMEFT) WC makes the constraint on SMEFT ($\nu$SMEFT) WC more stringent than the case with zero $\nu$SMEFT (SMEFT) WC in Tab.~\ref{tab:SMEFTdim67} (Tab.~\ref{tab:nuSMEFTdim67}). We also vary the sterile neutrino mass and display the constraint from the most severe process $n\to \pi^0 \bar{\nu}/\nu$ in Fig.~\ref{fig:bounds_SMEFT_nuSMEFT_mass}. As expected, the increase of sterile neutrino mass broadens the allowed parameter space and weakens the constraint on WCs.

\section{Conclusion}
\label{sec:Con}

The observation of BNV nucleon decay would imply the existence of NP beyond the SM. Future neutrino experiments will improve the sensitivity to BNV nucleon decay modes. The BNV nucleon decay also provides an intriguing probe of dark particles which escape from the detector. The massive dark particle induces similar signal but different kinematics compared to the conventional BNV nucleon decay with SM neutrino in final states.

In this work, we investigate the SMEFT/$\nu$SMEFT with baryon number violation and the impact of light sterile neutrino on BNV nucleon decays. We revisit the dimension-6 and dimension-7 operator bases in SMEFT/$\nu$SMEFT with $|\Delta (B-L)|=2$ or $|\Delta (B-L)|=0$. They are matched to the dimension-6 LEFT operators after the electroweak symmetry breaking. Based on the BChPT, we then obtain the effective chiral Lagrangian at low-energies and the BNV interactions of the sterile neutrino with baryons and mesons. The RGEs between different energy scales are also implemented. We then calculate the rates of nucleon decay to SM neutrinos or a sterile neutrino. Our main results are summarized as follows.
\begin{itemize}
\item We correct some longstanding mistakes of the quark index symmetry properties in the SMEFT/$\nu$SMEFT operator bases. We also derive the leading contribution to the RGEs for the dimension-7 BNV WCs in $\nu$SMEFT for the first time.
\item Assuming one dimension-6 (dimension-7) SMEFT operator at a time, as shown in Tab.~\ref{tab:SMEFTdim67}, the most stringent bound on the UV scale is $4.58\times 10^{15}~{\rm GeV}$ ($1.29\times 10^{11}~{\rm GeV}$) from $p\to K^+ \bar{\nu}$ ($p\to K^+ \nu$). For $\nu$SMEFT operator with massless sterile neutrino, it is $5.6\times 10^{15}~{\rm GeV}$ ($1.22\times 10^{11}~{\rm GeV}$) from $p\to K^+ \nu$ ($p\to K^+ \bar{\nu}$) as seen in Tab.~\ref{tab:nuSMEFTdim67}.
\item The presence of two WCs leads to flat directions of constraint in the parameter space of logarithmic coordinates. The phase difference between them also has non-negligible impact on the allowed parameter space.
\item The UV scale bound for $\nu$SMEFT operator generally decreases as the sterile neutrino mass increases.
\item For the case in which both SMEFT and $\nu$SMEFT operators are present, the involvement of non-zero $\nu$SMEFT (SMEFT) WC makes the constraint on SMEFT ($\nu$SMEFT) WC more stringent than the case with zero $\nu$SMEFT (SMEFT) WC.
\end{itemize}

\acknowledgments
We would like to thank Zhe Ren for useful discussions. T.~L. is supported by the National Natural Science Foundation of China (Grant No. 12375096, 12035008, 11975129). M.~S. acknowledges support by the Australian Research Council Discovery Project DP200101470. C.~Y.~Y. is supported by the STFC Consolidated Grant ST/X000583/1.

\appendix

\section{Electroweak contributions to dimension-7 $\nu$SMEFT operators}
\label{app:EWdim7nuSMEFT}

The electroweak contributions to the field normalization factors $Z=1+\delta Z$
in $R_\xi$ gauge are
\begin{align}
    \delta Z_{\rm fermion} &  = \frac{1}{16\pi^2\epsilon} \left[ - 2 Y_f^2 \xi_1 g_1^2   - 2 C_f \xi_2 g_2^2 \right] \;,
    \\
    \delta Z_{\rm scalar} &  = \frac{1}{16\pi^2\epsilon} \left[ 2 Y_s^2 ( 3- \xi_1) g_1^2   + 2 C_s (3-\xi_2) g_2^2 \right]\;,
\end{align}
where we defined the dimension $D=4-\epsilon$ and $C_f$ denotes the SU(2) Casimir operator which evaluates to $C_2=3/4$ for the fundamental representation. The SM hypercharges are $Y_Q=1/6$, $Y_u=2/3$, $Y_d=-1/3$, $Y_L=-1/2$, $Y_e=-1$ and $Y_H=1/2$.
The result for $C_{HdNQ}$ can be obtained from $C_{HduNQ}$ by substituting $Y_u\to Y_d$ and $Y_H\to -Y_H$, while the calculation of the counter term for $C_{HQNQ}$ requires more care.
The first-order in the $1/\epsilon$ expansion of the counterterms are
\begin{align}
    \delta C_{HdNQ,prst} &
    = \frac{C_{HdNQ,prst}}{16\pi^2 \epsilon } \left[
     2 ( 2 Y_Q Y_d \xi_1 + Y_d^2 (3+\xi_1) + Y_H (Y_Q+2Y_d)\xi_1  ) g_1^2 -2 C_2
    \xi_2 g_2^2  \right]\;,
\\
    \delta C_{HQNQ,prst} &
    = \frac{C_{HQNQ,prst}}{16\pi^2 \epsilon }
    \left[
    6\left(Y_Q^2 (1+\xi_1) -Y_Q Y_H \xi_1\right) g_1^2
    -\left(\frac52 + 3 \xi_2\right)g_2^2
    \right]
    \\\nonumber &
    -3 g_2^2\frac{C_{HQNQ,rpst}}{16\pi^2\epsilon}
    -2 g_2^2\frac{C_{HQNQ,ptsr}}{16\pi^2\epsilon}
    + g_2^2\frac{C_{HQNQ,rtsp}}{16\pi^2\epsilon}
    - g_2^2\frac{C_{HQNQ,trsp}}{16\pi^2\epsilon} \;,
\\
\delta C_{HduNQ,prst} &
    = \frac{C_{HduNQ,prst}}{16\pi^2 \epsilon } \left[ 2 (Y_Q (Y_u +Y_d) \xi_1 + Y_u Y_d (3+\xi_1) - Y_H (Y_Q+Y_u+Y_d)\xi_1  ) g_1^2
    \right.\nonumber\\ &\left. \qquad\qquad\qquad\qquad
    - 2 C_2  \xi_2 g_2^2  \right]
    \;.
\end{align}
The RGEs can be obtained from the general expression in Ref.~\cite{Antusch:2001ck}.
We define the bare WC $C_B$ in terms of the renormalized WC $C$, the counterterm $\delta C$ and field normalization factors $Z_{\phi_i}$
\begin{equation}
    C_B = \prod_{i \in I} Z_{\phi_i}^{n_i} [C+\delta C] \mu^{D_C \epsilon} \prod_{j \in J} Z_{\phi_j}^{n_j}
    \;.
\end{equation}
Then the contribution from gauge interactions with gauge coupling $g$ is~\cite{Antusch:2001ck}
\begin{align}
\dot C = 16\pi^2 \mu 
\frac{dC}{d\mu}
=    \delta C_{,1} - \frac12  \sum_{i \in I |J} \delta Z_{\phi_i,1} C
    +\dots\;,
\end{align}
where we used that the contribution of gauge interactions is universal and that gauge couplings enter quadratically in the leading order counterterm $\delta C_{,1}$ and field renormalization $\delta Z_{,1}$ in the $1/\epsilon$ expansion.
The dots indicate other contributions. From the field renormalization factors and the counterterms listed above, we obtain the electroweak contributions to the RGEs
of three BNV dimension-7 $\nu$SMEFT operators
\begin{align}
   \dot C_{HdNQ,prst} &
    =  - \left[ \frac94 g_2^2 + \frac{1}{12} g_1^2 
 \right] C_{HdNQ,prst}\;,
    \\
    \dot C_{HQNQ,prst} &
      =
    -\left[
    \frac{19}{4} g_2^2
    +\frac{7}{12} g_1^2
    \right]C_{HQNQ,prst}
    \\\nonumber &\qquad
    - g_2^2 \left[ 3 C_{HQNQ,rpst} + 2 C_{HQNQ,ptsr} - C_{HQNQ,rtsp} + C_{HQNQ,trsp} \right]\;,
    \\
    \dot C_{HduNQ,prst} &
    =  - \left[ \frac94 g_2^2   + \frac{25}{12} g_1^2 
    \right] C_{HduNQ,prst}
    \;.
\end{align}

\section{Dependence of sterile neutrino mass on the WC bounds in $\nu$SMEFT}
\label{app:mNdep}

\begin{figure}[H]
    \centering
\includegraphics[width=0.45\linewidth]{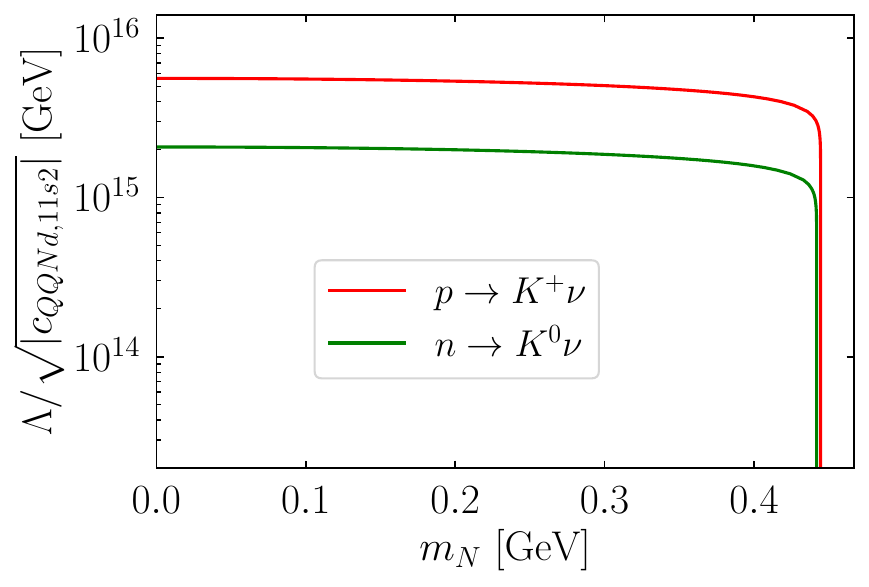}
\includegraphics[width=0.45\linewidth]{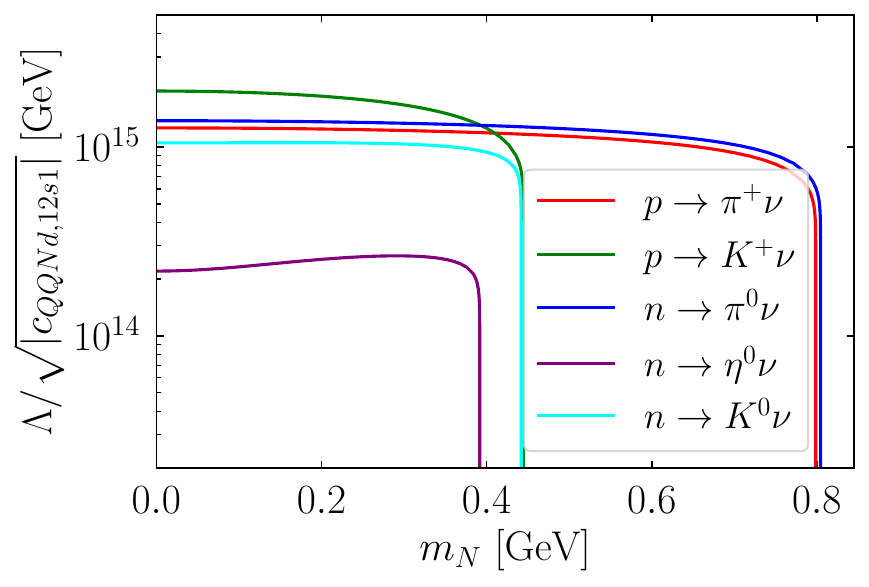}
\includegraphics[width=0.45\linewidth]{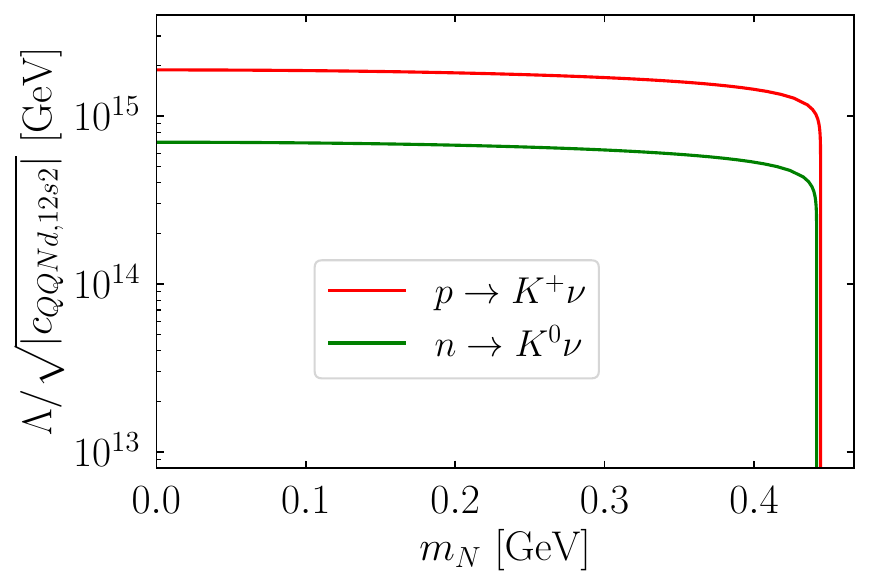}
\includegraphics[width=0.45\linewidth]{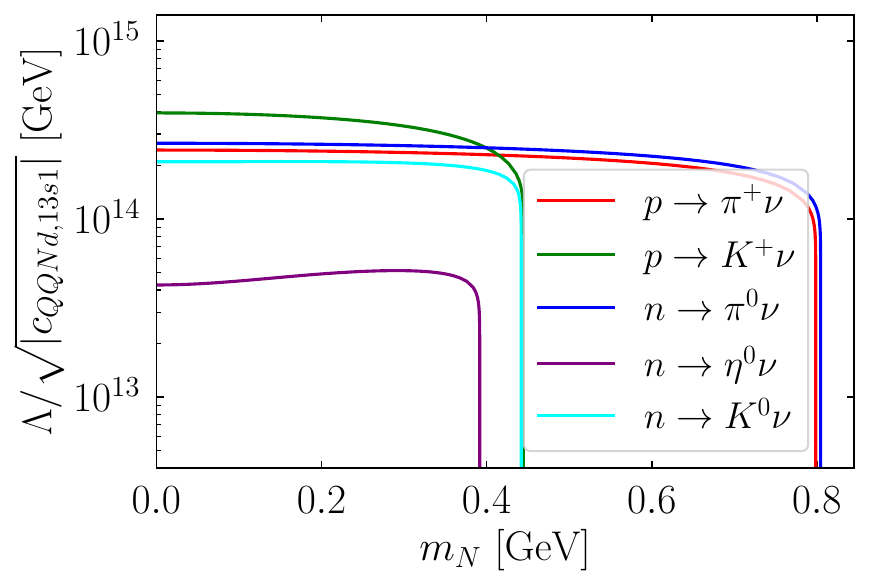}
\includegraphics[width=0.45\linewidth]{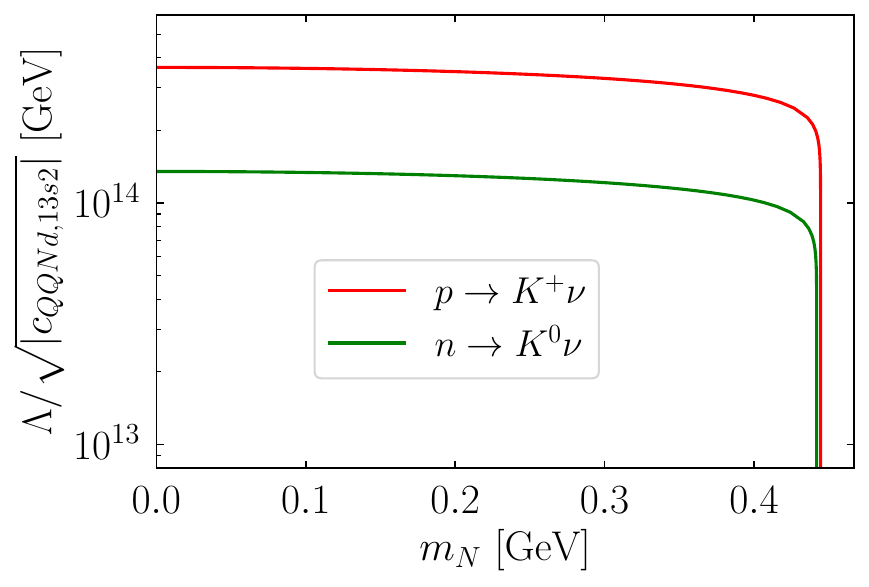}
\includegraphics[width=0.45\linewidth]{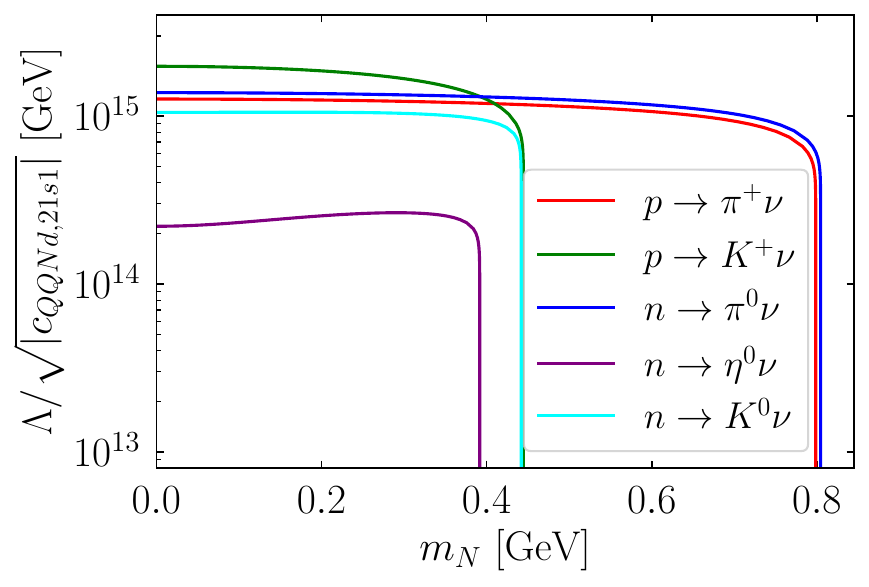}
    \caption{Bounds on dimension-6 WCs in $\nu$SMEFT derived from different processes as functions of the sterile neutrino mass $m_N$.}
    \label{fig:bounds_numas_app1}
\end{figure}

\begin{figure}[H]
    \centering
\includegraphics[width=0.45\linewidth]{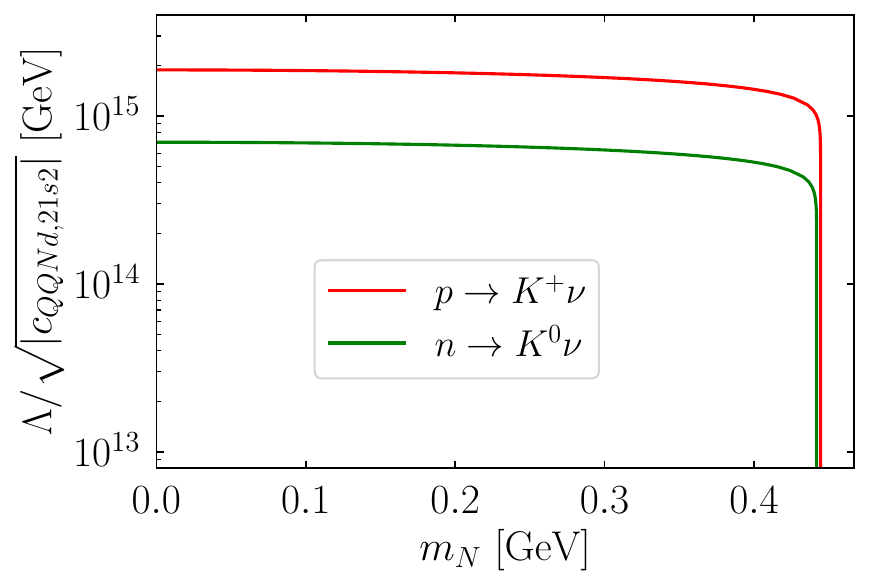}
\includegraphics[width=0.45\linewidth]{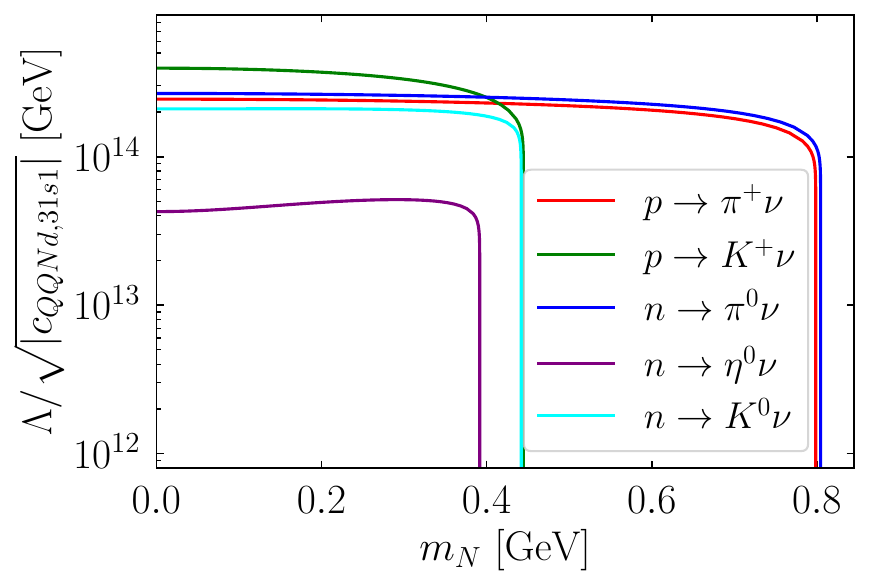}
\includegraphics[width=0.45\linewidth]{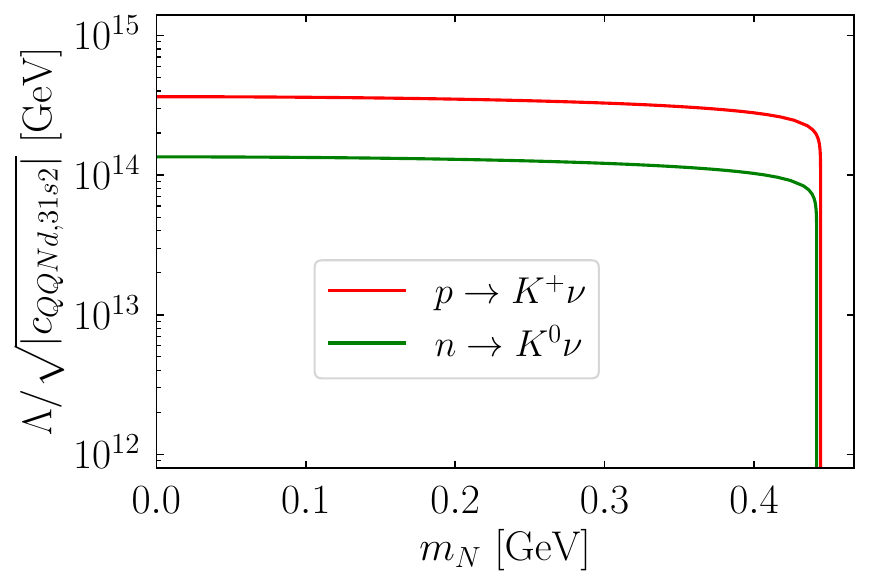}
\includegraphics[width=0.45\linewidth]{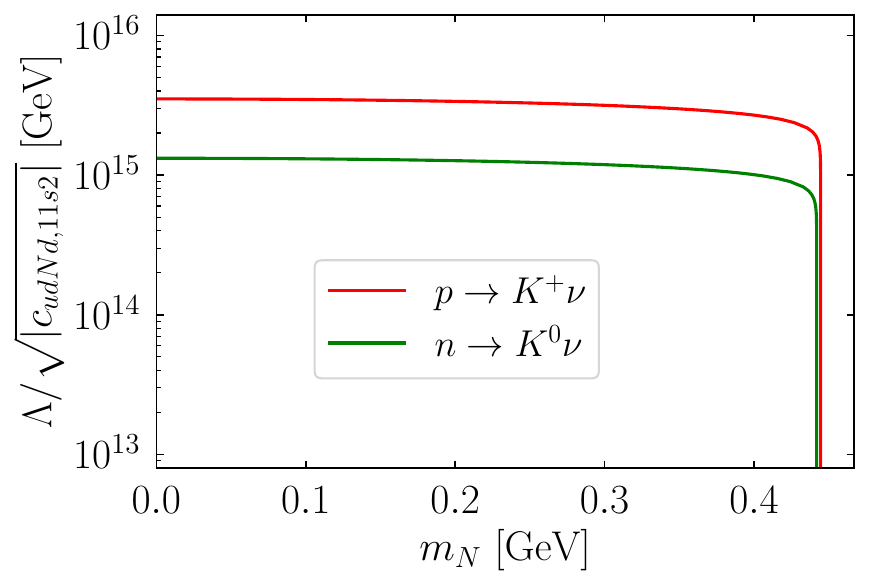}
\includegraphics[width=0.45\linewidth]{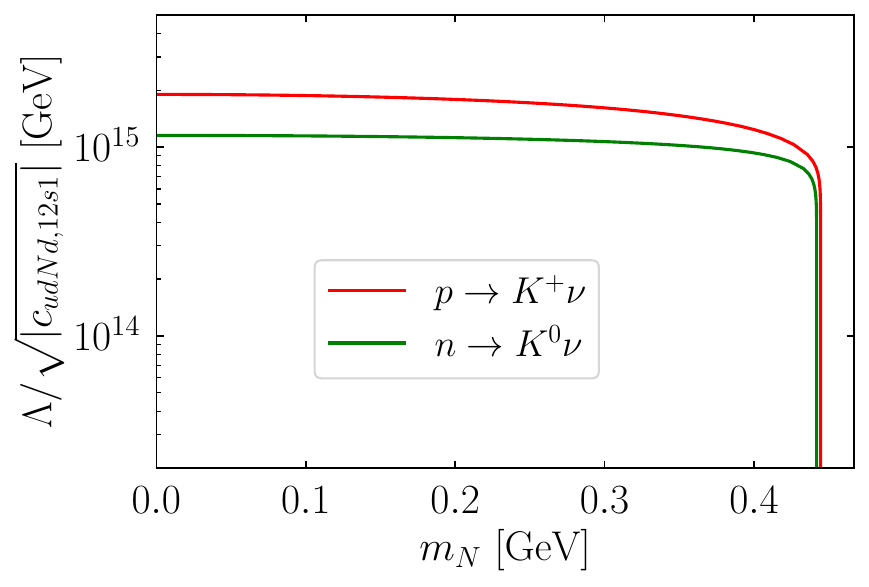}
    \caption{Bounds on dimension-6 WCs in $\nu$SMEFT derived from different processes as functions of the sterile neutrino mass $m_N$. [Fig.~\ref{fig:bounds_numas_app1} continued]}
    \label{fig:bounds_numas_app2}
\end{figure}

\begin{figure}[H]
    \centering
\includegraphics[width=0.4\linewidth]{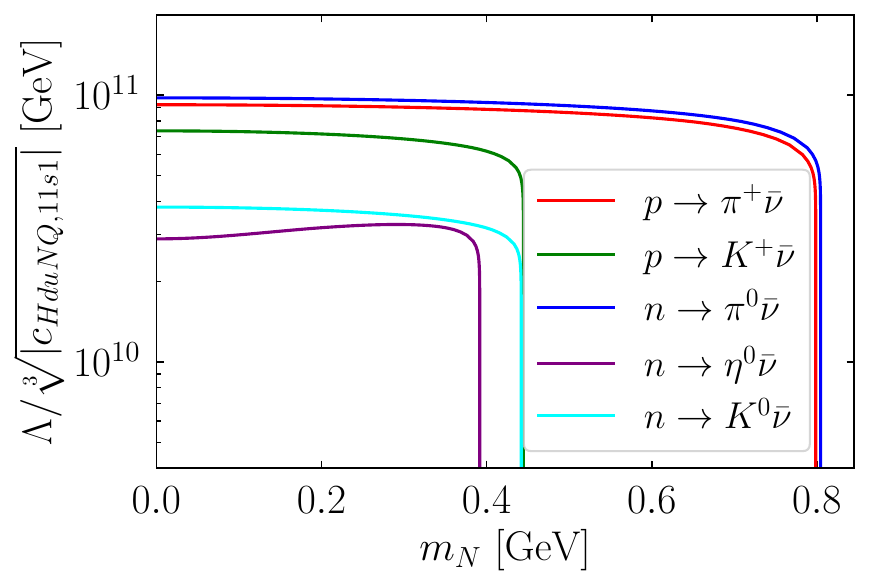}
\includegraphics[width=0.4\linewidth]{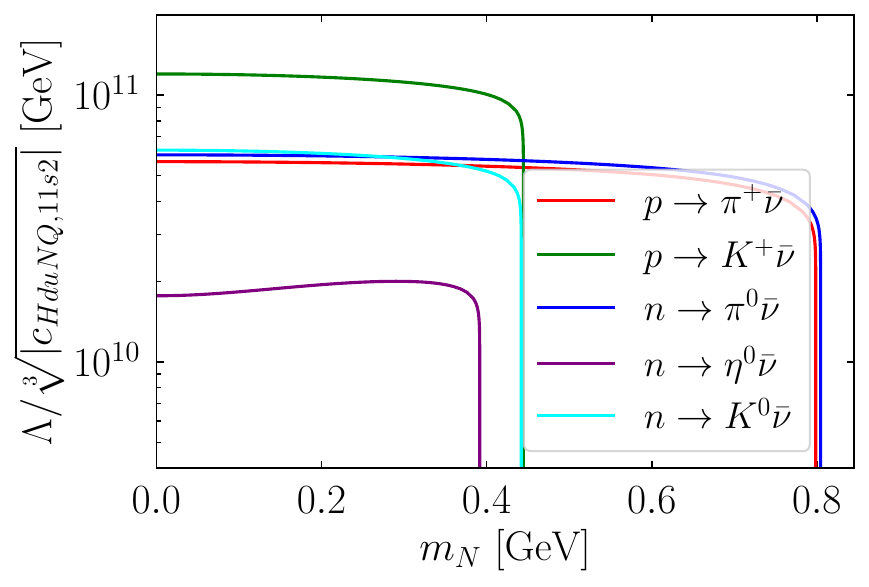}
\includegraphics[width=0.4\linewidth]{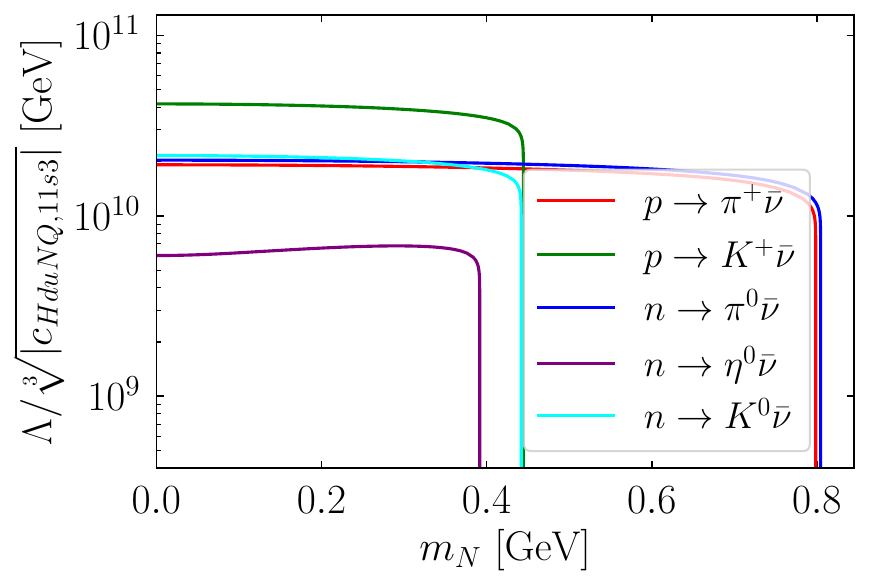}
\includegraphics[width=0.4\linewidth]{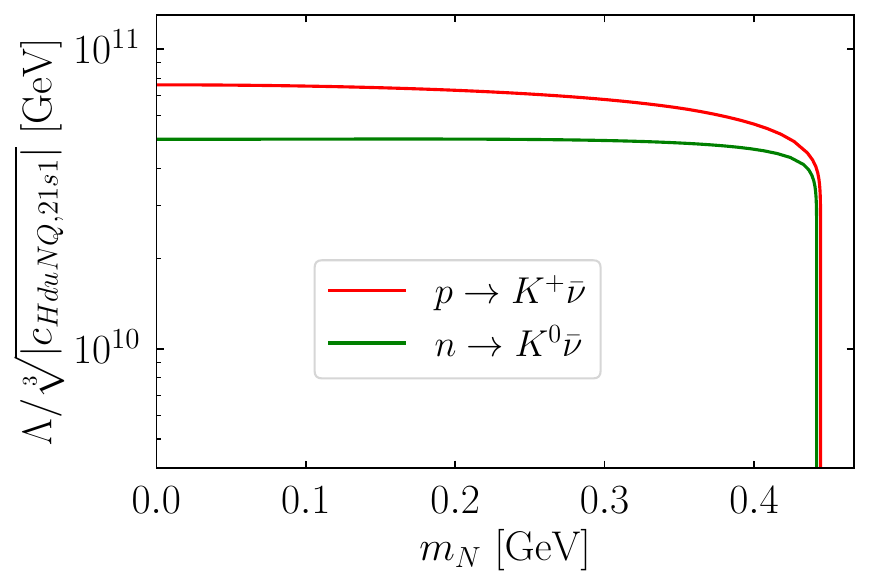}
\includegraphics[width=0.4\linewidth]{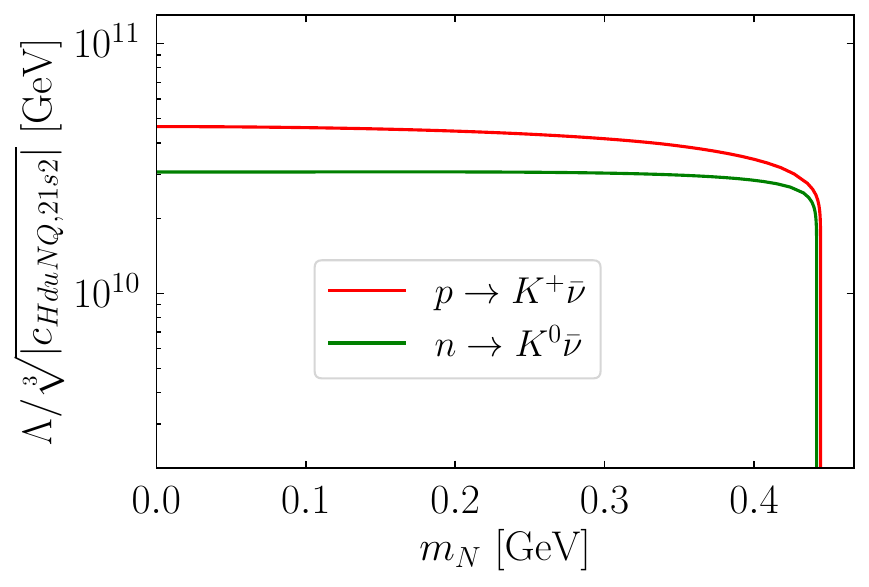}
\includegraphics[width=0.4\linewidth]{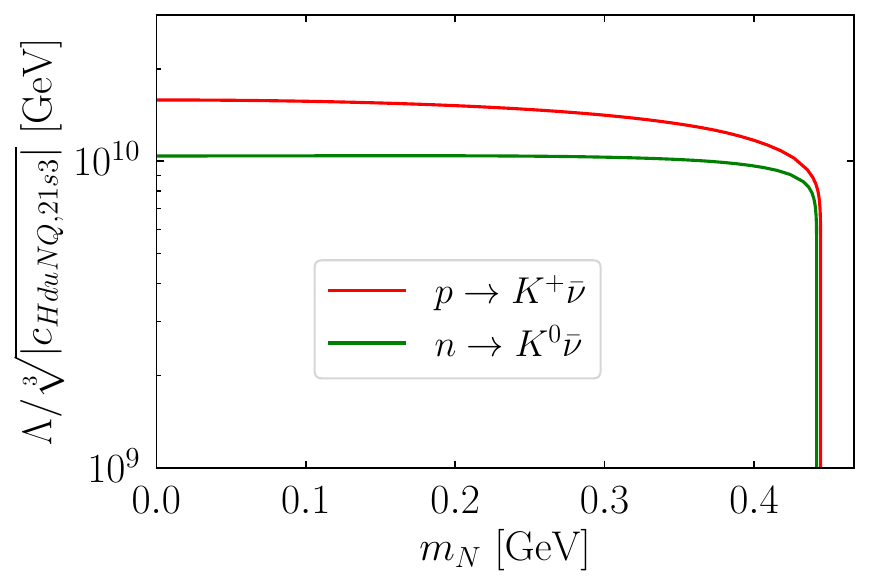}
\includegraphics[width=0.4\linewidth]{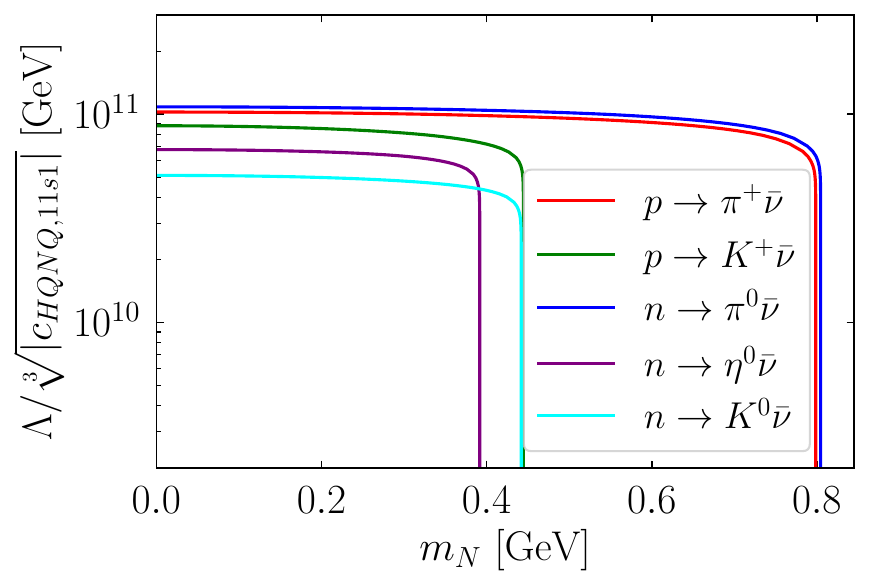}
\includegraphics[width=0.4\linewidth]{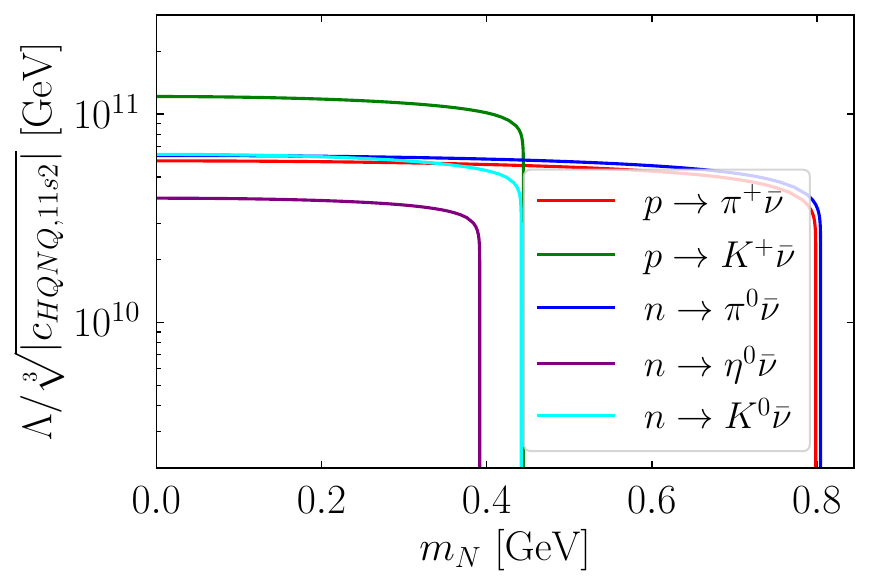}
\includegraphics[width=0.4\linewidth]{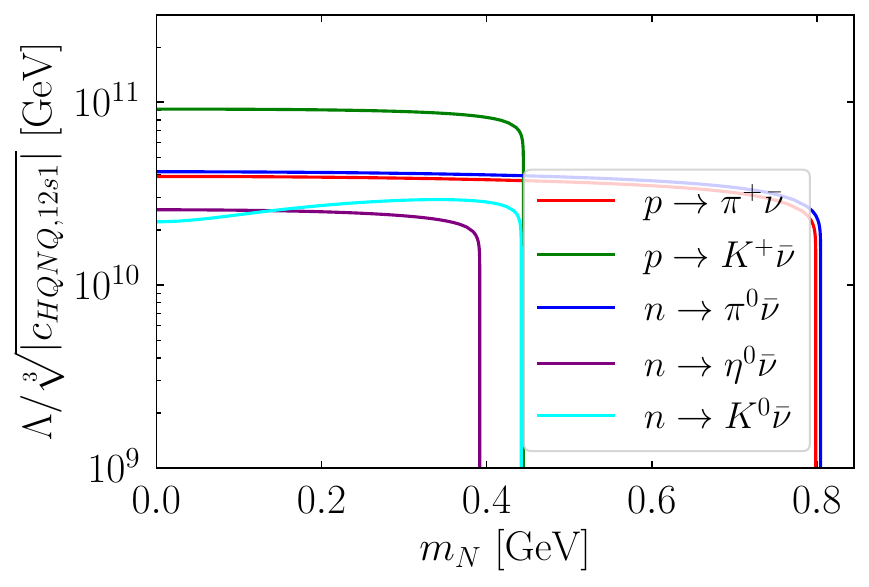}
\includegraphics[width=0.4\linewidth]{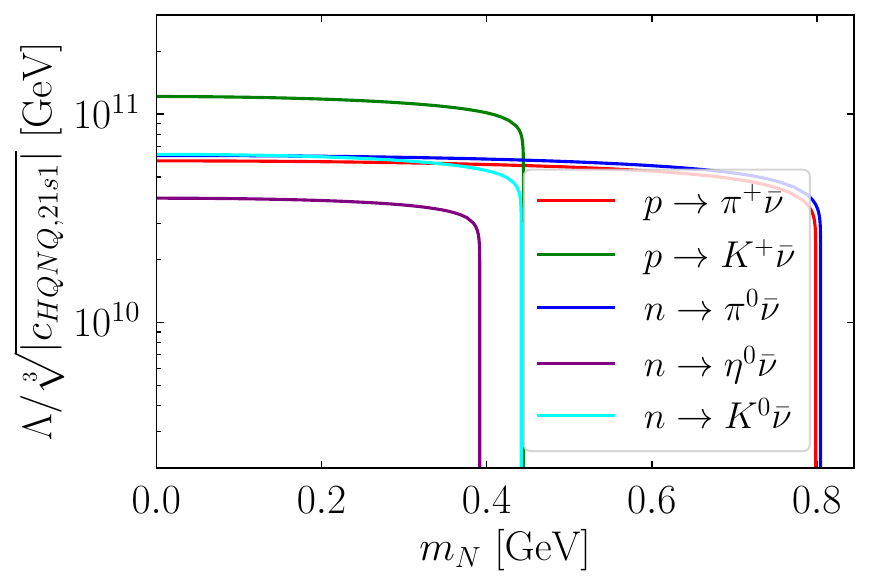}
    \caption{Bounds on dimension-7 WCs in $\nu$SMEFT derived from different processes as functions of the sterile neutrino mass $m_N$.
    }
\label{fig:bounds_numas_app3}
\end{figure}

\begin{figure}[H]
    \centering
\includegraphics[width=0.4\linewidth]{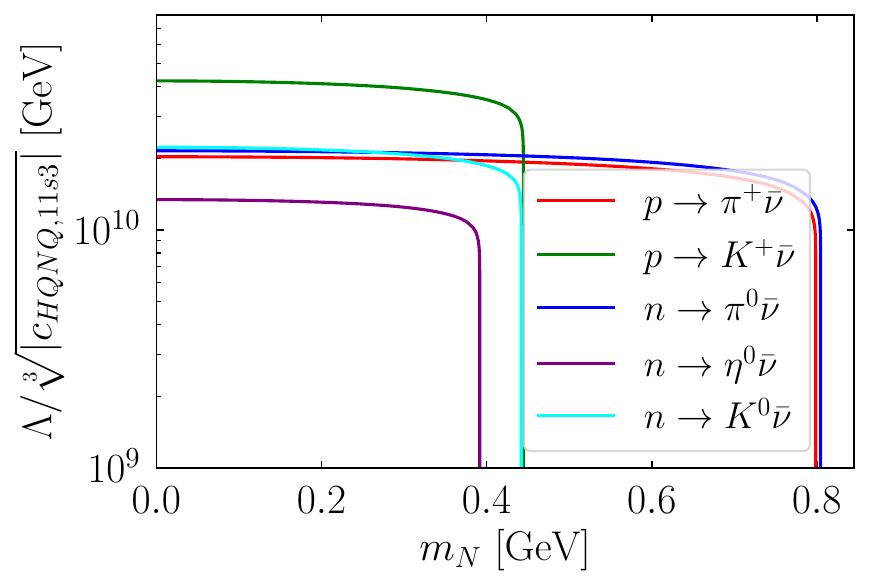}
\includegraphics[width=0.4\linewidth]{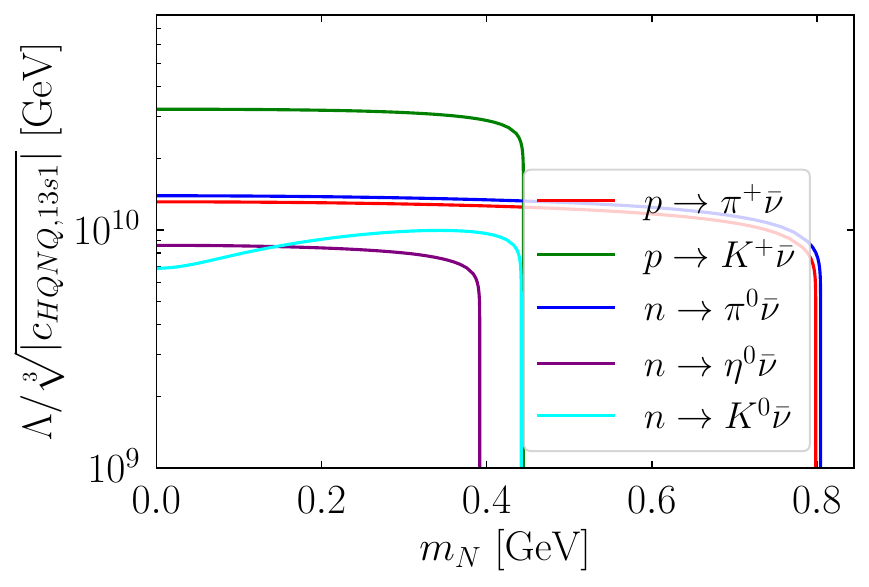}
\includegraphics[width=0.4\linewidth]{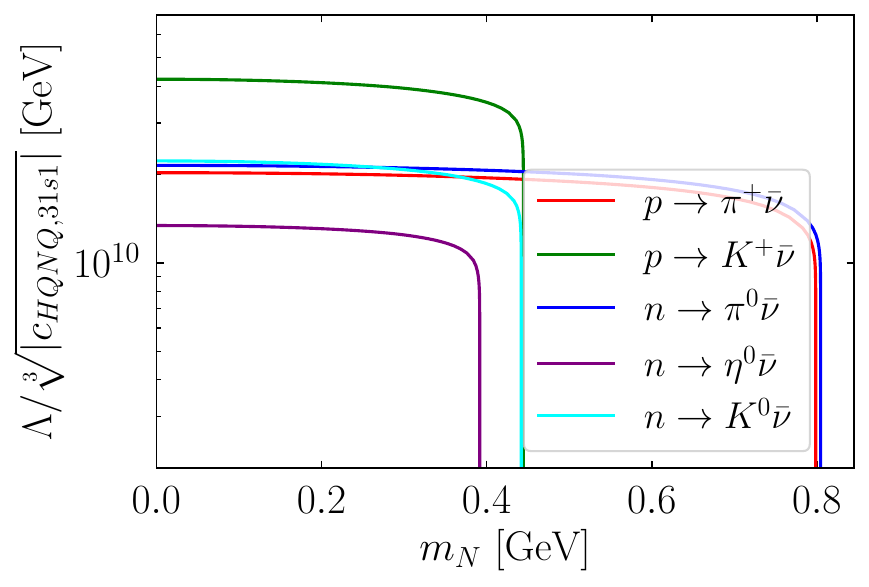}
\includegraphics[width=0.4\linewidth]{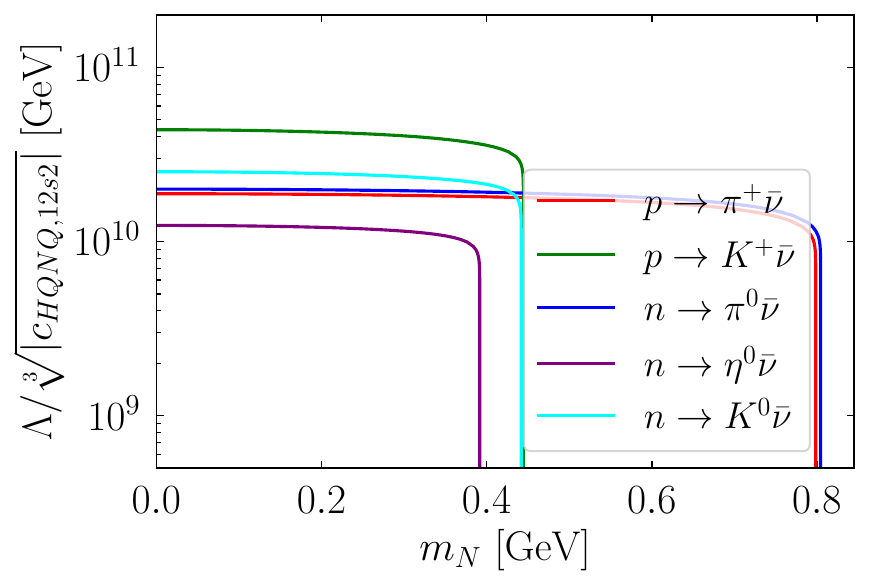}
\includegraphics[width=0.4\linewidth]{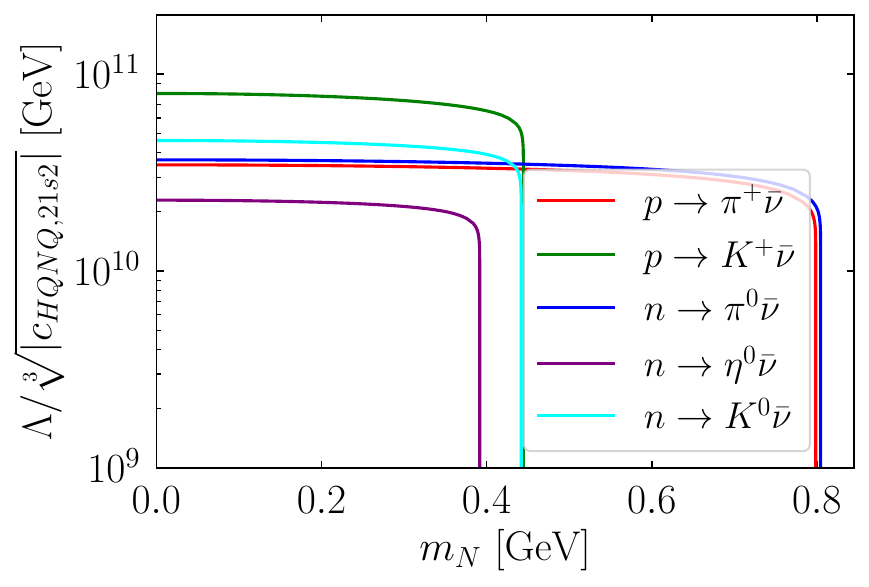}
\includegraphics[width=0.4\linewidth]{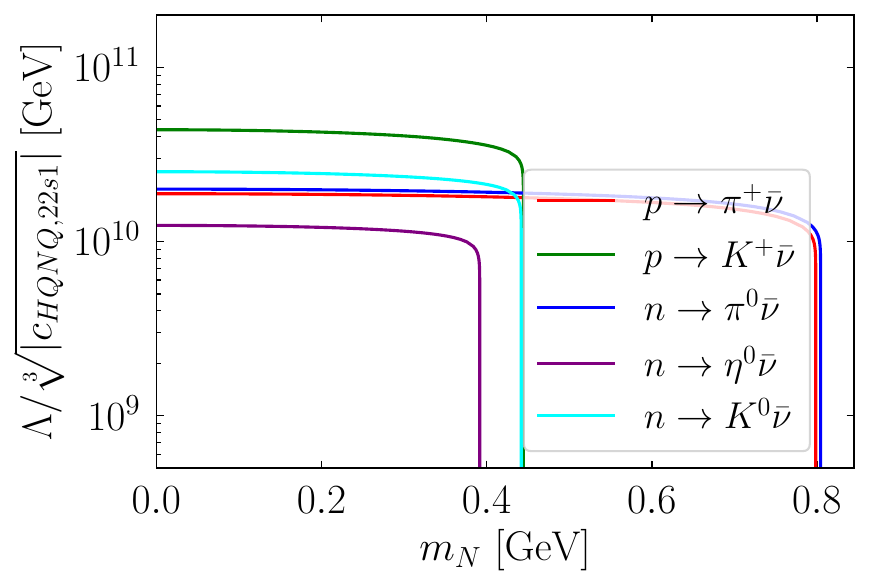}
\includegraphics[width=0.4\linewidth]{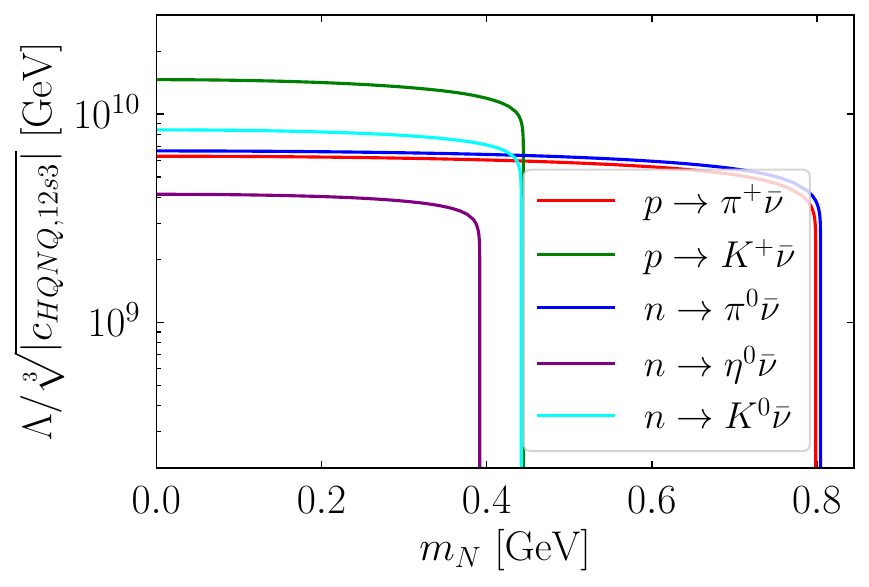}
\includegraphics[width=0.4\linewidth]{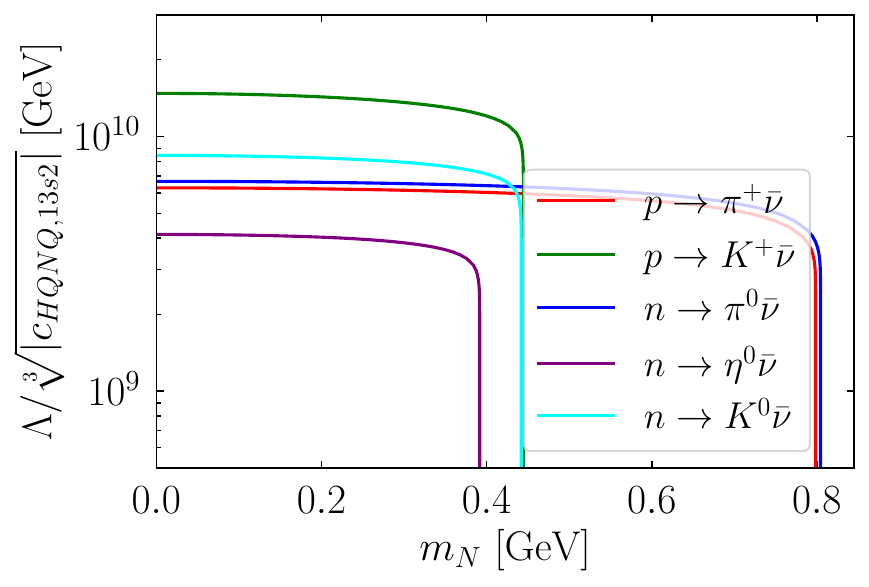}
\includegraphics[width=0.4\linewidth]{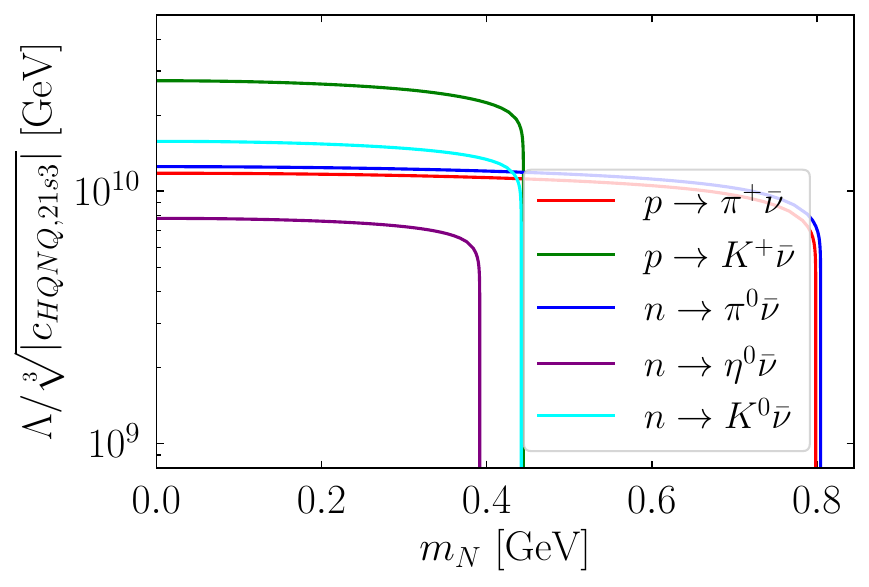}
\includegraphics[width=0.4\linewidth]{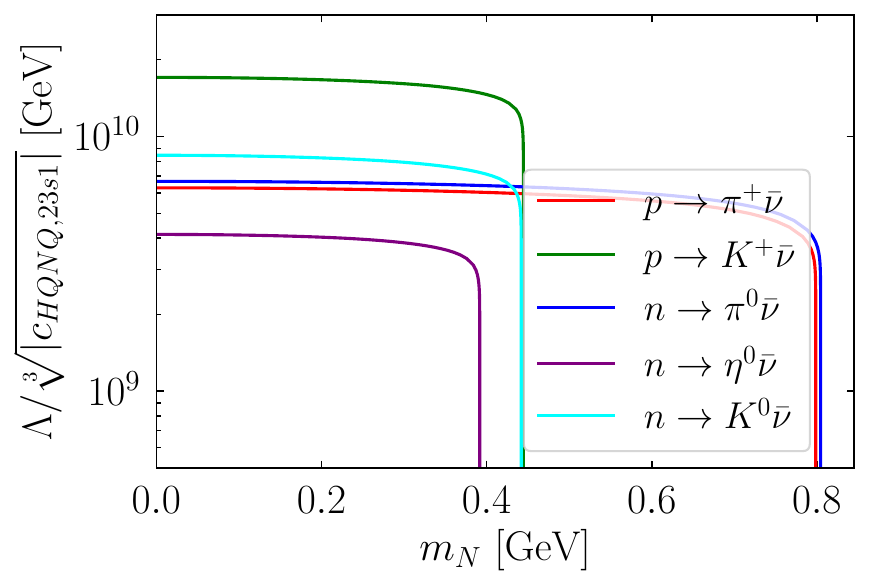}
    \caption{Bounds on dimension-7 WCs in $\nu$SMEFT derived from different processes as functions of the sterile neutrino mass $m_N$. [Fig.~\ref{fig:bounds_numas_app3} continued]
    }
    \label{fig:bounds_numas_app4}
\end{figure}

\begin{figure}[H]
    \centering
\includegraphics[width=0.4\linewidth]{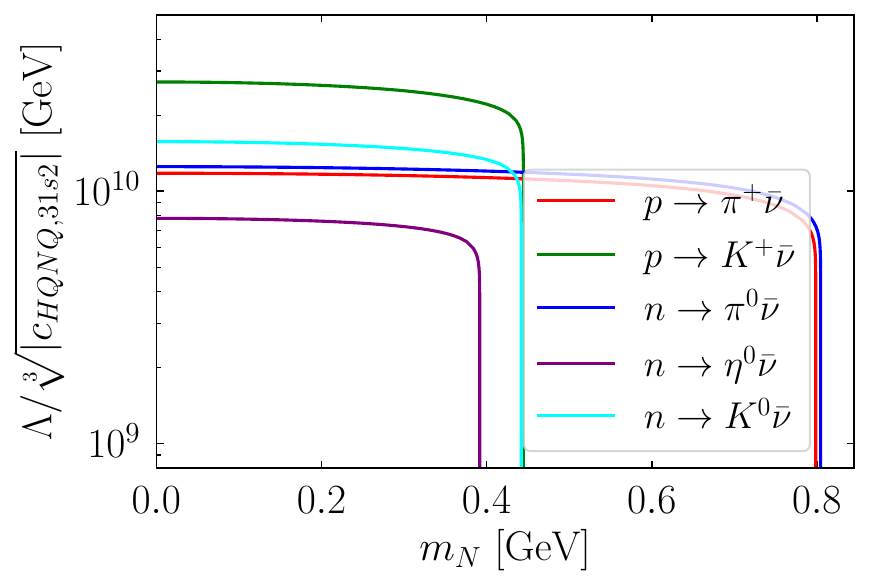}
\includegraphics[width=0.4\linewidth]{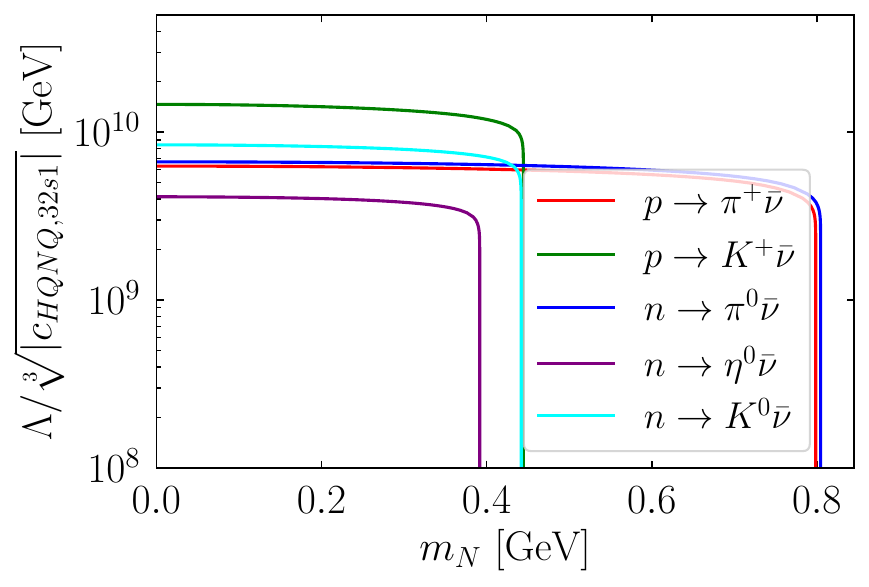}
\includegraphics[width=0.4\linewidth]{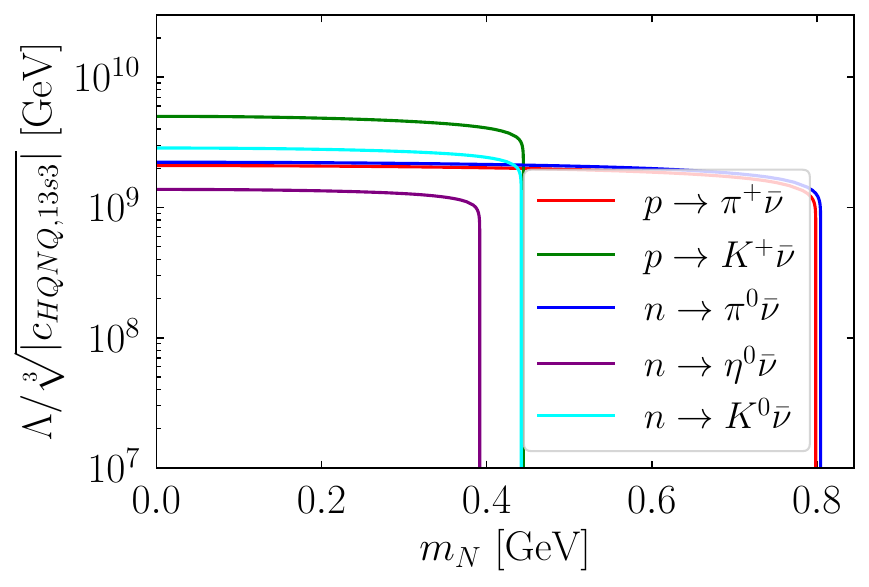}
\includegraphics[width=0.4\linewidth]{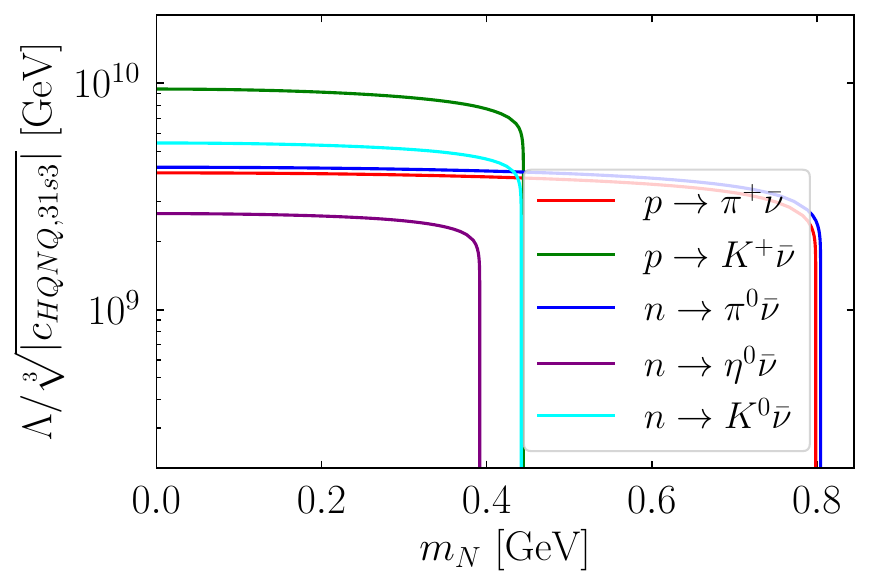}
\includegraphics[width=0.4\linewidth]{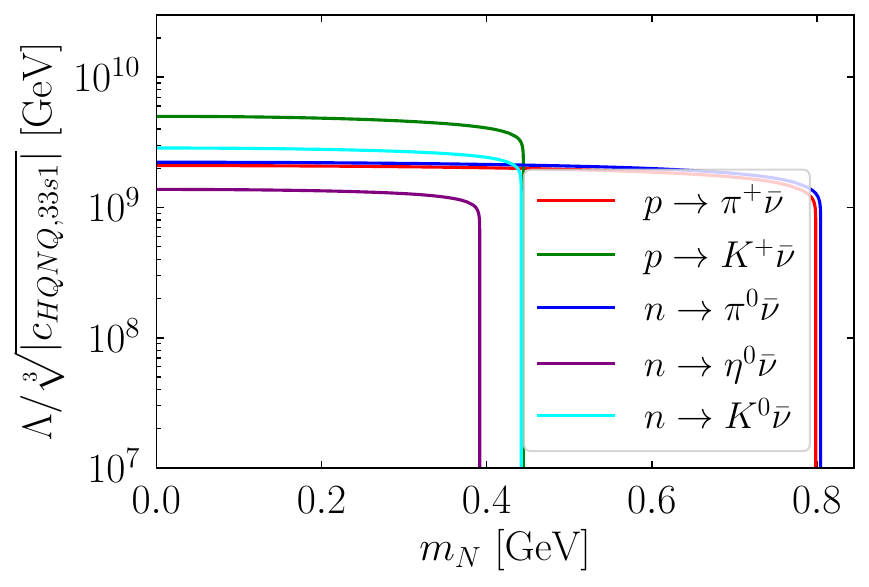}
    \caption{Bounds on dimension-7 WCs in $\nu$SMEFT derived from different processes as functions of the sterile neutrino mass $m_N$. [Fig.~\ref{fig:bounds_numas_app4} continued]
    }
    \label{fig:bounds_numas_app5}
\end{figure}

We collect the figures showing the dependence of sterile neutrino mass on the UV energy scale for the remaining  $\nu$SMEFT operators in Figs.~\ref{fig:bounds_numas_app1}--\ref{fig:bounds_numas_app5}.

\bibliographystyle{JHEP}
\bibliography{refs}

\providecommand{\href}[2]{#2}\begingroup\raggedright\begin{thebibliography}{10}

\bibitem{Takhistov:2016eqm}
{\scshape Super-Kamiokande} collaboration, V.~Takhistov, \emph{{Review of
  Nucleon Decay Searches at Super-Kamiokande}},  in \emph{{51st Rencontres de
  Moriond on EW Interactions and Unified Theories}}, pp.~437--444, 2016.
\newblock \href{http://arxiv.org/abs/1605.03235}{{\tt 1605.03235}}.

\bibitem{Super-Kamiokande:2020wjk}
{\scshape Super-Kamiokande} collaboration, A.~Takenaka et~al., \emph{{Search
  for proton decay via $p\to e^+\pi^0$ and $p\to \mu^+\pi^0$ with an enlarged
  fiducial volume in Super-Kamiokande I-IV}},
  \href{http://dx.doi.org/10.1103/PhysRevD.102.112011}{\emph{Phys. Rev. D} {\bf
  102} (2020) 112011}, [\href{http://arxiv.org/abs/2010.16098}{{\tt
  2010.16098}}].

\bibitem{Super-Kamiokande:2014otb}
{\scshape Super-Kamiokande} collaboration, K.~Abe et~al., \emph{{Search for
  proton decay via $p\to\nu K^+$ using 260 kiloton\textperiodcentered{}year
  data of Super-Kamiokande}},
  \href{http://dx.doi.org/10.1103/PhysRevD.90.072005}{\emph{Phys. Rev. D} {\bf
  90} (2014) 072005}, [\href{http://arxiv.org/abs/1408.1195}{{\tt 1408.1195}}].

\bibitem{Super-Kamiokande:2017gev}
{\scshape Super-Kamiokande} collaboration, K.~Abe et~al., \emph{{Search for
  nucleon decay into charged antilepton plus meson in 0.316 megaton$\cdot$years
  exposure of the Super-Kamiokande water Cherenkov detector}},
  \href{http://dx.doi.org/10.1103/PhysRevD.96.012003}{\emph{Phys. Rev. D} {\bf
  96} (2017) 012003}, [\href{http://arxiv.org/abs/1705.07221}{{\tt
  1705.07221}}].

\bibitem{Super-Kamiokande:2013rwg}
{\scshape Super-Kamiokande} collaboration, K.~Abe et~al., \emph{{Search for
  Nucleon Decay via $n \to \bar{\nu} \pi^{0}$ and $p \to \bar{\nu} \pi^{+}$ in
  Super-Kamiokande}},
  \href{http://dx.doi.org/10.1103/PhysRevLett.113.121802}{\emph{Phys. Rev.
  Lett.} {\bf 113} (2014) 121802}, [\href{http://arxiv.org/abs/1305.4391}{{\tt
  1305.4391}}].

\bibitem{Super-Kamiokande:2005lev}
{\scshape Super-Kamiokande} collaboration, K.~Kobayashi et~al., \emph{{Search
  for nucleon decay via modes favored by supersymmetric grand unification
  models in Super-Kamiokande-I}},
  \href{http://dx.doi.org/10.1103/PhysRevD.72.052007}{\emph{Phys. Rev. D} {\bf
  72} (2005) 052007}, [\href{http://arxiv.org/abs/hep-ex/0502026}{{\tt
  hep-ex/0502026}}].

\bibitem{Super-Kamiokande:2012zik}
{\scshape Super-Kamiokande} collaboration, C.~Regis et~al., \emph{{Search for
  Proton Decay via $p -> \mu^+ K^0$ in Super-Kamiokande I, II, and III}},
  \href{http://dx.doi.org/10.1103/PhysRevD.86.012006}{\emph{Phys. Rev. D} {\bf
  86} (2012) 012006}, [\href{http://arxiv.org/abs/1205.6538}{{\tt 1205.6538}}].

\bibitem{Hyper-Kamiokande:2018ofw}
{\scshape Hyper-Kamiokande} collaboration, K.~Abe et~al.,
  \emph{{Hyper-Kamiokande Design Report}},
  \href{http://arxiv.org/abs/1805.04163}{{\tt 1805.04163}}.

\bibitem{DUNE:2016evb}
{\scshape DUNE} collaboration, J.~Strait et~al., \emph{{Long-Baseline Neutrino
  Facility (LBNF) and Deep Underground Neutrino Experiment (DUNE)}: {Conceptual
  Design Report, Volume 3: Long-Baseline Neutrino Facility for DUNE June 24,
  2015}},  \href{http://arxiv.org/abs/1601.05823}{{\tt 1601.05823}}.

\bibitem{DUNE:2020ypp}
{\scshape DUNE} collaboration, B.~Abi et~al., \emph{{Deep Underground Neutrino
  Experiment (DUNE), Far Detector Technical Design Report, Volume II: DUNE
  Physics}},  \href{http://arxiv.org/abs/2002.03005}{{\tt 2002.03005}}.

\bibitem{JUNO:2015zny}
{\scshape JUNO} collaboration, F.~An et~al., \emph{{Neutrino Physics with
  JUNO}}, \href{http://dx.doi.org/10.1088/0954-3899/43/3/030401}{\emph{J. Phys.
  G} {\bf 43} (2016) 030401}, [\href{http://arxiv.org/abs/1507.05613}{{\tt
  1507.05613}}].

\bibitem{Georgi:1974sy}
H.~Georgi and S.~L. Glashow, \emph{{Unity of All Elementary Particle Forces}},
  \href{http://dx.doi.org/10.1103/PhysRevLett.32.438}{\emph{Phys. Rev. Lett.}
  {\bf 32} (1974) 438--441}.

\bibitem{Fritzsch:1974nn}
H.~Fritzsch and P.~Minkowski, \emph{{Unified Interactions of Leptons and
  Hadrons}}, \href{http://dx.doi.org/10.1016/0003-4916(75)90211-0}{\emph{Annals
  Phys.} {\bf 93} (1975) 193--266}.

\bibitem{Langacker:1980js}
P.~Langacker, \emph{{Grand Unified Theories and Proton Decay}},
  \href{http://dx.doi.org/10.1016/0370-1573(81)90059-4}{\emph{Phys. Rept.} {\bf
  72} (1981) 185}.

\bibitem{Nath:2006ut}
P.~Nath and P.~Fileviez~Perez, \emph{{Proton stability in grand unified
  theories, in strings and in branes}},
  \href{http://dx.doi.org/10.1016/j.physrep.2007.02.010}{\emph{Phys. Rept.}
  {\bf 441} (2007) 191--317}, [\href{http://arxiv.org/abs/hep-ph/0601023}{{\tt
  hep-ph/0601023}}].

\bibitem{Farrar:1978xj}
G.~R. Farrar and P.~Fayet, \emph{{Phenomenology of the Production, Decay, and
  Detection of New Hadronic States Associated with Supersymmetry}},
  \href{http://dx.doi.org/10.1016/0370-2693(78)90858-4}{\emph{Phys. Lett. B}
  {\bf 76} (1978) 575--579}.

\bibitem{Dimopoulos:1981zb}
S.~Dimopoulos and H.~Georgi, \emph{{Softly Broken Supersymmetry and SU(5)}},
  \href{http://dx.doi.org/10.1016/0550-3213(81)90522-8}{\emph{Nucl. Phys. B}
  {\bf 193} (1981) 150--162}.

\bibitem{Sakai:1981gr}
N.~Sakai, \emph{{Naturalness in Supersymmetric Guts}},
  \href{http://dx.doi.org/10.1007/BF01573998}{\emph{Z. Phys. C} {\bf 11} (1981)
  153}.

\bibitem{deGouvea:2014lva}
A.~de~Gouvea, J.~Herrero-Garcia and A.~Kobach, \emph{{Neutrino Masses, Grand
  Unification, and Baryon Number Violation}},
  \href{http://dx.doi.org/10.1103/PhysRevD.90.016011}{\emph{Phys. Rev. D} {\bf
  90} (2014) 016011}, [\href{http://arxiv.org/abs/1404.4057}{{\tt 1404.4057}}].

\bibitem{Heeck:2019kgr}
J.~Heeck and V.~Takhistov, \emph{{Inclusive Nucleon Decay Searches as a
  Frontier of Baryon Number Violation}},
  \href{http://dx.doi.org/10.1103/PhysRevD.101.015005}{\emph{Phys. Rev. D} {\bf
  101} (2020) 015005}, [\href{http://arxiv.org/abs/1910.07647}{{\tt
  1910.07647}}].

\bibitem{Girmohanta:2019xya}
S.~Girmohanta and R.~Shrock, \emph{{Improved Lower Bounds on Partial Lifetimes
  for Nucleon Decay Modes}},
  \href{http://dx.doi.org/10.1103/PhysRevD.100.115025}{\emph{Phys. Rev. D} {\bf
  100} (2019) 115025}, [\href{http://arxiv.org/abs/1910.08106}{{\tt
  1910.08106}}].

\bibitem{Antusch:2020ztu}
S.~Antusch, C.~Hohl and V.~Susi\v{c}, \emph{{Employing nucleon decay as a
  fingerprint of SUSY GUT models using SusyTCProton}},
  \href{http://dx.doi.org/10.1007/JHEP06(2021)022}{\emph{JHEP} {\bf 06} (2021)
  022}, [\href{http://arxiv.org/abs/2011.15026}{{\tt 2011.15026}}].

\bibitem{He:2021sbl}
X.-G. He and X.-D. Ma, \emph{{$\Delta B=2$ neutron decay into antiproton mode
  $n\to \bar pe^+\nu(\bar\nu)$}},
  \href{http://dx.doi.org/10.1016/j.physletb.2021.136298}{\emph{Phys. Lett. B}
  {\bf 817} (2021) 136298}, [\href{http://arxiv.org/abs/2101.01405}{{\tt
  2101.01405}}].

\bibitem{Fajfer:2023gfi}
S.~Fajfer and M.~Sadl, \emph{{Decay dynamics of $N\to \ell \pi $ and $N\to \ell
  \gamma$}}, \href{http://dx.doi.org/10.1103/PhysRevD.108.015011}{\emph{Phys.
  Rev. D} {\bf 108} (2023) 015011},
  [\href{http://arxiv.org/abs/2304.00825}{{\tt 2304.00825}}].

\bibitem{Beneito:2023xbk}
A.~B. Beneito, I, J.~Gargalionis, J.~Herrero-Garcia, A.~Santamaria and M.~A.
  Schmidt, \emph{{An EFT approach to baryon number violation: lower limits on
  the new physics scale and correlations between nucleon decay modes}},
  \href{http://dx.doi.org/10.1007/JHEP07(2024)004}{\emph{JHEP} {\bf 07} (2024)
  004}, [\href{http://arxiv.org/abs/2312.13361}{{\tt 2312.13361}}].

\bibitem{Gargalionis:2024nij}
J.~Gargalionis, J.~Herrero-Garc\'\i{}a and M.~A. Schmidt,
  \emph{{Model-independent estimates for loop-induced baryon-number-violating
  nucleon decays}},
  \href{http://dx.doi.org/10.1007/JHEP06(2024)182}{\emph{JHEP} {\bf 06} (2024)
  182}, [\href{http://arxiv.org/abs/2401.04768}{{\tt 2401.04768}}].

\bibitem{Beneke:2024hox}
M.~Beneke, G.~Finauri and A.~A. Petrov, \emph{{Indirect constraints on third
  generation baryon number violation}},
  \href{http://dx.doi.org/10.1007/JHEP09(2024)090}{\emph{JHEP} {\bf 09} (2024)
  090}, [\href{http://arxiv.org/abs/2404.09642}{{\tt 2404.09642}}].

\bibitem{Gisbert:2024sjw}
H.~Gisbert, A.~Rodr\'\i{}guez-S\'anchez and L.~Vale~Silva, \emph{{Baryon number
  violation with top quark operators in the SMEFT}},
  \href{http://arxiv.org/abs/2409.00218}{{\tt 2409.00218}}.

\bibitem{FileviezPerez:2022ypk}
P.~Fileviez~Perez et~al., \emph{{On Baryon and Lepton Number Violation}},
  \href{http://arxiv.org/abs/2208.00010}{{\tt 2208.00010}}.

\bibitem{Dev:2022jbf}
P.~S.~B. Dev et~al., \emph{{Searches for baryon number violation in neutrino
  experiments: a white paper}},
  \href{http://dx.doi.org/10.1088/1361-6471/ad1658}{\emph{J. Phys. G} {\bf 51}
  (2024) 033001}, [\href{http://arxiv.org/abs/2203.08771}{{\tt 2203.08771}}].

\bibitem{Ohlsson:2023ddi}
T.~Ohlsson, \emph{{Proton decay}},
  \href{http://dx.doi.org/10.1016/j.nuclphysb.2023.116268}{\emph{Nucl. Phys. B}
  {\bf 993} (2023) 116268}, [\href{http://arxiv.org/abs/2306.02401}{{\tt
  2306.02401}}].

\bibitem{Davoudiasl:2014gfa}
H.~Davoudiasl, \emph{{Nucleon Decay into a Dark Sector}},
  \href{http://dx.doi.org/10.1103/PhysRevLett.114.051802}{\emph{Phys. Rev.
  Lett.} {\bf 114} (2015) 051802}, [\href{http://arxiv.org/abs/1409.4823}{{\tt
  1409.4823}}].

\bibitem{Helo:2018bgb}
J.~C. Helo, M.~Hirsch and T.~Ota, \emph{{Proton decay and light sterile
  neutrinos}}, \href{http://dx.doi.org/10.1007/JHEP06(2018)047}{\emph{JHEP}
  {\bf 06} (2018) 047}, [\href{http://arxiv.org/abs/1803.00035}{{\tt
  1803.00035}}].

\bibitem{Barducci:2018rlx}
D.~Barducci, M.~Fabbrichesi and E.~Gabrielli, \emph{{Neutral Hadrons
  Disappearing into the Darkness}},
  \href{http://dx.doi.org/10.1103/PhysRevD.98.035049}{\emph{Phys. Rev. D} {\bf
  98} (2018) 035049}, [\href{http://arxiv.org/abs/1806.05678}{{\tt
  1806.05678}}].

\bibitem{Heeck:2020nbq}
J.~Heeck, \emph{{Light particles with baryon and lepton numbers}},
  \href{http://dx.doi.org/10.1016/j.physletb.2020.136043}{\emph{Phys. Lett. B}
  {\bf 813} (2021) 136043}, [\href{http://arxiv.org/abs/2009.01256}{{\tt
  2009.01256}}].

\bibitem{Davoudiasl:2023peu}
H.~Davoudiasl, \emph{{Stellar signals of a baryon-number-violating long-range
  force}}, \href{http://dx.doi.org/10.1103/PhysRevD.108.015023}{\emph{Phys.
  Rev. D} {\bf 108} (2023) 015023},
  [\href{http://arxiv.org/abs/2304.06071}{{\tt 2304.06071}}].

\bibitem{Liang:2023yta}
J.-H. Liang, Y.~Liao, X.-D. Ma and H.-L. Wang, \emph{{Dark sector effective
  field theory}}, \href{http://dx.doi.org/10.1007/JHEP12(2023)172}{\emph{JHEP}
  {\bf 12} (2023) 172}, [\href{http://arxiv.org/abs/2309.12166}{{\tt
  2309.12166}}].

\bibitem{Fridell:2023tpb}
K.~Fridell, C.~Hati and V.~Takhistov, \emph{{Noncanonical nucleon decays as
  window into light new physics}},
  \href{http://dx.doi.org/10.1103/PhysRevD.110.L031701}{\emph{Phys. Rev. D}
  {\bf 110} (2024) L031701}, [\href{http://arxiv.org/abs/2312.13740}{{\tt
  2312.13740}}].

\bibitem{Li:2024liy}
T.~Li, M.~A. Schmidt and C.-Y. Yao, \emph{{Baryon-number-violating nucleon
  decays in ALP effective field theories}},
  \href{http://dx.doi.org/10.1007/JHEP08(2024)221}{\emph{JHEP} {\bf 08} (2024)
  221}, [\href{http://arxiv.org/abs/2406.11382}{{\tt 2406.11382}}].

\bibitem{Song:2023lxf}
H.~Song, H.~Sun and J.-H. Yu, \emph{{Effective field theories of axion, ALP and
  dark photon}}, \href{http://dx.doi.org/10.1007/JHEP01(2024)161}{\emph{JHEP}
  {\bf 01} (2024) 161}, [\href{http://arxiv.org/abs/2305.16770}{{\tt
  2305.16770}}].

\bibitem{Song:2023jqm}
H.~Song, H.~Sun and J.-H. Yu, \emph{{Complete EFT operator bases for dark
  matter and weakly-interacting light particle}},
  \href{http://dx.doi.org/10.1007/JHEP05(2024)103}{\emph{JHEP} {\bf 05} (2024)
  103}, [\href{http://arxiv.org/abs/2306.05999}{{\tt 2306.05999}}].

\bibitem{Grojean:2023tsd}
C.~Grojean, J.~Kley and C.-Y. Yao, \emph{{Hilbert series for ALP EFTs}},
  \href{http://dx.doi.org/10.1007/JHEP11(2023)196}{\emph{JHEP} {\bf 11} (2023)
  196}, [\href{http://arxiv.org/abs/2307.08563}{{\tt 2307.08563}}].

\bibitem{delAguila:2008ir}
F.~del Aguila, S.~Bar-Shalom, A.~Soni and J.~Wudka, \emph{{Heavy Majorana
  Neutrinos in the Effective Lagrangian Description: Application to Hadron
  Colliders}},
  \href{http://dx.doi.org/10.1016/j.physletb.2008.11.031}{\emph{Phys. Lett. B}
  {\bf 670} (2009) 399--402}, [\href{http://arxiv.org/abs/0806.0876}{{\tt
  0806.0876}}].

\bibitem{Aparici:2009fh}
A.~Aparici, K.~Kim, A.~Santamaria and J.~Wudka, \emph{{Right-handed neutrino
  magnetic moments}},
  \href{http://dx.doi.org/10.1103/PhysRevD.80.013010}{\emph{Phys. Rev. D} {\bf
  80} (2009) 013010}, [\href{http://arxiv.org/abs/0904.3244}{{\tt 0904.3244}}].

\bibitem{Bhattacharya:2015vja}
S.~Bhattacharya and J.~Wudka, \emph{{Dimension-seven operators in the standard
  model with right handed neutrinos}},
  \href{http://dx.doi.org/10.1103/PhysRevD.94.055022}{\emph{Phys. Rev. D} {\bf
  94} (2016) 055022}, [\href{http://arxiv.org/abs/1505.05264}{{\tt
  1505.05264}}].

\bibitem{Liao:2016qyd}
Y.~Liao and X.-D. Ma, \emph{{Operators up to Dimension Seven in Standard Model
  Effective Field Theory Extended with Sterile Neutrinos}},
  \href{http://dx.doi.org/10.1103/PhysRevD.96.015012}{\emph{Phys. Rev. D} {\bf
  96} (2017) 015012}, [\href{http://arxiv.org/abs/1612.04527}{{\tt
  1612.04527}}].

\bibitem{Chala:2020vqp}
M.~Chala and A.~Titov, \emph{{One-loop matching in the SMEFT extended with a
  sterile neutrino}},
  \href{http://dx.doi.org/10.1007/JHEP05(2020)139}{\emph{JHEP} {\bf 05} (2020)
  139}, [\href{http://arxiv.org/abs/2001.07732}{{\tt 2001.07732}}].

\bibitem{Li:2020lba}
T.~Li, X.-D. Ma and M.~A. Schmidt, \emph{{General neutrino interactions with
  sterile neutrinos in light of coherent neutrino-nucleus scattering and meson
  invisible decays}},
  \href{http://dx.doi.org/10.1007/JHEP07(2020)152}{\emph{JHEP} {\bf 07} (2020)
  152}, [\href{http://arxiv.org/abs/2005.01543}{{\tt 2005.01543}}].

\bibitem{Li:2021tsq}
H.-L. Li, Z.~Ren, M.-L. Xiao, J.-H. Yu and Y.-H. Zheng, \emph{{Operator bases
  in effective field theories with sterile neutrinos: d \ensuremath{\leq} 9}},
  \href{http://dx.doi.org/10.1007/JHEP11(2021)003}{\emph{JHEP} {\bf 11} (2021)
  003}, [\href{http://arxiv.org/abs/2105.09329}{{\tt 2105.09329}}].

\bibitem{Hall:1983id}
L.~J. Hall and M.~Suzuki, \emph{{Explicit R-Parity Breaking in Supersymmetric
  Models}}, \href{http://dx.doi.org/10.1016/0550-3213(84)90513-3}{\emph{Nucl.
  Phys. B} {\bf 231} (1984) 419--444}.

\bibitem{Chang:1996sw}
D.~Chang and W.-Y. Keung, \emph{{New limits on R-parity breakings in
  supersymmetric standard models}},
  \href{http://dx.doi.org/10.1016/S0370-2693(96)01271-3}{\emph{Phys. Lett. B}
  {\bf 389} (1996) 294--298}, [\href{http://arxiv.org/abs/hep-ph/9608313}{{\tt
  hep-ph/9608313}}].

\bibitem{Domingo:2024qoj}
F.~Domingo, H.~K. Dreiner, D.~K\"ohler, S.~Nangia and A.~Shah, \emph{{A novel
  proton decay signature at DUNE, JUNO, and Hyper-K}},
  \href{http://dx.doi.org/10.1007/JHEP05(2024)258}{\emph{JHEP} {\bf 05} (2024)
  258}, [\href{http://arxiv.org/abs/2403.18502}{{\tt 2403.18502}}].

\bibitem{Choi:1996nk}
K.~Choi, E.~J. Chun and J.~S. Lee, \emph{{Proton decay with a light gravitino
  or axino}}, \href{http://dx.doi.org/10.1103/PhysRevD.55.R3924}{\emph{Phys.
  Rev. D} {\bf 55} (1997) 3924--3926},
  [\href{http://arxiv.org/abs/hep-ph/9611285}{{\tt hep-ph/9611285}}].

\bibitem{Choi:1998ak}
K.~Choi, K.~Hwang and J.~S. Lee, \emph{{Constraints on R-parity and B violating
  couplings in gauge mediated supersymmetry breaking models}},
  \href{http://dx.doi.org/10.1016/S0370-2693(98)00371-2}{\emph{Phys. Lett. B}
  {\bf 428} (1998) 129--135}, [\href{http://arxiv.org/abs/hep-ph/9802323}{{\tt
  hep-ph/9802323}}].

\bibitem{Claudson:1981gh}
M.~Claudson, M.~B. Wise and L.~J. Hall, \emph{{Chiral Lagrangian for Deep Mine
  Physics}}, \href{http://dx.doi.org/10.1016/0550-3213(82)90401-1}{\emph{Nucl.
  Phys. B} {\bf 195} (1982) 297--307}.

\bibitem{Jenkins:1991ne}
E.~E. Jenkins and A.~V. Manohar, \emph{{Baryon chiral perturbation theory}},
  in \emph{{Workshop on Effective Field Theories of the Standard Model}}, 10,
  1991.

\bibitem{McGrew:1999nd}
C.~McGrew et~al., \emph{{Search for nucleon decay using the IMB-3 detector}},
  \href{http://dx.doi.org/10.1103/PhysRevD.59.052004}{\emph{Phys. Rev. D} {\bf
  59} (1999) 052004}.

\bibitem{Atre:2009rg}
A.~Atre, T.~Han, S.~Pascoli and B.~Zhang, \emph{{The Search for Heavy Majorana
  Neutrinos}},
  \href{http://dx.doi.org/10.1088/1126-6708/2009/05/030}{\emph{JHEP} {\bf 05}
  (2009) 030}, [\href{http://arxiv.org/abs/0901.3589}{{\tt 0901.3589}}].

\bibitem{Abdullahi:2022jlv}
A.~M. Abdullahi et~al., \emph{{The present and future status of heavy neutral
  leptons}}, \href{http://dx.doi.org/10.1088/1361-6471/ac98f9}{\emph{J. Phys.
  G} {\bf 50} (2023) 020501}, [\href{http://arxiv.org/abs/2203.08039}{{\tt
  2203.08039}}].

\bibitem{Minkowski:1977sc}
P.~Minkowski, \emph{{$\mu \to e\gamma$ at a Rate of One Out of $10^{9}$ Muon
  Decays?}}, \href{http://dx.doi.org/10.1016/0370-2693(77)90435-X}{\emph{Phys.
  Lett. B} {\bf 67} (1977) 421--428}.

\bibitem{Boyarsky:2020dzc}
A.~Boyarsky, M.~Ovchynnikov, O.~Ruchayskiy and V.~Syvolap, \emph{{Improved big
  bang nucleosynthesis constraints on heavy neutral leptons}},
  \href{http://dx.doi.org/10.1103/PhysRevD.104.023517}{\emph{Phys. Rev. D} {\bf
  104} (2021) 023517}, [\href{http://arxiv.org/abs/2008.00749}{{\tt
  2008.00749}}].

\bibitem{Bondarenko:2021cpc}
K.~Bondarenko, A.~Boyarsky, J.~Klaric, O.~Mikulenko, O.~Ruchayskiy, V.~Syvolap
  et~al., \emph{{An allowed window for heavy neutral leptons below the kaon
  mass}}, \href{http://dx.doi.org/10.1007/JHEP07(2021)193}{\emph{JHEP} {\bf 07}
  (2021) 193}, [\href{http://arxiv.org/abs/2101.09255}{{\tt 2101.09255}}].

\bibitem{Jenkins:2017jig}
E.~E. Jenkins, A.~V. Manohar and P.~Stoffer, \emph{{Low-Energy Effective Field
  Theory below the Electroweak Scale: Operators and Matching}},
  \href{http://dx.doi.org/10.1007/JHEP03(2018)016}{\emph{JHEP} {\bf 03} (2018)
  016}, [\href{http://arxiv.org/abs/1709.04486}{{\tt 1709.04486}}].

\bibitem{Liao:2020zyx}
Y.~Liao, X.-D. Ma and Q.-Y. Wang, \emph{{Extending low energy effective field
  theory with a complete set of dimension-7 operators}},
  \href{http://dx.doi.org/10.1007/JHEP08(2020)162}{\emph{JHEP} {\bf 08} (2020)
  162}, [\href{http://arxiv.org/abs/2005.08013}{{\tt 2005.08013}}].

\bibitem{Fonseca:2017lem}
R.~M. Fonseca, \emph{{The Sym2Int program: going from symmetries to
  interactions}},
  \href{http://dx.doi.org/10.1088/1742-6596/873/1/012045}{\emph{J. Phys. Conf.
  Ser.} {\bf 873} (2017) 012045}, [\href{http://arxiv.org/abs/1703.05221}{{\tt
  1703.05221}}].

\bibitem{Fonseca:2019yya}
R.~M. Fonseca, \emph{{Enumerating the operators of an effective field theory}},
  \href{http://dx.doi.org/10.1103/PhysRevD.101.035040}{\emph{Phys. Rev. D} {\bf
  101} (2020) 035040}, [\href{http://arxiv.org/abs/1907.12584}{{\tt
  1907.12584}}].

\bibitem{Dreiner:2008tw}
H.~K. Dreiner, H.~E. Haber and S.~P. Martin, \emph{{Two-component spinor
  techniques and Feynman rules for quantum field theory and supersymmetry}},
  \href{http://dx.doi.org/10.1016/j.physrep.2010.05.002}{\emph{Phys. Rept.}
  {\bf 494} (2010) 1--196}, [\href{http://arxiv.org/abs/0812.1594}{{\tt
  0812.1594}}].

\bibitem{Alonso:2014zka}
R.~Alonso, H.-M. Chang, E.~E. Jenkins, A.~V. Manohar and B.~Shotwell,
  \emph{{Renormalization group evolution of dimension-six baryon number
  violating operators}},
  \href{http://dx.doi.org/10.1016/j.physletb.2014.05.065}{\emph{Phys. Lett. B}
  {\bf 734} (2014) 302--307}, [\href{http://arxiv.org/abs/1405.0486}{{\tt
  1405.0486}}].

\bibitem{Abbott:1980zj}
L.~F. Abbott and M.~B. Wise, \emph{{The Effective Hamiltonian for Nucleon
  Decay}}, \href{http://dx.doi.org/10.1103/PhysRevD.22.2208}{\emph{Phys. Rev.
  D} {\bf 22} (1980) 2208}.

\bibitem{Celis:2017hod}
A.~Celis, J.~Fuentes-Martin, A.~Vicente and J.~Virto, \emph{{DsixTools: The
  Standard Model Effective Field Theory Toolkit}},
  \href{http://dx.doi.org/10.1140/epjc/s10052-017-4967-6}{\emph{Eur. Phys. J.
  C} {\bf 77} (2017) 405}, [\href{http://arxiv.org/abs/1704.04504}{{\tt
  1704.04504}}].

\bibitem{Fuentes-Martin:2020zaz}
J.~Fuentes-Martin, P.~Ruiz-Femenia, A.~Vicente and J.~Virto, \emph{{DsixTools
  2.0: The Effective Field Theory Toolkit}},
  \href{http://dx.doi.org/10.1140/epjc/s10052-020-08778-y}{\emph{Eur. Phys. J.
  C} {\bf 81} (2021) 167}, [\href{http://arxiv.org/abs/2010.16341}{{\tt
  2010.16341}}].

\bibitem{Aebischer:2018bkb}
J.~Aebischer, J.~Kumar and D.~M. Straub, \emph{{Wilson: a Python package for
  the running and matching of Wilson coefficients above and below the
  electroweak scale}},
  \href{http://dx.doi.org/10.1140/epjc/s10052-018-6492-7}{\emph{Eur. Phys. J.
  C} {\bf 78} (2018) 1026}, [\href{http://arxiv.org/abs/1804.05033}{{\tt
  1804.05033}}].

\bibitem{Liao:2016hru}
Y.~Liao and X.-D. Ma, \emph{{Renormalization Group Evolution of Dimension-seven
  Baryon- and Lepton-number-violating Operators}},
  \href{http://dx.doi.org/10.1007/JHEP11(2016)043}{\emph{JHEP} {\bf 11} (2016)
  043}, [\href{http://arxiv.org/abs/1607.07309}{{\tt 1607.07309}}].

\bibitem{Zhang:2023ndw}
D.~Zhang, \emph{{Revisiting renormalization group equations of the SMEFT
  dimension-seven operators}},
  \href{http://dx.doi.org/10.1007/JHEP02(2024)133}{\emph{JHEP} {\bf 02} (2024)
  133}, [\href{http://arxiv.org/abs/2310.11055}{{\tt 2310.11055}}].

\bibitem{Jenkins:2017dyc}
E.~E. Jenkins, A.~V. Manohar and P.~Stoffer, \emph{{Low-Energy Effective Field
  Theory below the Electroweak Scale: Anomalous Dimensions}},
  \href{http://dx.doi.org/10.1007/JHEP01(2018)084}{\emph{JHEP} {\bf 01} (2018)
  084}, [\href{http://arxiv.org/abs/1711.05270}{{\tt 1711.05270}}].

\bibitem{Yoo:2021gql}
J.-S. Yoo, Y.~Aoki, P.~Boyle, T.~Izubuchi, A.~Soni and S.~Syritsyn,
  \emph{{Proton decay matrix elements on the lattice at physical pion mass}},
  \href{http://dx.doi.org/10.1103/PhysRevD.105.074501}{\emph{Phys. Rev. D} {\bf
  105} (2022) 074501}, [\href{http://arxiv.org/abs/2111.01608}{{\tt
  2111.01608}}].

\bibitem{deBlas:2017xtg}
J.~de~Blas, J.~C. Criado, M.~Perez-Victoria and J.~Santiago, \emph{{Effective
  description of general extensions of the Standard Model: the complete
  tree-level dictionary}},
  \href{http://dx.doi.org/10.1007/JHEP03(2018)109}{\emph{JHEP} {\bf 03} (2018)
  109}, [\href{http://arxiv.org/abs/1711.10391}{{\tt 1711.10391}}].

\bibitem{Li:2022abx}
H.-L. Li, Y.-H. Ni, M.-L. Xiao and J.-H. Yu, \emph{{The bottom-up EFT: complete
  UV resonances of the SMEFT operators}},
  \href{http://dx.doi.org/10.1007/JHEP11(2022)170}{\emph{JHEP} {\bf 11} (2022)
  170}, [\href{http://arxiv.org/abs/2204.03660}{{\tt 2204.03660}}].

\bibitem{Li:2023cwy}
X.-X. Li, Z.~Ren and J.-H. Yub, \emph{{Complete tree-level dictionary between
  simplified BSM models and SMEFT d\ensuremath{\leq}7 operators}},
  \href{http://dx.doi.org/10.1103/PhysRevD.109.095041}{\emph{Phys. Rev. D} {\bf
  109} (2024) 095041}, [\href{http://arxiv.org/abs/2307.10380}{{\tt
  2307.10380}}].

\bibitem{Beltran:2023ymm}
R.~Beltr\'an, R.~Cepedello and M.~Hirsch, \emph{{Tree-level UV completions for
  $N_R$SMEFT $d=6$ and $d=7$ operators}},
  \href{http://dx.doi.org/10.1007/JHEP08(2023)166}{\emph{JHEP} {\bf 08} (2023)
  166}, [\href{http://arxiv.org/abs/2306.12578}{{\tt 2306.12578}}].

\bibitem{Aebischer:2024csk}
J.~Aebischer, T.~Kapoor and J.~Kumar, \emph{{wilson: A package for
  renormalization group running in the SMEFT with Sterile Neutrinos}},
  \href{http://arxiv.org/abs/2411.07220}{{\tt 2411.07220}}.

\bibitem{Mihaila:2012pz}
L.~N. Mihaila, J.~Salomon and M.~Steinhauser, \emph{{Renormalization constants
  and beta functions for the gauge couplings of the Standard Model to
  three-loop order}},
  \href{http://dx.doi.org/10.1103/PhysRevD.86.096008}{\emph{Phys. Rev. D} {\bf
  86} (2012) 096008}, [\href{http://arxiv.org/abs/1208.3357}{{\tt 1208.3357}}].

\bibitem{Bali:2022qja}
{\scshape RQCD} collaboration, G.~S. Bali, S.~Collins, W.~S\"oldner and
  S.~Weish\"aupl, \emph{{Leading order mesonic and baryonic SU(3) low energy
  constants from Nf=3 lattice QCD}},
  \href{http://dx.doi.org/10.1103/PhysRevD.105.054516}{\emph{Phys. Rev. D} {\bf
  105} (2022) 054516}, [\href{http://arxiv.org/abs/2201.05591}{{\tt
  2201.05591}}].

\bibitem{Workman:2022ynf}
{\scshape Particle Data Group} collaboration, R.~L. Workman et~al.,
  \emph{{Review of Particle Physics}},
  \href{http://dx.doi.org/10.1093/ptep/ptac097}{\emph{PTEP} {\bf 2022} (2022)
  083C01}.

\bibitem{Antusch:2001ck}
S.~Antusch, M.~Drees, J.~Kersten, M.~Lindner and M.~Ratz, \emph{{Neutrino mass
  operator renormalization revisited}},
  \href{http://dx.doi.org/10.1016/S0370-2693(01)01127-3}{\emph{Phys. Lett. B}
  {\bf 519} (2001) 238--242}, [\href{http://arxiv.org/abs/hep-ph/0108005}{{\tt
  hep-ph/0108005}}].

\end{thebibliography}\endgroup

\end{document}